\documentclass[aps,prd,longbibliography,nofootinbib,square,sort&compress,10pt,fleqn,showpacs]{revtex4-1}
\usepackage[titletoc]{appendix}
\usepackage[english]{babel}
\usepackage{amssymb}
\usepackage{centernot}
\usepackage{upgreek}
\usepackage{yfonts}
\usepackage[fleqn]{amsmath}
\usepackage{empheq}
\usepackage[T1]{fontenc}
\usepackage[x11names]{xcolor}
\usepackage{empheq}
\usepackage{color}
\allowdisplaybreaks
\usepackage{mathtools}
\usepackage{eucal}
\usepackage{graphicx}
\usepackage{float}
\usepackage[pdfencoding=auto,psdextra]{hyperref}
\usepackage{bookmark}
\usepackage{url}
\usepackage{breakurl}
\usepackage{epsfig}
\usepackage{bm}
\usepackage[utf8x,latin1]{inputenc} 
\usepackage{emerald}
\usepackage{slantsc}
\usepackage{array}

\let\tempcal \mathcal
\usepackage{dutchcal}
\let\dutchcal\mathcal
\let\mathcal\tempcal

\makeatletter
\newcommand{\mathleft}{\@fleqntrue\@mathmargin0pt}
\newcommand{\mathcenter}{\@fleqnfalse}
\makeatother

\def\sst{\scriptscriptstyle}
\def\be{\mathleft\begin{equation}}
\def\ee{\end{equation}}
\def\ba{\mathleft\begin{eqnarray}}
\def\ea{\end{eqnarray}}
\def\Asu{\begin{subequations}}
\def\esu{\end{subequations}}

\def\F{{\mathcal F}}
\def\I{{\mathcal I}}
\def\L{{\mathcal L}}

\def\R{{\mathcal R}}

\def\Sc{{\mathcal S}}

\def\A{{\sst{\rm A}}}
\def\B{{\sst{\rm B}}}
\def\C{{\sst{\rm C}}}

\def\M{{\cal M}}
\def\N{{\cal N}}

\def\R{{\cal R}}

\def\a{\alpha}
\def\b{\beta}
\def\g{\gamma}     

\def\d{\delta}

\def\la{\label}
\def\pd{\partial}
\def\le{\left}
\def\ri{\right}

\def\Min{{\dutchcal M}}

\def\ss{{\mathsf{s}}}
\def\mm{{\mathsf{m}}}
\def\nn{{\mathsf{n}}}
\def\mM{{\mathfrak{M}}}

\begin{document}
\title{Post-Newtonian Lagrangian of an $N$-body System with Arbitrary Mass and Spin Multipoles}
\author{Sergei M. Kopeikin}
\affiliation{Department of Physics \& Astronomy, University of Missouri, 322 Physics Bldg., Columbia, Missouri 65211, USA}
\email{E-mail: kopeikins@missouri.edu}
\begin{abstract}
The present paper derives the post-Newtonian Lagrangian of translational motion of N arbitrary-structured bodies with all mass and spin multipoles in a scalar-tensor theory of gravity. The multipoles depend on time and evolve in accordance with their own dynamic equations of motion. The Lagrangian is retrieved from the post-Newtonian equations of motion by solving the inverse problem of the Lagrangian mechanics and generalizes a well-known Lagrangian of pole-dipole-quadrupole massive particles to the particles of higher multipolarity. Analytic treatment of the higher-order multipole contributions is important for more rigorous computation of gravitational waveform of inspiralling compact binaries at the latest stage of their orbital evolution before merger when tidal and rotational deformations of stars are no longer small and rapidly change in time. The Lagrangian of an $N$-body system with arbitrary mass and spin multipoles is instrumental for formulation of the post-Newtonian conservation laws of energy, momenta and the integrals of the center of mass.   
\end{abstract}
%\keywords{}
\maketitle
\tableofcontents
\date{\today}
%\newpage
\section{Introduction}

Modern astrophysics has reached a significant progress in understanding the nature of ultra-relativistic objects like neutron stars and black holes  through precise and accurate measurements of compact binary systems conducted in different bandwidths of electromagnetic spectrum spreading out from radio to gamma-ray wavelength \citep{kramer2009CQGra,Hilditch2001,Casares2017,Dubus2013}. Gravitational wave astronomy had opened a new window to study dynamics of the binary systems with compact stars having fast intrinsic rotation and ultra-short orbital periods \citep{thorne_nobel}. Such binary systems consist either of two neutron stars or a neutron star (NS) and a black hole (BH) or two black holes which are significantly deformed by centrifugal and tidal forces. Moreover, the gravitational radiation-reaction force causes a rapid secular decrease of the orbital period and separation between the two bodies while the intrinsic angular momentum (spin) of each body is subject to wide precession which changes the rotational and tidal deformations of the bodies in the course of time. Precise measurement of these deformations by means of gravitational wave detectors is a key to more profound understanding of the interior structure of compact astrophysical objects and the equation of state (EoS) of hadronic matter at super-critical density and pressure of the degenerate Bose condensate of baryons \citep{gruber2016} and/or quark-gluon plasma \citep{SATZ20114} just before the merger. It also provides a  deeper insight into non-linear geometrodynamics of binary black hole \citep{thorne2014PhyU,destounis2019}.

Information about the equation of state and other fundamental physical parameters of matter of the tidally-disturbed bodies in the binary system (like the Love numbers, viscosity, shear stress, etc.) is mapped through the gravity field equations to a set of tidally-induced multipole moments and their time derivatives \citep{Nagar_2009PhRvD,Poisson_2009PhRvD}. There are two, physically-meaningful types of the multipole moments in general relativity called mass and spin multipoles \citep{bld1986}. Scalar-tensor theory of gravity contains one more type of multipoles called scalar-field multipoles \citep{DamEsFar,kovl_2004}. The multipoles parameterize the equations of orbital evolution of the inspiralling binary systems. The mass multipole moments appear already in the Newtonian approximation while the spin and scalar field multipoles come up for the first time at the first post-Newtonian (PN) approximation. Translational equations of motion of extended, arbitrary-structured bodies in an $N$-body system with the full account for the intrinsic mass and spin multipoles of the bodies have been derived in general relativity \citep{racine_2005PhRvD,racine2013PhRvD} and in scalar-tensor theory of gravity \citep{kovl_2004,xie_2010AcPSl,k2019PRD}. These equations extend the domain of analytic applicability of equations of motion of pole-dipole massive particles in higher PN approximations from inspiral to coalescence stage of the orbital evolution of compact binary systems. 

General physical characteristics of a gravitational $N$-body system can be, in principle, understood without explicit integration of equations of motion if the first integrals of the equations are known. The first integrals are functions of dynamic variables and their time derivatives in the configuration space of the $N$-body system. The dynamic variables are coordinates of the center of mass and multipole moments of the bodies which are functions of time. The first integrals can be obtained by applying the N\"other theorem to the Lagrangian of an $N$-body system \citep{Petrov_2017book}. The knowledge of the analytic form of the first integrals is important for formulating the conservation laws of energy, angular and linear momenta, and the integrals of the center of mass of the $N$-body system. Knowledge of their analytic form is critical for minimization of the errors of numerical integration of the equations of motion of inspiralling compact binaries when computing the tidal and tidal-resonant effects \citep{Kokkotas_1995MNRAS} as well as for refining the statistical estimates of the parameters of gravitational waveforms \citep{babak_2008CQGra}.  

In principle, the Lagrangian of an $N$-body system can be derived directly from the Hilbert action of general relativity (or any other, viable alternative gravity theory) as proposed by \citet{infeld_RRMP1957} who extended similar idea of \citet{fokker_1929ZPhy} from electromagnetism to general relativity. Fokker-Infeld method was applied by \citet{vab} who derived the post-Newtonian Lagrangian of an $N$-body system consisting of rigidly rotating and oblated bodies possessing the intrinsic spin-dipole and mass quadrupole moments.
The Fokker-Infeld Lagrangian of an $N$-body system was also derived for structureless (non-spinning) point-particles \footnote{The structureless and non-spinning point-particles correspond to spherically-symmetric and non-rotating extended bodies.} in the second PN approximation of general relativity with different techniques \citep{OhtaOKH1,OhtaOKH2,OhtaOKH3,OhtaOKH4,Ohta1988PThPh,Damour_1985GReGr,Damour_1987book}. Similar method was used to derive the Arnovitt-Deser-Misner (ADM) Hamiltonian of point-like particles of an $N$-body system in the third \citep{Andrade_2001,Blanchet_2003} and forth PN approximations \citep{Damour_PRD2014,Bernard_PRD2016,Damour_2016PhRvD93h4014D}. Corresponding ADM Hamiltonian of an $N$-body problem for spinning point-like particles has been derived by \citet{Jaranowski_PRD2015,Schaefer2018} who employed the regularization technique to deal with the formally-divergent integrals from distributions which are used to model the point-like bodies of the $N$-body system \citep{spin_Hamiltonian_schaefer}.  

An independent method to get the Lagrangian ${\cal L}$ of an $N$-body system is based on solving the inverse problem of the Lagrangian mechanics by writing down a system of self-adjoint differential equations in partial derivatives for, yet unknown, function ${\cal L}$ such that its solution, if it exists, yields the variational derivative corresponding to the dynamical equations of motion of the $N$-body system. The Lagrangian is defined not uniquely but up to a total time derivative from an arbitrary scalar function \citep{Petrov_2017book}. Mathematical foundations justifying the existence and uniqueness of the solution of the inverse problem for a system of ordinary differential equations of the second order have been worked out by \citet{douglas_1941}, significantly rectified and improved by \citet{Santilli_1978} and further refined and enhanced by other researchers -- see, for instance, \citep{Saunders_2010,Zenkov_2015} and reference therein. Solution of the system of the self-adjoint differential equations relies upon a rigorous mathematical background but is hardly applicable for practical computation of the Lagrangian of a gravitational $N$-body problem in the post-Newtonian approximations due to the complexity of the equations. Technically, it is much simpler to look for the Lagrangian in the form of a linear combination of some scalar functions with yet undetermined numerical coefficients, then, to take the variational derivative from this combination and equate it to the original equations of motion of the $N$-body problem. Identification and comparison of similar terms yield a system of algebraic equations for the above-mentioned numerical coefficients which can be solved by applying a standard linear algebra technique. If such solution exists it immediately yields the post-Newtonian Lagrangian of the $N$-body system. This "linear algebra" approach to solving the inverse problem of the Lagrangian mechanics was applied in general relativity to build the Lagrangian of an $N$-body system made up of pole-dipole particles \citep{gk86,Damour_1985GReGr,Damour_1989MG5,Blanchet_2003}.

We notice that in spite of the enormous success in developing the post-Newtonian approximations for solving two-body problem of point-like masses \citep{Will_2011PNAS} the post-Newtonian Lagrangian of a gravitational $N$-body system consisting of arbitrary-structured, rotating and dynamically-evolving extended bodies has not been formulated so far. The problem is that such a Lagrangian should contain contributions from all mass and spin multipoles of the bodies and match analytically with the Lagrangian of the point-like pole-dipole particles. Some ideas of how to fulfill this task have been proposed at the Newtonian level in \citep{ANANDAN_2003,Zilio_2010CQG}. Brumberg \citep{vab} had employed the Fokker-Infeld approach combined together with Fock's method \citep{fockbook} in order to get the post-Newtonian Lagrangian for particles endowed with dipole and quadrupole moments defined in the global coordinates of an $N$-body system. These multipoles are not directly measurable and must be converted to the body-adapted local coordinates in order to get physically meaningful results \citep{fockbook,Damour_1987book,kopeikin_2011book,soffel_2019book}. Definite progress was made in deriving the Lagrangian for particles with spin \citep{damour_1982CR,Barker1986_acceleration,Barker_Oconnell1987JMP,spin_Hamiltonian_schaefer,wang_2011PhRvDW} and quadrupole moments \citep{Vines_Flanagan_PRD024046,steinhoff_2010PhRvD,Steinhoff_etal_PRD104028,Steinhoff_2015}. However, we are not aware of any other attempt in scientific literature aimed to build the post-Newtonian Lagrangian of an isolated gravitating system comprised of N arbitrary-structured extended bodies possessing all time-dependent mass and spin multipole moments beyond the quadrupole order. Therefore, the present paper focuses on solution of this problem in the first PN approximation of scalar-tensor theory of gravity. 

We find out the Lagrangian by solving the inverse problem of the Lagrangian mechanics with the post-Newtonian equations of motion of $N$ bodies derived in our previous paper \citep{k2019PRD}. Section \ref{ue345c} is a brief introduction to the post-Newtonian formalism of the local and global coordinate charts used for description of the dynamics of an astronomical $N$-body system. We also define in this section the post-Newtonian multipole moments as well as the mass and center of mass of an extended body from the $N$-body system. Section \ref{oq23x5} explains the configuration space of an $N$-body problem, describes the post-Newtonian equations of motion of arbitrary structured bodies and discusses the method of solving variational equations of the inverse problem of the Lagrangian mechanics. Section \ref{mx5ql1} describes preparatory transformations of the gravitational force from its original form given in section \ref{oq23x5} to that which is required for retrieving the Lagrangian by solving the inverse problem. Section \ref{z2a8gc} brings the Newtonian terms in the equations of motion to the Lagrangian form. The post-Newtonian gravitational force is transformed to the Lagrangian form in section \ref{in2q9ze}. A general formula for the Lagrangian of an $N$-body system is derived in section \ref{ct39z6}. Section \ref{jhru747} specifies  the generic Lagrangian of the previous section to the case of a binary system and analyzes its structure in monopole, dipole, and quadrupole approximations. It also compares our Lagrangian in the quadrupole-particle approximation with the results of previous researchers. Conclusions are given in section \ref{concl73746}. Appendix \ref{app1a} explains the virial theorems which are used for interconnecting the external and internal dynamic variables in the configuration space of an $N$-body system in the process of derivation of the Lagrangian. 

Mathematical symbols and notations used in the present paper are as follows:
\begin{itemize}
\item[--] the small Greek letters $\a,\b,\g,\ldots$ denote spacetime indices of tensors and run through values $0,1,2,3$, 
\item[--] the small Roman indices $i,j,k,\ldots$ denote spatial tensor indices and take values $1,2,3$,
\item[--] the capital Roman letters $L,K,N,S$ denote spatial tensor multi-indices, for example, the spatial tensor $T^L=T^{i_1i_2\ldots i_l}$, $T^{L-1}=T^{i_1i_2\ldots i_{l-1}}$, $T^{LN}=T^{i_1i_2\ldots i_lj_1j_2\ldots j_n}$, etc.,
\item[--] the capital Roman indices A,B,C label the bodies of an $N$-body system. Each of them takes values from the set $\{1,2,\ldots,{\rm N}\}$,  
\item[--] the Einstein summation rule is applied for repeated (dummy) indices and multi-indices, for example,  ${\cal P}^\a {\cal Q}_\a\equiv {\cal P}^0 {\cal Q}_0+{\cal P}^1 {\cal Q}_1+{\cal P}^2 {\cal Q}_2 + {\cal P}^3 {\cal Q}_3$, ${\cal P}^i {\cal Q}_i\equiv {\cal P}^1 {\cal Q}_1+{\cal P}^2 {\cal Q}_2 + {\cal P}^3 {\cal Q}_3$, ${\cal P}^L {\cal Q}_L={\cal P}^{i_1i_2\ldots i_l}{\cal Q}_{i_1i_2\ldots i_l}$, etc.,
\item[--] the Kronecker symbol $\delta_{ij}=\delta^{ij}=\delta^i_j=\delta^j_i$ is a unit matrix in 3-dimensional space,
\item[--] the Levi-Civita symbol $\varepsilon_{ijk}=\varepsilon^{ijk}$, is a fully antisymmetric tensor in 3-dimensional space with $\varepsilon_{123}=+1$, 
\item[--] $g_{\a\b}$ is a metric tensor on spacetime manifold,
\item[--] $\eta_{\a\b}={\rm diag}\{-1,+1,+1,+1\}$ is the Minkowski metric,
\item[--] $h_{\a\b}=g_{\a\b}-\eta_{\a\b}$ is the metric perturbation of the Minkowski spacetime,
\item[--] $w^\a=(u,w^i)$ are the local coordinates adapted to body A$\in\{1,2,\ldots,{\rm N}\}$ with $u$ being the local coordinate time. 
\item[--] $x^\a=\{t,x^i\}$ are the global (single chart) coordinates covering the entire spacetime in a single chart. 
\item[--] $\pd_i=\pd/\pd x^i$ is a partial derivative with respect to a spatial coordinate $x^i$,
\item[--] the multi-index partial derivatives with respect to coordinates $x^\a$ are denoted as $\partial_L\equiv\partial_{i_1\ldots i_l}=\pd_{i_1}\pd_{i_2}...\pd_{i_l}$; $\partial_{L-1}\equiv\partial_{i_1\ldots i_{l-1}}$; $\partial_{pL-1}\equiv\partial_{pi_1...i_{l-1}}$, etc.,
\item[--] tensor (Greek) indices of geometric objects on spacetime manifold are raised and lowered with the full metric $g_{\a\b}$,
\item[--] tensor (Greek) indices of the metric tensor perturbation $h_{\a\b}$ are raised and lowered with the Minkowski metric $\eta_{\a\b}$,
\item[--] the spatial (Roman) indices of geometric objects are raised and lowered with the Kronecker symbol $\delta^{ij}$,
\item[--] summation over all bodies of the $N$-body system is denoted as $\sum\limits_\A\equiv\sum\limits_{\A=1}^{\rm N}$, or $\sum\limits_{\B}\equiv\sum\limits_{{\B}=1}^{\rm N}$, etc.,
\item[--] summation over N-1 bodies of the $N$-body system that excludes, let say, body B is denoted $\sum\limits_{\A\not={\B}}$, 
\item[--] the factorial is $l!=l(l-1)(l-2)...2\cdot 1$, 
\item[--] the double factorial $l!!=l(l-2)...4\cdot 2$ if $l$ is even, and $l!!=l(l-2)...3\cdot 1$ if $l$ is odd,
\item[--] the round parentheses embracing a group of tensor indices denote full symmetrization, for example, $T_{(\a\b\g)}=\displaystyle\frac1{3!}\left(T_{\a\b\g}+T_{\b\g\a}+T_{\g\a\b}+T_{\b\a\g}+T_{\a\g\b}+T_{\g\b\a}\right)$, etc.\;,
\item[--] the square parentheses around a pair of tensor indices denote anti-symmetrization, for example, $T^{[\a\b]\g}=\displaystyle\frac12\left(T^{\a\b\g}-T^{\b\a\g}\right)$\;,etc.
\item[--] the angular brackets around spatial tensor indices denote a symmetric trace-free (STF) projection. For example, the STF projection $T_{<L>}$ of tensor $T_L$ is constructed from its symmetric part $S_L= T_{(L)}=T_{(i_1i_2...i_l)}$ by subtracting all the permissible traces \citep{thor,bld1986}
\ba\label{stfformula}
T_{<L>}&=&\sum_{n=0}^{[l/2]}\frac{(-1)^n}{2^nn!}\frac{l!}{(l-2n)!}\frac{(2l-2n-1)!!}{(2l-1)!!}\delta_{(i_1i_2...}\delta_{i_{2n-1}i_{2n}}S_{i_{2n+1}...i_l)j_1j_1...j_nj_n}\;,
\ea
where $[l/2]$ is the largest integer less than or equal to $l/2$.
\item[--] the STF spatial derivative is denoted by the angular parentheses around the STF indices, for example, $\pd_{<L>}\equiv\pd_{<i_1i_2...i_l>}$ or $\pd_{<K>}\equiv \pd_{<i_1i_2...i_k>}$\;.
\end{itemize}
Other notations will be introduced and explained in the main text of the paper as they appear. In particular, in most equations of the present paper we use a geometic system of units in which the fundamental speed, $c$, and the universal gravitational constant, $G$, are set equal to unity.
    
\section{Coordinate systems}\label{ue345c}

We consider an isolated astronomical system consisting of N extended, arbitrary-structured bodies which are bound by attractive gravitational forces. We label the bodies with the capital letters A,B,C each of which takes values from the set of integers $\{$1,2,...,N$\}$.
Each body A is characterized by yet unspecified but physically-admissible distribution of mass density $\rho$, internal thermodynamic energy $\Pi$ and stress tensor $\sigma^{ij}$. The bodies are assumed to be well-separated in space and consist of matter occupied a finite volume with a distinctive boundary. The present paper does not admit physical singularities of spacetime like black holes and assume that the bodies have weak self-gravity and move sufficiently slow in space. This is the realm of the PN approximations which are applied to solve the gravity field equations in relativistic celestial mechanics of the solar system \citep{kopeikin_2011book,soffel_2019book}, binary pulsars \citep{kramer2009CQGra} and for calculating templates of gravitational waves emitted by inspiralling compact binaries \citep{Blanchet_2002LRR,Will_2011PNAS}.  

An adequate mathematical description of dynamics of an $N$-body system requires introduction of N+1 coordinate systems \citep{bk89,bk-nc,dsx1}. The primary coordinate chart, $x^\a=(t,x^i)$, is global and covers the entire spacetime manifold while a local coordinate chart is introduced in the neighborhood of world line of each body A$\in\{1,2,...,{\rm N}\}$. From the point of view of the Lagrangian mechanics the global coordinates are necessary to parameterize the orbital motion of each body A in the $N$-body system in terms of coordinates $x^i_\A$, velocity $v^i_\A=\dot{x}^i_\A$ and acceleration $a^i_\A$ of its  center of mass which are treated as external dynamic variables in the configuration space. The local coordinates, $w^\a_\A=(u_\A,w^i_\A)$, are introduced for each body A in order to describe the temporal evolution of its mass and spin multipole moments considered as internal dynamic variables. The global and local coordinates of body A overlap in a neighborhood of the body and can be transformed one to another \citep{Kopejkin_1988CeMec,kopeikin_2011book,soffel_2019book} so that the local coordinates can be represented as function of the global coordinates, $w^\a_\A=w^\a_\A(t,{\bm x})$, and vice versa, $x^\a=x^\a(u_\A,{\bm w}_\A)$. In what follows, we shall drop the body label A attached to the local coordinates of the body and identify $w^\a\equiv w^\a_\A$ if there is no confusion with the local coordinates of another body B$\neq$A. In all other cases, the local coordinates are tagged with the label of the corresponding body. 

The global coordinates specify the orbital motion of the center of mass of each body. The local coordinates characterize the changes in the orientation of the body and its rotational and tidal deformations which cause temporal changes of the multipole moments of the body as 
the body moves along its orbit. Conversely, the intrinsic changes of the body's multipole moments affect the body's orbital motion. It is desirable to separate the body's internal and external degrees of freedom as much as possible. It can be achieved through a suitable construction of the local coordinates adapted to each body and their subsequent splicing with the global coordinates by the method of matched asymptotic expansions.

We utilize a scalar-tensor theory of gravity where the scalar field $\Phi$ and metric tensor $g_{\a\b}$ are solutions of the corresponding field equations of the theory -- for more details see \citep{k2019PRD}. We assume that the bodies of an $N$-body system move slowly, and gravitational and scalar fields are weak everywhere. The field equations are solved  iteratively by the method of post-Newtonian (PN) approximations \citep{kopeikin_2011book} with a small parameter $\epsilon\sim v/c\sim (U/c^2)^{1/2}$ where $v$ is a characteristic velocity of matter, $U$ is a typical value of gravitational potential, and $c$ is the ultimate speed of gravity which is equal to the speed of light in vacuum \citep{Kopeikin_2004CQGra,kopeikin_2006IJMPD}. The scalar field $\Phi$ and the metric tensor $g_{\a\b}$ are expanded in Taylor series around their background values with respect to the small parameter $\epsilon$. In particular, the post-Newtonian expansion of scalar field $\Phi$ is 
\ba\label{j7fx41z}
\Phi=\Phi_0\le(1+\phi\ri)+{\cal O}\le(\phi^2\ri)\;,
\ea
where $\Phi_0$ is the background value of the scalar field and $\phi$ is a dimensionless perturbation of the field. The scalar field is assumed to be self-interacting and the strength of the self-interaction is given by function $\omega(\Phi)$. This function is also expanded in a Taylor series with respect to the perturbation of the field,
\ba\label{per77c1}
\omega(\Phi)&=&\omega_0+\omega'_0\phi+{\cal O}\le(\phi^2\ri)\;,
\ea
where $\omega_0\equiv\omega(\Phi_0)$ is the background value of the self-interaction, and $\omega'_0\equiv\le(d\omega/d\ln\Phi\ri)_{\Phi_0}$ is its first derivative.

The metric tensor $g_{\a\b}$ is expanded in the post-Newtonian series around the background Minkowski metric, $\eta_{\a\b}$, 
\ba\label{iq3z90} 
g_{\a\b}&=&\eta_{\a\b}+{h_{\a\b}}+{\cal O}\le(\le|h_{\a\b}\ri|^2\ri)\;,
\ea
where ${h_{\a\b}}$ is the perturbation of the metric tensor. The perturbations $\phi$ and $h_{\a\b}$ are of the order of $\epsilon^2$. 

Exact form of the perturbations $\phi$ and $h_{\a\b}$ is given below in next sections. The residual terms in \eqref{j7fx41z}--\eqref{iq3z90} can be found in \citep{DamEsFar,kopeikin_2011book,k2019PRD} but they are not required in the present paper. Specific functional form of the perturbations, $\phi$ and $h_{\a\b}$, depends on the type of the coordinates introduced in spacetime of an $N$-body system. We distinguish the global coordinates covering the entire spacetime manifold from the local coordinates adapted to each body and covering a close neighborhood of the body all along the world line of its center of mass.  

\subsection{The global coordinates}

In the global coordinates $x^\a=(t,x^i)$ the post-Newtonian perturbation of the metric tensor reads
\begin{eqnarray}
  \label{12.6}
  h_{00}(t,\bm{x}) & = & \phantom{-}2\,U(t,\bm{x})+{\cal O}(4)\;,\\
  \label{12.8}
  h_{0i}(t,\bm{x})&=& -2(1+\gamma)\,U^i(t,\bm{x})+{\cal O}(5)\;,\\
  \label{12.7}
    h_{ij}(t,\bm{x})&= &\phantom{-}2\gamma\delta_{ij}U(t,\bm{x})+{\cal O}(4)\;,
  \ea
where 
\ba\label{96x51z} 
\g&=&1-(\omega_0+2)^{-1}\;,
\ea
is the parameter of the scalar-tensor theory, and here and everywhere else the symbol ${\cal O}(n)$ denotes the post-Newtonian terms of the order of $\epsilon^n$. Parameter $\g$ is formally similar but not physically-equivalent to the parameter $\g_{\rm PPN}$ of the parametrized post-Newtonian (PPN) formalism \citep{willbook} which is introduced in the PPN metric tensor formally to characterize the curvature of space-like hypersurface of constant time $t$. On the other hand, the covariant parameter $\g$ defined in the scalar-tensor theory of gravity by \eqref{96x51z}, is a measure of the strength of self-coupling interaction of the long-range scalar field $\Phi$ given in terms of the background value $\omega_0$ of function $\omega(\Phi)$.  

The other functions appearing in \eqref{12.6}--\eqref{12.7} include scalar (Newtonian-type) and vector gravitational potentials of the $N$-body system,
\ba\label{asz9t2}
U(t,\bm{x})&=&\sum_{\A} U_{\A}(t,\bm{x})\;,\qquad\qquad U^i(t,\bm{x})=\sum_{\A} U^i_{\A}(t,\bm{x})\;,
\ea
which are linear superpositions of scalar and vector gravitational potentials of each body of the $N$-body system, and
\ba
\label{12.10rtfz}
U_{\A}(t,\bm{x})=\int\limits_{{\cal V}_{\A}}\frac{\rho^{\ast}(t,\bm{x}')}{|\bm{x}-\bm{x}'|}d^3x'\;,\qquad\qquad
U^i_{\A}(t,\bm{x}) =\int\limits_{{\cal V}_{\A}}\frac{\rho^{\ast}(t,\bm{x}')v^i(t,{\bm x}')}{|\bm{x}-\bm{x}'|}d^3x'\;.
\ea  
Here ${\cal V}_{\A}$ denotes the spatial volume occupied by body A, $\rho^*\equiv\sqrt{-g}u^0\rho$ is the invariant mass density \citep{fockbook}, $u^0$ is time component of 4-velocity of matter $u^\a$, $g={\rm det}|g_{\a\b}|$ is a determinant of the metric tensor, $\rho$ is the local mass density entering the energy-momentum tensor, and $v^i=dx^i/dt=u^i/u^0$ is 3-dimensional velocity of matter in the global coordinates. The invariant mass density $\rho^*$ is more convenient for calculations as its product with the volume element is invariant, $\rho^*d^3x={\rm inv}$, with respect to coordinate transformations as well as to the Lie transport along matter worldlines  \citep{fockbook,kopeikin_2011book}. 

Scalar field perturbation is mathematically identical to the Newtonian gravitational potential \citep{k2019PRD},
\ba\label{u3vxz6dx}
\phi(t,{\bm x})=U(t,\bm{x})\;.
\ea
Nonetheless, the reader should keep in mind that the laws of transformation of the scalar field and the metric tensor are different. It requires to carefully distinguish transformations of the scalar potential in \eqref{asz9t2} from that in \eqref{u3vxz6dx} which is a part of the metric tensor component $g_{00}$, as they carry out information about geometrically-different types of the field.  

Subsequent derivation requires to single out one of the bodies, let say a body A, and split the gravitational potentials in two parts -- internal and external,
\be\la{nx4z9d5}
U(t,\bm{x})=U_{\A}(t,\bm{x})+\bar U(t,\bm{x})\;,\qquad\qquad  U^i(t,\bm{x})=U^i_{\A}(t,\bm{x})+\bar{U}^i(t,\bm{x})\;,  
\ee  
where $U_{\A}$ and $U^i_{\A}$ denote the internal gravitational potentials produced by body A alone,   
\be\la{tb54vd}
\bar U(t,\bm{x})=\sum_{{\B}\not=\A} U_{\B}(t,\bm{x})\;,\qquad\qquad \bar{U}^i(t,\bm{x})=\sum_{{\B}\not=\A} U^i_{\B}(t,\bm{x})\;,
\ee
denote the external gravitational potentials of all other (external) bodies of the $N$-body system but the body A,
and 
\ba
U_{\B}(t,\bm{x}) =\int\limits_{{\cal V}_{\B}}\frac{\rho^{\ast}(t,\bm{x}')}{|\bm{x}-\bm{x}'|}d^3x'\;,\qquad\qquad
U^i_{\B}(t,\bm{x}) =\int\limits_{{\cal V}_{\B}}\frac{\rho^{\ast}(t,\bm{x}')v^i(t,\bm{x}')}{|\bm{x}-\bm{x}'|}d^3x'\;,
  \ea   
denote the gravitational potentials of the external body B.

We do not impose any specific limitations either on the distribution of mass and velocities inside each body of the $N$-body system or on the shape of the volume of integration. They remain arbitrary to the extent admitted by the gravitational field equations of the PN approximations. In particular, the mass density and velocity must obey the microscopic equations of motion of matter explained in Appendix \ref{kuku876}. The boundary of the volume integration ${\cal V}_\A$ of each body A is determined in accordance with the solution of the internal problem for this body in the local coordinates -- see, for example, \citep{chandr87,kopeikin_2016PhRvD}. 

\subsection{The local coordinates}

There are N local coordinate charts each of which is adapted to a particular body in the $N$-body system \citep{kopeikin_2011book,soffel_2019book}. By default, we denote the local coordinates of body A as $w^\a=(u,w^i)$. If one needs to distinguish the local coordinates adapted to body A from those adapted to body B, we use a corresponding label, for example, $w^\a_\A=(u_\A,w^i_\A)$ are the local coordinates adapted to body A, $w^\a_\B=(u_\B,w^i_\B$ are the local coordinates adapted to body B, and so on.

In the local coordinates of body A$\in\{1,2,...,{\rm N}\}$ the metric tensor perturbation takes on the following form \citep{k2019PRD},
\begin{eqnarray}
  \label{1.8}
  {h}_{00}(u,\bm{w}) & = &\phantom{-} 2 {U}_{\A}(u,\bm{w})+2\sum_{l=1}^{\infty}\frac{1}{l!}{\cal Q}_Lw^{L}+{\cal O}(4)\;,\\
  \label{1.9}
  {h}_{0i}(u,\bm{w}) & = & -2(1+\gamma){U}^i_{\A}(u,\bm{w})+\frac13(1-\gamma)\dot{\cal P} w^i+ \\\nonumber
&&+\sum_{l=1}^{\infty}\frac{l+1}{(l+2)!}\varepsilon_{ipq}{\cal C}_{pL}w^{qL}
  +2\sum_{l=1}^{\infty}\frac{2l+1}{(2l+3)(l+1)!}\bigl[2\dot{\cal Q}_L+(\gamma-1)\dot{\cal P}_L\bigr]w^{iL}+{\cal O}(5)\;,\\
  \label{1.10}
  h_{ij}(u,\bm{w}) & = &\phantom{-} 2\gamma\delta_{ij}{h}_{00}(u,\bm{w})  +2(\gamma-1)\delta_{ij}\sum_{l=1}^{\infty}\frac{1}{l!}{\cal P}_L w^{L}+{\cal O}(4)\;,
  \ea
where ${\cal Q}_L$ and ${\cal C}_L$ are the external gravitational multipoles generated by the bodies being external to body A, ${\cal P}_L$ are the external multipoles moments of the scalar field, and $U_\A$ and $U^i_\A$ are the internal gravitational potentials generated solely by the matter of body A. The multipoles ${\cal Q}_L$ and ${\cal C}_L$ are also called gravitoelectric and gravitomagnetic multipoles as they are connected with the corresponding electric-type and magnetic-type components of the curvature tensor \citep{kop_2019EPJP}.

We did not show explicitly in the local metric \eqref{1.8}--\eqref{1.10} the residual post-Newtonian terms though the terms of the order of ${\cal O}(4)$ in $h_{00}$ are requited for calculation of the post-Newtonian equations of motion \citep{dsx1,dsx2,kovl_2004,racine_2005PhRvD,k2019PRD} and for defining the post-Newtonian mass multipole moments of the bodies -- see section \ref{j2x10t} below. These terms can be found in \citep{k2019PRD}. 

The internal gravitational potentials are defined as integrals over volume ${\cal V}_{\A}$ occupied by the body,
\begin{eqnarray}
  \label{1.11}
  {U}_{\A}(u,\bm{w})  =  \int\limits_{{\cal V}_{\A}}\frac{\rho^*(u,\bm{w}')}{|\bm{w}-\bm{w}'|}d^3w'\;,\qquad\qquad
  \la{qw4cx}
  {U}^{i}_{\A}(u,\bm{w})  = \int\limits_{{\cal V}_{\A}}\frac{\rho^*(u,\bm{w}')\nu^i(u,\bm{w}')}{|\bm{w}-\bm{w}'|}d^3w'\;,
  \ea
where $\nu^i=dw^i/du$ is velocity of matter of body A with respect to the local coordinates adapted to this body.
Definitions of the internal potentials in \eqref{1.11} look similar to those given in \eqref{12.10rtfz} in the global coordinates. However, since the local and global coordinates are interrelated by the post-Newtonian transformation, the numerical values of the internal potentials ${U}_{\A}(u,\bm{w})$, ${U}^i_{\A}(u,\bm{w})$ coincide with the numerical values of ${U}_{\A}(t,\bm{x})$, ${U}^i_{\A}(t,\bm{x})$ only in the Newtonian approximation -- see more details about the correspondence between the gravitational potentials in \citep[\S5.2.3]{kopeikin_2011book}.  
  
The internal potentials are expanded in the multipolar series outside the body A as explained in detail in \citep{Blanchet_1989AIHPA,di}
\ba\label{ha4zw1}
U_{\A}(u,\bm{w})&=&\sum_{l=0}^\infty\frac{(-1)^l}{l!}\M^L_\A\pd_L \le(\frac1r\ri)\;,\\\label{m3dz21}
U^i_{\A}(u,\bm{w})&=&\sum_{l=0}^\infty\frac{(-1)^l}{(l+1)!}\dot\M^{iL}_\A\pd_L \le(\frac1r\ri)+\sum_{l=0}^\infty\frac{(-1)^l}{l!(l+2)}\varepsilon^{ipq}\Sc^{pL}_\A\pd_{qL} \le(\frac1r\ri)\;,
\ea
where $\M^L_\A$ are the mass multipoles, and $\Sc^L_\A$ are the spin multipoles of body A, $r=|{\bm w}|$ is the radial distance from the origin of the local coordinates of body A to the field point $w^i$, and the overdot denotes the total time derivative with respect to the local time $u$. The mass and spin multipoles are defined in terms of the matter variables below in section \ref{j2x10t}.  

The metric tensor perturbation $h_{\a\b}(u,{\bm w})$ in the local coordinates is a solution of the field equations of the scalar-tensor theory of gravity. Therefore, it must be consistent with the metric tensor perturbation $h_{\a\b}(t,{\bm x})$ in the global coordinates as they represent exactly the same gravitational field written down in terms of different variables. This fact allows us to establish a transformation from the global to local coordinates of body A (and its inverse) by applying the technique of matched asymptotic expansions \citep{Lagerstrom1988} to the metric perturbations. This technique also allows to find out the explicit form of the external multipoles ${\cal Q}_L$, ${\cal C}_L$ in terms of partial derivatives from the gravitational potentials $\bar{U}(t,{\bm x})$, $\bar{U}^i(t,{\bm x})$ of the external (with respect to body A) bodies of the $N$-body system \citep{kopeikin_2011book}. 

In case of $l\ge 2$ the {\it gravitoelectric} multipoles ${\cal Q}_L$ represent physically a tidal gravitational field in the neighborhood of body A \citep{k2019PRD}
\ba\label{jje8c}
{\cal Q}_L&=&\pd_L{\bar U}({\bm x}_\A)+{\cal O}(2)\;,\qquad\qquad (l\ge 2)\;,
\ea
where we have used multi-index notation for the STF partial derivative, $\pd_L\equiv\pd_{i_1i_2...i_l}$.
For $l=1$, the gravitoelectric dipole moment \citep{k2019PRD}
\ba\label{nrt7b}
{\cal Q}_i&=&\pd_i\bar U({\bm x}_\A)-a^i_\A+{\cal O}(2)\;,
\ea 
is the difference between the Newtonian gravity force and acceleration of the center of mass of body A, $a^i_\A=\ddot{x}^i_\A$, where the overdot is time derivative with respect to time $t$. There is no monopole gravitoelectric multipole, ${\cal Q}\equiv 0$, as it is eliminated by re-scaling of local coordinates.

The {\it gravitomagnetic} multipoles ${\cal C}_\L$ are \citep{k2019PRD}
\begin{eqnarray}
  \label{3.29}
  \varepsilon_{ipk}{\cal C}_{pL} & = & 4(1+\gamma)\le[ v_{\A}^{[i}\pd^{k]<L>}\bar U({\bm x}_\A) +\pd^{<L>[i}\bar{U}^{k]}({\bm x}_\A)
   -\frac{l}{l+1}\delta^{<i_{l}[i}\pd^{k]L-1>}\dot{\bar{U}}({\bm x}_\A)\ri]+{\cal O}(2)\;,\qquad\qquad (l\ge 1)\;,
\end{eqnarray}
where we have used notation for the STF partial derivative, the overdot denotes the time derivative with respect to time $t$, and the square parentheses around a pair of indices denote anti-symmetrization. The external gravitomagentic dipole ${\cal C}_i$ has a physical meaning of the {\it dynamic} angular velocity of rotation of spatial axes of the local coordinates \citep{bk89}. We assume that the local coordinates of body A$\in\{1,2,...,{\rm N}\}$ do not rotate dynamically, that is ${\cal C}_i\equiv 0$. This assumption is equivalent to the statement that the local triad of spatial axes of the local coordinates undergoes the Fermi-Walker transport along the world line of the center of mass of body A.

The external multipoles of the scalar field $\Phi$ look similar to the gravitoelectric multipole,
\ba\label{jjec54}
{\cal P}_L&=&\pd_L{\bar U}({\bm x}_\A)+{\cal O}(2)\qquad\qquad (l\ge 0).
\ea 
However, it should be emphasized that ${\cal P}_L={\cal Q}_L+{\cal O}(2)$, if and only if, the index $l\ge 2$. Monopole of the scalar field 
\be\label{ju3cz41}
{\cal P}={\bar U}({\bm x}_\A)\;,
\ee 
defines the value of the gravitational potential of external bodies on the world line of the center of mass of body A, and dipole of the scalar field, 
\be\label{okne712}
{\cal P}_i=\pd_i{\bar U}({\bm x}_\A)\;,
\ee 
formally coincides with the Newtonian gravity force computed on the world line of the center of mass of body A. Dipoles ${\cal P}_i$ and ${\cal Q}_i$ are linked one to another by equation \eqref{nrt7b} that is, 
\be\label{unayzc234}
{\cal P}_i={\cal Q}_i+a^i_\A\;.
\ee

Post-Newtonian corrections of the order ${\cal O}(2)$ are required only in \eqref{jje8c} for derivation of the post-Newtonian equations of motion. They can can be found in \citep{kovl_2004,k2019PRD} but we don't need their explicit form in the present paper 

\subsection{The internal mass and spin multipoles of a single body from an $N$-body system}\label{j2x10t}

Like in general relativity, the metric tensor in scalar-tensor theory of gravity depends on two types of {\it canonical} STF multipoles -- the mass multipoles $\M^L_\A$ and the spin multipoles $\Sc^L_\A$. However, in contrast to general relativity there are two possible definitions of the multipoles affiliated respectively to the Einstein and Jordan frames of the scalar-tensor theory. It turns out that the Jordan-frame multipoles define the force of gravitational interaction between massive bodies in the equations of motion of an $N$-body system \citep{kovl_2004,k2019PRD}. On the other hand, the principle of the effacing of internal structure of the bodies \citep{kovl_2008} from the equations of motion suggests that it is the Einstein-frame monopole and dipole moments which are to be used for definition of the inertial mass and the center of mass of each body. The mass multipoles in the Jordan frame are called {\it active} multipoles and those defined in the Einstein frame are termed {\it conformal} multipoles \citep{kopeikin_2011book}.  

The {\it active} mass multipoles referred to the local coordinates of body A, read \citep{k2019PRD}
\begin{eqnarray}\label{act29}
  {\M}_\A^{L} & = & \int\limits_{{\cal V}_{\A}}\left[\sigma(u,{\bm w})w^{<L>}+\frac{1}{2(2l+3)}\pd^2_u\sigma(u,{\bm w})w^2w^{<L>}
  -2\frac{\gamma+1}{l+1}\frac{2l+1}{2l+3}\pd_u\sigma^i(u,{\bm w})w^{<iL>}\right]d^3w\;,
  \ea
where the {\it active} post-Newtonian mass and current density distributions \citep{kovl_2004}
\begin{eqnarray}
  \label{pz3}
  \sigma(u,\bm{w}) & = & \rho^{\ast}(u,\bm{w})\left[1+(\gamma+\frac12)\nu^2(u,\bm{w})+\Pi(u,\bm{w})
  -  (2\beta-1){U}_{\A}(u,\bm{w})  \right]\\\nonumber
  &&\hspace{1.8cm}+\gamma{\sigma}^{kk}(u,\bm{w})-(2\beta-\gamma-1){\cal P}-\sum_{k=1}^{\infty}\frac{1}{k!}\Big[{\cal Q}_{K}+2(\beta-1) {\cal P}_{K}\Big]w^{K}\;,\\
  \label{opc52xz41}
 \sigma^i(u,\bm{w}) & = & \rho^{\ast}(u,\bm{w})\nu^i(u,\bm{w})\;,
  \end{eqnarray}
depend on the invariant density $\rho^*$, velocity $\nu^i$, kinetic and potential energy density of matter as well as on trace ${\sigma}^{kk}=\delta_{ik}\sigma^{ik}$ of the stress tensor ${\sigma}^{ik}$ of matter of body A. The stress tensor and the internal structure of bodies in the $N$-body system are  constrained by equation of state which is not required to be specified for our calculations. Thus, the bodies can be either solid or liquid or made of some other kind of ordinary or condensed matter. Definition \eqref{act29} of the mass multipoles includes post-Newtonian terms which come from the post-Newtonian terms ${\cal O}(4)$ in the metric \eqref{1.8} -- see more details in \citep[Chapter 4]{kopeikin_2011book}. 

The {\it active} mass multipoles depend on two parameters of the scalar-tensor theory - $\g$ and $\b$. Parameter $\g$ has been explained above in \eqref{96x51z}. Parameter $\b$ is a measure of the dependence of self-interaction of the scalar field $\Phi$ on its strength \citep{k2019PRD}
\ba\label{q8ve43l}
\beta&=&1+\frac14\omega'_0(2\omega_0+3)^{-1}(\omega_0+2)^{-2}\;,
\ea
where $\omega'_0=(d\omega/d\ln\Phi)_{\Phi=\Phi_0}$ is a derivative of the self-coupling function $\omega(\Phi)$ with respect to the dimensional scalar field perturbation $\phi=\Phi/\Phi_0-1$, where $\Phi_0$ is the background value of the scalar field \citep{k2019PRD}. 
In the present paper $\b$ is formally similar to but not fully identical to the parameter $\b_{\rm PPN}$ of the PPN formalism \citep{willbook} where it was introduced to the metric tensor phenomenologically to parameterize the possible deviation of the strength of self-interaction of the metric tensor perturbations from its general-relativistic value, $\b_{\rm PPN}=1$. Notice that since $\Phi_0$ depends on time due to the expansion of universe the parameters $\beta$ and $\gamma$ are functions of time as well, and their secular evolution can be measured, at least in principle, with precise astronomical techniques like pulsar timing, lunar laser ranging, etc. \citep{galkop_2016PhRvD}.   

The mass multipole \eqref{act29} also depends on the gravitoelectric external multipoles ${\cal Q}_L$ and the scalar field multipoles ${\cal P}_L$ which are expressed in terms of the partial derivatives from the gravitational potential of external bodies. The reason for inclusion these terms to the definition of the multipoles is that it fully eliminates {\it non-canonical} multipoles \eqref{NL15a} from equations of motion and validates the effacing principle \citep{kovl_2008}. It is also required by the law of conservation of energy in tidal interaction \citep{purdue_1999PRD}.  
Making use of equations \eqref{jje8c}, \eqref{nrt7b} and \eqref{jjec54} we can bring the mass multipoles \eqref{act29} to the form that is more suitable for derivation of the Lagrangian of the $N$-body problem. It reads 
\ba\label{klz2a1}
 {\M}_\A^L &=&\I_\A^{L}+a_\A^p{\M}^{pL}_\A+\gamma\bar U({\bm x}_\A){\M}_\A^L+(1-2\b)\sum_{k=0}^{\infty}\frac{1}{k!}\pd_K\bar U({\bm x}_\A)\mathfrak{I}_\A^{KL}+\frac{l}{2l+1}a_\A^{<i_l}\N_\A^{L-1>}\;,
\end{eqnarray}
where 
\ba\label{wvzx35x}
\I_\A^{L}&=&\int\limits_{{\cal V}_{\A}}\tilde{\sigma}(u,{\bm w})w^{<L>} d^3w+\frac{1}{(2l+3)}\left[\frac12\ddot{\cal N}_\A^{L}-2(1+\gamma)\frac{2l+1}{l+1} \dot{\cal R}_\A^{L}\right]\;,
\ea
are the Blanchet-Damour mass multipoles of body A \citep{Blanchet_1989AIHPA} in scalar-tensor theory of gravity that are functionals of the {\it active} mass density distribution
\ba\label{vv44hh66aa}
\tilde{\sigma}(u,\bm{w}) & = & \rho^{\ast}(u,\bm{w})\left[1+(\gamma+\frac12)\nu^2(u,\bm{w})+\Pi(u,\bm{w})
  -  (2\beta-1){U}_{\A}(u,\bm{w})\right]+\gamma{\sigma}^{kk}(u,\bm{w})\;,
\ea
and {\it non-canonical} multipoles
\Asu\label{NL15a}
\ba\label{74gfcvdu39}
  \N^L_\A &=& \int\limits_{{\cal V}_{\A}}\rho^*(u,{\bm w})|{\bm w}|^2 w^{<L>}d^3w\;,\\
  \R^L_\A&=&\int\limits_{{\cal V}_{\A}}\rho^*\nu^p(u,\bm{w}) w^{<pL>}d^3w\;.
  \ea
  \esu
The symbol
  \ba\label{unc62d}
  \mathfrak{I}^L_\A&=&\int\limits_{{\cal V}_{\A}}\rho^*(u,{\bm w})w^{(i_1i_2...i_l)}d^3w\;,
  \ea
denotes a {\it symmetric} mass multipole of body A in the body-adapted local coordinates which is required only in the post-Newtonian terms. It is worth noticing some useful relationships between various multipoles,
\Asu\label{iop34x4}
\ba 
\mathfrak{I}^{<L>}_\A&=&\M^L_\A\;,\\
\mathfrak{I}^{p<L>}_\A&=& {\M}^{pL}_\A+\frac{l}{2l+1}\delta^{p<i_l}\N_\A^{L-1>}\;,\\
\mathfrak{I}^{kk<L>}_\A&=&{\cal N}_\A^L\;.
\ea 
\esu 
The mass multipoles of body A are functions of the local coordinate time $u$ which can be also interpreted as the proper time of a fictitious observer placed at the origin of the local coordinates adapted to body A. It is useful to notice that the multipoles $\M^L_\A$ and $\I^L_\A$ differ by terms of the post-Newtonian order of magnitude and can be used interchangeably in the post-Newtonian terms. We prefer to use in the post-Newtonian terms the multipoles $\M^L_\A$ like in \eqref{klz2a1}.

It should be emphasized that the mass multipoles \eqref{klz2a1} are not simply the integrals of the internal dynamic variables which appear in the definition of the Blanchet-Damour mass multipoles \eqref{wvzx35x} and the {\it non-canonical} multipoles \eqref{NL15a}, \eqref{unc62d}. The {\it active} mass multipoles also depend on the external dynamic variables -- coordinates $x^i_\A$ of the centers of mass of the bodies and their accelerations $a^i_\A$. This dependence complicates derivation of the Lagrangian of the $N$-body problem since the partial derivatives from the mass multipoles with respect to the external variables do not vanish -- $\pd{\M}_\A^L/\pd x^i_\A\neq 0$ and $\pd{\M}_\A^L/\pd a^i_\A\neq 0$. Hence, the variational derivative from the {\it active} mass multipoles do not vanish either -- see section \ref{uq376xv2}.  

The active multipoles of body B are defined similarly to those of body A. Namely, we have
\ba\label{klzsd1}
 {\M}_\B^N &=&\I_\B^{N}+a_\B^p{\M}^{pN}_\B+\gamma\bar U({\bm x}_\B){\M}_\B^N+ (1-2\b)\sum_{k=0}^{\infty}\frac{1}{k!}\pd_K\bar U({\bm x}_\B)\mathfrak{I}_\B^{KN}+\frac{n}{2n+1}a_\B^{<i_n}\N_\B^{N-1>} \;,
 \ea 
 where
 \ba\label{jj77kk88}
\I_\B^{N}&=&\int\limits_{{\cal V}_{\B}}\tilde{\sigma}(u,{\bm w})w^{<N>} d^3w+\frac{1}{(2n+3)}\left[\frac12\ddot{\cal N}_\B^{N}-2(1+\gamma)\frac{2n+1}{n+1} \dot{\cal R}_\B^{N}\right]\;, 
\end{eqnarray}
are the Blanchet-Damour active mass multipoles of body B.
The mass multipoles of body B are functions of the coordinate time of the local coordinates of body B. The local coordinate time of body A does not coincide with that of body B as the bodies move with respect to each other and are exposed to different gravitational potentials.  We shall discuss the correspondence between the local times of the bodies in section \ref{ii58}.  

In addition to the internal mass multipoles, ${\cal M}^L_\A$ of body A, there exists a set of internal STF spin multipoles ${\cal S}^L_\A$ characterizing vector potential \eqref{m3dz21} of body A \citep{kovl_2004}
\begin{equation}
  \label{1.32}
  \mathcal{S}^{L}_\A = \int\limits_{{\cal V}_{\A}}\varepsilon^{pq<i_l} w^{i_{l-1}...i_1>p}\sigma^q(u,{\bm w})d^3w\;,
\end{equation}
where $\varepsilon^{kpq}$ is the fully antisymmetric Levi-Civita symbol. The internal spin multipoles depend on the local coordinate time of the corresponding body.
Definition \eqref{1.32} is sufficient for deriving the post-Newtonian translational motion of extended bodies in the $N$-body system. Derivation of the post-Newtonian equations of rotational motion of the bodies requires extension of \eqref{1.32} to the post-Newtonian domain. We don't discuss it in the present paper but details can be found in \citep{k2019PRD}.

Both the mass and spin multipole moments of each body are subject to tidal and rotational deformations which progressively change as the bodies of the $N$-body system move in the course of time. The time evolution of multipole moments is determined by corresponding equations of motion for these moments. We assume that these equations of motion are known and can be solved, at least, in principle. In the present paper we will need three equations of motion which govern the time evolution of mass monopole, mass quadrupole and spin dipole. These equations are given in equations \eqref{aa55zz77}, \eqref{ec5} of Appendix \ref{app1a} and in \eqref{vw291z3} respectively. 

\subsection{Definition of the inertial mass and the center of mass of a single body from an $N$-body system}\label{op3vz1a}

There are two distinctive world lines inside the world tube covered by the local coordinates adopted to body A. They are representing a world line of the origin of the local coordinates and that of the body's center of mass. Naturally, the origin of the local coordinates adapted to body A has spatial coordinates $w^i=0$ which are identified with the point $x^i_\A$ in the global coordinates. The center of mass of body A can move along a different world line and, as a rule, the correspondence between the two world lines is established by imposing an additional vector-like supplementary condition \citep{Costa2015}.

We shall denote the local coordinates of the center of mass of body A as $w^i_\A$. Definition of the center of mass in the post-Newtonian approximation depends on our choice of the mass density function describing the post-Newtonian distribution of matter inside the body. We want to find out the most optimal definition of the center of mass which eliminates as much spurious effects in the equations of motion of the body as possible. As mentioned above, in scalar-tensor theory of gravity there are two mass density distribution functions corresponding to the {\it conformal} (Einstein's frame) and {\it active} (Jordan's frame) multipole moments. Our study \citep{kovl_2004,k2019PRD,xie_2010AcPSl} shows that in the scalar-tensor theory of gravity the center of mass of body A is defined the most optimally if one uses the {\it conformal} mass density distribution inside the body, 
\ba\label{ik34nbz}
\varrho(u,{\bm w})&=&\rho^{\ast}(u,{\bm w})\le[1+\frac{3}{2}\nu^2(u,{\bm w})+\Pi(u,{\bm w})-{U}_{\A}(u,{\bm w})\ri]+\sigma^{kk}\;,
\ea
which depends on the invariant mass density $\rho^*=\sqrt{-g}u^0\rho$, internal velocity of matter $\nu^i=dw^i/du$, its internal energy $\Pi$, and gravitational potential $U_\A$ of the body. The {\it conformal} mass density $\varrho$ does not depend on the PPN parameters $\b$, $\g$ 

It is instructive to introduce a general-relativistic post-Newtonian mass of body A by formula
\ba
  \label{grmass}
  {\mathfrak m}_\A&=&\int\limits_{{\cal V}_{\A}}\hat\sigma(u,{\bm w})d^3w\;,
\ea
where the general-relativistic mass density in the post-Newtonian approximation is \citep{fockbook,willbook}
\ba\label{ik34ux5s}
\hat\sigma(u,{\bm w})&=&\rho^{\ast}(u,{\bm w})\le[1+\frac{1}{2}\nu^2(u,{\bm w})+\Pi(u,{\bm w})-\frac{1}{2}{U}_{\A}(u,{\bm w})\ri]\;.
\ea
The general-relativistic mass \eqref{grmass} depends exclusively on the internal dynamic variables of body A and does not depend on the external variables like coordinates and velocity of the centers of mass of bodies in the $N$-body system. The post-Newtonian {\it inertial} mass of body A is defined as a {\it conformal} monopole moment of the body by the following equation \citep[Eq. 164]{k2019PRD} 
\begin{eqnarray}
  \label{confmass}
  m_\A 
  & = &{\mathfrak m}_\A\left[1+(\gamma-1){\cal P}\right]-\sum_{l=1}^{\infty}\frac{l+1}{l!}{\cal Q}_L{\M}_\A^{L}\;,
\end{eqnarray}
where ${\cal P}=\bar U({\bm x}_\A)$ is the external monopole moment of the scalar field and ${\cal Q}_L$ are the gravitoelectric multipole moments \eqref{jje8c}. Notice that the inertial mass \eqref{confmass} depends on the external dynamic variables through ${\cal P}$ and ${\cal Q}_L$.  

Position of the center of mass $w^i_\A$ of body A in the local coordinates adapted to the body is defined by the {\it conformal} dipole moment of the body \citep{k2019PRD},
\ba\label{yxc36cf}
{\cal J}^i_\A=m_\A w^i_\A+I^i_\A\;,
\ea
where (see \citep[Eq. 170]{k2019PRD}) 
\begin{eqnarray}
  \label{confdipole}
  m_\A w^i_\A& = & \int\limits_{{\cal V}_{\A}} \hat\sigma(u,\bm{w})w^id^3w+(\gamma-1){\cal P}{\mathfrak m}_\A w^i_\A -\sum_{l=1}^{\infty}\frac{l+1}{l!}{\cal Q}_L{\cal M}^{iL}_\A
  - \frac{1}{2}\sum_{l=0}^{\infty}\frac{1}{(2l+3)l!}{\cal Q}_{iL}\mathcal{N}^L_\A\;,
\end{eqnarray}
and a vector function $I^i_\A=I^i_\A(t)$ characterizes an additional post-Newtonian displacement of the center of mass with respect to the origin of the local frame of body A which can depend on time. The quantity $m_\A$ in \eqref{confdipole} is the {\it inertial} mass of body A \eqref{confmass} and the STF {\it non-canonical} multipole, $\mathcal{N}^L_\A$, is given in \eqref{74gfcvdu39}. The displacement vector function $I^i_\A$ should be specified by imposing a particular supplementary condition which choice is dictated by computational circumstances but, in principle, is arbitrary. Notice that in accordance to the computational scheme of derivation of the equations of motion, the point on spacetime manifold having spatial global coordinate $x^i_\A$ is always identified with the origin of the local coordinates adapted to body A (the point with $w^i=0$) independently of the value of the residual displacement $I^i_\A$ of the center of mass with respect to the origin of the local coordinates. 

We postulate that at any instant of time the {\it conformal} dipole moment vanishes at the origin of the local coordinates adapted to body A. In other words, the condition ${\cal J}^i_\A\equiv 0$ defines the point with the local coordinates $w^i=0$. The center of mass of body A, $w^i_\A$, coincides with the origin of the local coordinates, $w^i=0$, if the residual displacement, $I^i_\A= 0$. It is evident that the condition, $I^i_\A= 0$, can be always fulfilled, at least, at some fixed instant of time. However, this condition should be propagated in time as derivation of the equations of motion requires to know behavior of $I^i_\A$ along the entire worldline of the origin of the local coordinates. It turns out that the condition $I^i_\A(t)=0$ can indeed be satisfied continuously. The proof was furnished in the Newtonian approximation in \citep{Kopejkin_1988CeMec,bk-nc} and in the first PN approximation by \citet{dsx2,kovl_2004} in the local coordinates. Extension of this result to the global coordinates involves more comprehensive computations which have been conducted by \citet{racine_2005PhRvD,racine2013PhRvD} in general relativity and, independently, by \citet{k2019PRD} in scalar-tensor theory of gravity. These computations yield translational post-Newtonian equations of motion of extended bodies in the global coordinates with all multipoles taken into account and with the center of mass of each body lying at the origin of the local coordinates. We have discovered (see discussion in section IX.A.6 of \citep{k2019PRD}) that the constrain, $I^i_\A(t)=0$, may be not the most optimal one as a sufficiently large number of terms with second time derivatives entering equations of translational motion can be eliminated by choosing another condition, $I^i_\A=-3a_A^p{\cal M}^{ip}_A$ \citep[Eq. 289]{k2019PRD}. This simplifies the equations of motion. For the time being we leave the vector function $I^i_\A$ unconstrained and discuss its specific choice in section \ref{uq376xv2} more explicitly. The fact of the matter is that derivation of the Lagrangian from the equations of motion produces terms with the second time derivatives which are not reduced to the form of the Lagrangian (variational) derivative and must be associated with the specific choice of function $I^i_\A$ -- see equation \eqref{op1qz4}. 

It is worthwhile to notice that the {\it conformal} dipole moment ${\cal J}^i_\A$ of body A is not equivalent to the {\it active} dipole moment $\M^i_\A$ of the body. Thus, $\M^i_\A\neq 0$ even if ${\cal J}^i_\A=0$ which means that extended bodies in scalar-tensor theory of gravity possess a non-vanishing {\it active} mass dipole moment as contrasted to general relativity \citep{kovl_2004,Bernard_2020PRD}. This is why the lowest order of gravitational waves emitted by self-gravitating systems in scalar-tensor theory of gravity is dipole with multipole index $l=1$ \citep{willbook} while in general relativity it is quadrupole with $l=2$  \citep{Landau1975}. For the reader's convenience we provide a formula for the {\it active} dipole moment of body A in case when the conformal moment ${\cal J}^i_\A=0$. This formula is derived by making use of a vector virial theorem \eqref{vect46z3a} in the definition of the {\it active} dipole moment and reads \citep[Eq. 171]{k2019PRD}
\ba\label{iom3wvx6}
{\cal M}^i_\A&=&-\frac12\eta\int_{{\cal V}_\A}\rho^*U_\A w^id^3w+\frac15(\g-1)\le(3\dot{\cal R}^i_\A-\frac12\ddot{\cal N}_\A\ri)
-2(\b-1)\le({\cal P}_j-{\cal Q}_j\ri)\le({\cal M}^{ij}_\A+\frac13\d^{ij}{\cal N}_\A\ri)\\\nonumber
&&-\frac12\eta\sum_{l=0}^\infty\frac1{l!}\le({\cal Q}_L{\cal M}^{iL}_\A+\frac1{2l+3}{\cal Q}_{iL}{\cal N}^L_\A\ri)-\frac12(\g-1)\sum_{l=0}^\infty\frac{2l+1}{l!}{\cal Q}_L{\cal M}^{iL}_\A\;,
\ea
where $\eta=4\b-g-3$ is called the Nordtvedt parameter \citep{willbook}, ${\cal P}^i$ is the external dipole moment of the scalar field defined in \eqref{okne712}, the local non-geodesic acceleration ${\cal Q}_i$ is defined in \eqref{nrt7b}, the external gravitoelectric multipoles ${\cal Q}_L$ are shown in \eqref{jje8c} and the {\it non-canonical} multipole moments are given in \eqref{NL15a}. The reader can see that the {\it active} dipole ${\cal M}^i_\A$ vanishes in general relativity if the center of mass is chosen at the origin of the local frame of body A. In scalar-tensor theory the {\it active} dipole of a single body differs from the {\it conformal} dipole even for a single, non-spherical and/or non-stationary body by terms entering the first line of equation \eqref{iom3wvx6}. 

Equation \eqref{iom3wvx6} is also useful for discussing the dipolar tidal effects in scalar-tensor theories which cause a peer interest for their measurement may be possible in coalescing binary systems with an advanced third generation of gravitational wave detectors \citep{Bernard_2018PRD,Bernard_2020PRD}. The tidal {\it active} dipole moment of each body in the binary is caused by the external dipole moment ${\cal P}_i$ of the scalar field induced by its companion. Assuming that tides are stationary and neglecting higher-order multipoles in \eqref{iom3wvx6} we get the tidal part of the {\it active} dipole of body A
\ba\label{gt3vz41}
{\cal D}^i_\A&=& \frac23(1-\b){\cal N}_\A{\cal P}_i\;,
\ea  
where ${\cal N}_\A$ is the moment of inertia of body A. Expression \eqref{gt3vz41} tells us that in general relativity there is no dipole effects as the {\it active} dipole vanishes in the limit when $\b\rightarrow 1$. 

Formula \eqref{gt3vz41} describes one of the numerous finite-size effects in scalar-tensor theory of gravity and agrees with the expectation that these effects are of the first post-Newtonian order of magnitude \citep{kovl_2004,espositofarese_2011mmgr}. In case, of black holes and/or neutron stars the moment of inertia ${\cal N}_\A\simeq G^2M^2_\A/c^4$, where $M_\A$ is relativistic mass of the object. It makes clear that for the compact relativistic objects the tidally-induced {\it active} dipole moment ${\cal D}^i_\A\simeq G^3c^{-6}$ that affects motion of the finite-size bodies at the third post-Newtonian approximation \citep{kovl_2004} unlike general relativity where the finite-size tidal effects appear only in the fifth ($\sim c^{-10}$) post-Newtonian order \citep{k85,Damour_1987book}. We also notice that the coefficient in the right hand side of equation \eqref{gt3vz41} can be interpreted as the scalar tidal deformability parameter similar to that having been introduced in paper by \citet[Eq. 2]{Bernard_2020PRD}  
\ba\label{bra4xz900}
\lambda^{\rm (s)}_\A&=&\frac23(1-\b){\cal N}_\A\;.
\ea
The scalar deformability parameter defines a dimensionless scalar tidal Love number (sTLN) $k^{(s)}_\A=(c^4/G^2M_A^2)\lambda^{(s)}_\A$ which extends the concept of the TLN in general relativity where tides are caused solely by the quadrupole and higher harmonics of tidal field \citep{Hinderer_2008PhRvD,Poisson_2009PhRvD,Nagar_2009PhRvD}. Of course, the sTLN are absent in general relativity.

\section{Post-Newtonian $N$-body Problem}\label{oq23x5}

\subsection{Configuration Space of Dynamic Variables}

We consider post-Newtonian dynamics of N extended, arbitrary-structured bodies with a continuous distribution of matter inside each body. The distribution of matter is fully characterized by the energy-momentum tensor $T^{\a\b}$ depending on the internal dynamic variables like the density of matter, its velocity, internal energy and stress/pressure. The internal variables of the matter's continuum obey a set of equations of motion which govern the dynamic evolution of the variables (see Appendix \ref{app1a} for the case of the Newtonian approximation). The $N$-body system is considered in the present paper as being completely isolated from other astrophysical systems in the universe and we assume that the background spacetime manifold is asymptotically-flat with the Minkowski metric $\eta_{\a\b}$ at infinity \footnote{Relativistic celestial mechanics of an $N$-body system residing on the expanding cosmological manifold is a non-trivial generalization of the asymptotically-flat case \citep{Kopeikin_2014AnPhy,galkop_2016PhRvD}.}. We postulate that each body of the $N$-body system is electrically neutral and does not exchange mass with the other bodies of the system. Thus, the only force of interaction between the bodies is the force of gravity carried out by tensor and scalar fields. 

Self-consistent description of celestial dynamics of an $N$-body system calls for N+1 coordinate charts which are used for solving the internal and external problems of motion. A single global coordinate chart $x^\a=(t, x^i)$ covers the entire spacetime manifold. It is introduced in order to characterize the relative motion of the centers of mass of the bodies with respect to each other. The other N local coordinate charts are introduced in a neighborhood of each body  A$\in\{1,2,...,{\rm N}\}$. The local coordinates $w^\a_\A=(u_\A,w^i_\A)$ adapted to the body A and are carried out along with the body. The local coordinates are used for the description of dynamics of the internal motions of matter inside the body and evolution of the multipole moments of the body. 

The independent dynamic variables of the external problem are coordinates $x^i_\A$, velocities $v^i_\A=\dot{x}^i_\A$, and accelerations $a^i_\A=\ddot{x}^i_\A$ of the center of mass of body A in the global coordinates. We have included the acceleration as it explicitly appears in the equations of motion of the $N$-body problem possessing all multipoles \citep{racine_2005PhRvD,racine2013PhRvD,k2019PRD}. The acceleration-dependent Lagrangians are fairly common in relativistic mechanics. For example, in case of description of the dynamic evolution of spin \citep{damour_1982CR,Barker1986_acceleration} and in the Lagrangians of point-like particles in the post-Newtonian approximations of higher-order \citep{Damour_1985GReGr,gk86,Ohta1988PThPh,Damour_1989MG5}. The configuration space of the external variables $x^i_\A$ has dimension $3N$. 

Choosing a set of the internal dynamic variables in the local coordinates of body A depends on the approach to the solution of the internal problems. In case of an extended bodies the most natural choice would be a set of hydrodynamics variables like density of matter, pressure, internal energy, etc. interconnected through the equations of state. This approach is clearly unsuitable in case of black holes which internal dynamic variables consists of a set of the mass and spin multipole moments. The multipole moments of the extended bodies can be also used as a set of basic dynamic variables of the internal problem of an extended body A because the multipole moments are defined as volume integrals from the distribution of hydrodynamic variables which can be considered as a mapping from one set of the dynamic variables to another. In any case the configuration space of the internal variables is infinite. 

In this paper we use the multipole moments of the body (both mass and spin type) as the primary internal dynamic variables. They are not supposed to change under a virtual variation of the external variables and should be defined in such a way that make them split distinctively from the external degrees of freedom. The mass and spin multipoles $\M^L_\A$ and $\Sc^L_\A$ which enter equations of motion of the center of mass (see section \ref{mau35zsx1} below), have been defined by equations \eqref{klz2a1} and \eqref{1.32} respectively. The spin multipoles enters only the post-Newtonian terms of the equations of motion and its Newtonian-like definition definition \eqref{1.32} is fully satisfactory and sufficient to consider it as a primary dynamic variable in the post-Newtonian Lagrangian of translational motion. This is not true for the mass multipole $\M^L_\A$ of body A enters the Newtonian force while its definition \eqref{klz2a1} includes explicitly post-Newtonian terms depending on the coordinates and accelerations of the centers of mass of external bodies. These external variables should be excluded so that only the first term ${\cal I}^L_\A$ in the right hand side of \eqref{klz2a1} can be considered as a primary internal dynamic variable. There are also secondary internal dynamic variables of body A which include other characteristics of matter distribution inside the body like the {\it non-canonical} multipole moments ${\cal N}^L_\A$ and ${\cal R}^L_\A$ defined in \eqref{NL15a} as well as various integrals defining the amount of the total internal energy of matter and/or gravitational field of the body, etc., discussed in Appendix \ref{app1a}. It is worth pointing out over here that the mass and spin multipoles of body A are residing on the world line of the center of mass of the body and can be treated as purely spatial Cartesian tensors in tangent spacetime being orthogonal to 4-velocity of the center of mass of the body. Time evolution of the multipoles is governed by the local equations of motion of the internal dynamic variables, e.g., the equation for time derivative of spin \eqref{vw291z3} or that for mass \eqref{aa55zz77}, etc. 

Dynamical behavior of an $N$-body system can be analyzed with the help of a Lagrangian which serves as an indispensable tool for deriving conserved quantities that are obtained by employing the property of relativistic invariance of the Lagrangian with respect to local diffeomorphisms in the configuration space of dynamic variables in accordance with N\"other's theorem \citep{Petrov_2017book}.  The Lagrangian of the $N$-body problem can be formulated directly starting from the full action of gravity theory \citep{infeld_book,vab,Ohta1988PThPh,Bernard_PRD2016,Damour_2016PhRvD93h4014D,Marchand_2018PhRvD}. It is known as the Fokker Lagrangian since the method of its derivation has been first formulated by Fokker \citep{fokker_1929ZPhy}. Especially successful for the derivation of the Fokker Lagrangian became the effective field theory approach applied to point-like massive particles \citep{goldberger_PRD,Foffa_2019PhRvD} and to covariant description of dynamics of extended bodies with spin and quadrupole moments \citep{Steinhoff_2015,Steinhoff_etal_PRD104028}.

The Lagrangian of an $N$-body system can be also derived independently by solving the inverse problem of the Lagrangian mechanics which comes down to the integration of a system of partial differential equations being adjoint to the equations of motion of the $N$-body problem \citep{Santilli_1978}. Of course, the Fokker Lagrangian and the Lagrangian of the inverse problem are to be identical up to the terms which do not affect the dynamics of the bodies. These are the total time derivatives and/or functions of the "double zero" \citep{Barker1986_acceleration}. Correspondence between the various  presentations of one and the same Lagrangian can be established by the method of redefinition of position variables \citep{Damour_Schafer1991JMP} which is based on application of contact transformations of dynamic variables in the jet prolongation of extended configuration space. In the present paper we derive the Lagrangian by solving the inverse problem for the post-Newtonian equations of motion of the $N$-body system taken from \citep{k2019PRD} and given explicitly in next section.  

\subsection{Post-Newtonian Equations of Motion}\label{mau35zsx1}

Translational equations of motion for each body A$\in\{1,2,...,{\rm N}\}$ from an $N$-body system in the framework of a scalar-tensor theory of gravity have the following form \citep[Eq. 290]{k2019PRD}
\begin{equation}
  \label{MBaB1}  
m^{\phantom{i}}_{\A}a^i_{\A} = {\dutchcal F}^i_{\rm N} + {\dutchcal F}^i_{\rm pN}+\ddot{I}^i_\A\;,
\end{equation}
where $m_{\A}$ is the inertial ({\it conformal}) mass\index{mass!inertial} of body A \eqref{confmass}, $a^i_\A=d^2x^i_\A/dt^2=\ddot{x}^i_\A$ is the coordinate acceleration of the center of mass of body A, and the net gravitational force in the right hand side of \eqref{MBaB1} is split in three parts -- the Newtonian gravitational force ${\dutchcal F}^i_{\rm N}$, the post-Newtonian gravitational force ${\dutchcal F}^i_{\rm pN}$, and the post-Newtonian term $\ddot{I}^i_\A$ depends on the post-Newtonian shift of the origin of the local coordinates of body A from its center of mass as explained above in section \ref{op3vz1a}. Vector-function $I^i_\A=I^i_\A(t)$ is not subject to dynamical equations as it characterizes the residual coordinate freedom in choosing position of the center of mass of the body in the local coordinates. The original derivation of the post-Newtonian equations of motion \eqref{MBaB1} of the $N$-body problem has been given in \citep{k2019PRD} under condition that the center of mass of body A is located at the origin of the local frame at any instant of time which yields $I^i_\A\equiv 0$. Below, we demonstrate that solution of the inverse problem of the Lagrangian mechanics does not allow to keep on this condition and requires to displace the body's center of mass from the origin of the local frame yielding $I^i_\A\neq 0$. 

The inertial mass $m_\A$ is not constant but depends on time. This is because body A is affected by the tidal forces from other N-1 bodies which interact with the multipole moments of body A causing post-Newtonian variation of the mass as the bodies move \citep{dsx1,dsx2,kovl_2004,racine_2005PhRvD}. Depending on the equation of state and viscosity the tidal deformation of the body causes dissipation of the internal energy of the body that produces a secular decrease of the body's mass. If the dissipation of the internal energy can be neglected the mass of the body is subject merely to periodic orbital variations. In any case, the time derivative of the body mass, $\dot{{m}}_\A\neq 0$. This effect should be computed analytically and taken into account in the subsequent derivation of the Lagrangian. 

The force components are represented in the form of a linear superposition of the STF partial derivatives from functions $R_{\A\B}$ and $R^{-1}_{\A\B}$ representing respectively the coordinate distance between bodies A and B and its inverse. More specifically, 
\begin{eqnarray}
  \label{w1q5}
  {\dutchcal F}^i_{\rm N}&=&\sum_{\B\neq\A}\sum_{l=0}^{\infty}\sum_{n=0}^{\infty}\frac{(-1)^n}{l!n!}{\cal M}_{\A}^{L}{\cal M}_\B^{N}\pd_{iLN}R_{\A\B}^{-1}\;
  \\
\label{eee3s}
{\dutchcal F}^i_{\rm pN}&=&
\sum_{\B\not=\A}\sum_{l=0}^{\infty}\sum_{n=0}^{\infty}\frac{(-1)^n}{l!n!}\Bigg[\Big(\a^{iLN}+\beta^{iLN}\Big)\pd_{LN}+\Big(\a^{ipLN}+\beta^{ipLN}\Big)\pd_{pLN}+\a^{ipqLN}\pd_{pqLN}\\\nonumber
&&+\Big(\a^{LN}+\beta^{LN}+\gamma^{LN}\Big)\pd_{iLN}+\Big(\alpha^{pLN}+\beta^{pLN}+\gamma^{pLN}\Big)\pd_{ipLN}+\alpha^{pqLN}\pd_{ipqLN}\Big]R_{\A\B}^{-1}\\\nonumber
&+&\frac12\sum_{\B\not=\A}\sum_{l=0}^\infty\sum_{n=0}^{\infty}\frac{(-1)^n}{l!n!}{\cal M}_\A^L\Big[\ddot{{\cal M}}_\B^N\pd_{<iLN>}- {{\cal M}}_\B^Na_\B^p\pd_{<ipLN>}+{{\cal M}}_\B^Nv_\B^pv_\B^q\pd_{<ipqLN>}\Big]R_{\A\B}\\\nonumber
&+&3\left(a_\A^k\ddot{\cal M}_{\A}^{ik}+2\dot{a}_\A^k\dot{\cal M}_{\A}^{ik}+\ddot{a}_\A^k{\cal M}_{\A}^{ik}\right)-\varepsilon_{ikq}\left(2a^k_\A\dot{\mathcal{S}}_\A^q+\dot{a}^k_\A\mathcal{S}_\A^q\right)
\;,
\ea
where ${\cal M}_{\A}^{L}={\cal M}_{\A}^{<a_1...a_l>}$ are the {\it active} STF multipoles of body A referred to the local frame of body A, ${\cal M}_\B^{N}={\cal M}_\B^{<b_1...b_n>}$ are the components of the {\it active} STF multipoles of external body B referred to the local frame of body B, 
\be\la{9b2c5}
R^i_{\A\B}\equiv x^i_{\A}-x^i_\B\;,\qquad\qquad R_{\A\B}=|{\bm x}_\A-{\bm x}_\B|\;,
\ee
is the coordinate distance between the centers of mass of body A and B, and repeated indices mean the Einstein summation rule. 

The mass and spin multipole moments of body A that appear in the force of gravity \eqref{w1q5}, \eqref{eee3s}, are given by equations \eqref{klz2a1}, \eqref{1.32} and similarly for body B. The multipoles of body A are spatial tensors defined on the world line ${\cal W}_\A$ of the center of mass of the body. They are functions of the coordinate time $u_\A$ of the local coordinates adapted to body A. We use in \eqref{w1q5}, \eqref{eee3s} the following notations: $\M^L_\A\equiv\M^L_\A(u_\A)$, $\Sc^L_\A\equiv\Sc^L_\A(u_\A)$. The multipoles of any other body $\B\neq\A$ are spatial tensors defined on the world line of the center of mass of body B and depending on the coordinate time $u_\B$ of the local coordinates adapted to the body B. We have denoted $\M^L_\B\equiv\M^N_\B(u_\B)$, $\Sc^N_\B\equiv\Sc^L_\B(u_\B)$ in \eqref{w1q5}, \eqref{eee3s}. Particular functional dependence of the multipoles on time is determined by solving equations of the time evolution of the multipoles of the corresponding body. 

The spatial partial derivatives from the coordinate distance are computed in accordance with definitions
\ba\label{acser243}
\pd_{L}R_{\A\B}^{-1}     &\equiv&\lim_{{\bm x}\rightarrow{\bm x}_{\A}}\pd_{L}R_{\B}^{-1}\;,\\\label{nrvx31w}
\pd_{L}R_{\A\B}    &\equiv&\lim_{{\bm x}\rightarrow{\bm x}_{\A}} \pd_{L}R_{\B} \;,
\ea
which are taken at point ${\bm x}_\A\equiv(x^i_\A)$ -- the center of mass of body A, and we understand that $R_\B\equiv|{\bm x}-{\bm x}_\B|$. In other words, the partial derivatives from $R_{\A\B}^{-1}$ and $R_{\A\B}$ are to be always understood as the derivatives taken with respect to the first argument of $R_{\A\B}$ or $R_{\A\B}^{-1}$. This convention means, for example, that the $l$-order partial derivative $\pd_LR_{\B\A}^{-1}$ is not equal to $\pd_LR_{\A\B}^{-1}$. Indeed, applying \eqref{acser243} to $R_{\B\A}^{-1}$, we observe that
\ba\label{ii44v}
\pd_{L}R_{\B\A}^{-1}&=&(-1)^l\pd_{L}R_{\A\B}^{-1}\;,
\ea
and similar equation is valid for $R_{\B\A}$.
It is also worth noticing that $\pd_{<L>}R_{\A\B}^{-1}=\pd_{L}R_{\A\B}^{-1}$ due to the fact that function $R^{-1}_\B$ is a fundamental solution of the Laplace equation, $\triangle R^{-1}_\B=0$, everywhere but the point $x^i=x^i_\B$. At the same time, $\triangle R_\B\not=0$ and, hence, $\pd_{<L>}R_{\A\B}\not=\pd_{L}R_{\A\B}$. Exact correspondence between $\pd_{<L>}R_{\A\B}$ and $\pd_{L}R_{\A\B}$ is given in \citep[Eq. (268)]{k2019PRD}

The coefficients of the post-Newtonian force \eqref{eee3s} are given below exactly as they appear in \citep[Eqs. (302)--(313)]{k2019PRD} with some typos corrected in Eqs. (304), (306), (309) and (310). We have
\ba
%%%%%%%%%%%%%%%%%%%%%%%%%%%%%%%%%%%
\label{oq7v3d}
%%%%%%%%%%%%%%%%%%%%%%%%%%%%%%
\a^{iLN}&=&\Bigg[v^i_{\A}-2(1+\g)v^i_{\A\B}\Bigg]{\cal M}^{L}_\A\dot{\cal M}^{N}_{\B}+\left[\frac{2(1+\g)}{n+1}-\frac1{2l+2n+3}\right]{\cal M}^{L}_\A\ddot{\cal M}^{iN}_\B\\\nonumber
&&+2(1+\g)\left[\frac1{n+1}\dot{\cal M}^{L}_\A\dot{\cal M}^{iN}_\B -\dot{\cal M}^{L}_\A{\cal M}^N_\B v^i_{\A\B}\right]-\frac{2l^2+3l+3+2\g}{l+1} \dot{\cal M}^{iL}_\A\dot{\cal M}^N_\B\\\nonumber
&&-\frac1{2l+3}\left[(l+2)(2l+1)+\frac{2n}{2l+2n+3}\right] {\cal M}^{iL}_\A\ddot{\cal M}^N_\B -\frac{l^2+l+2+2\g}{l+1} \ddot{\cal M}^{iL}_\A{\cal M}^N_\B\;,\\
\la{xxx5d6}
\a^{ipLN}&=&\left[\frac2{2l+2n+5}v^i_\B v^p_{\B}-2(1+\g)v^i_{\A\B} v^p_{\A\B}-v^i_\A v^p_{\B}\right]{\cal M}^{L}_\A{\cal M}^{N}_\B
+\frac{2(1+\g)}{n+1}{\cal M}^{L}_\A\dot{\cal M}^{iN}_\B v^p_{\A\B}\\\nonumber
&&+2\left[\frac{(l+2)(2l+1)}{2l+3}v^p_{\B}-(l+1)v^p_\A\right]{\cal M}^{iL}_\A\dot{\cal M}^{N}_\B
-\frac{2l^2+3l+3+2\g}{l+1}\dot{\cal M}^{iL}_\A{\cal M}^{N}_\B v^p_{\A\B}\\\nonumber
&&+\frac{2(1+\g)}{n+2}\varepsilon_{ipq}\left({\cal M}^L_\A\dot{\Sc}^{qN}_{\B}+\dot{\cal M}^L_\A{\Sc}^{qN}_\B\right)+\frac{2(1+\g)}{l+2}\varepsilon_{ipq}\left(\dot{\Sc}^{qL}_\A{\cal M}^N_{\B}+{\Sc}^{qL}_\A\dot{\cal M}^N_\B\right)\;,\\
\la{puk34w}
\a^{ipqLN}&=&\frac1{2l+2n+7}\Big({\cal M}^{iL}_\A{\cal M}^N_{\B}-{\cal M}^{L}_\A{\cal M}^{iN}_\B\Big)v^p_\B v^q_{\B}-(l+1){\cal M}^{iL}_\A{\cal M}^N_\B v^p_{\A\B} v^q_{\A\B}\\\nonumber
&&+\frac{2(1+\g)}{l+2}\varepsilon^{ipk}{\Sc}^{kL}_\A{\cal M}^N_\B v^q_{\A\B}+\frac{2(1+\g)}{n+2}\varepsilon^{ipk}{\cal M}^L_\A{\Sc}^{kN}_\B v^q_{\A\B}\;,
\\\nonumber\\
%%%%%%%%%%%%%%%%%%%%%%%%%%%%%%%%
\la{no23c}
\alpha^{LN}&=&\left[(1+\g)v^2_{\A\B}-\frac12\frac{2l+2n+3}{2l+2n+5}v^2_{\B}-v^2_\A\right]{\cal M}^L_\A{\cal M}^N_{\B}+{\cal M}^{kL}_\A\dot{\cal M}^N_\B v^k_{\A}-\frac{2(1+\g)}{n+1}{\cal M}^{L}_\A\dot{\cal M}^{kN}_\B v^k_{\A\B}\\\nonumber
&&-\frac1{2l+2n+5}{\cal M}^{kL}_\A\ddot{\cal M}^{kN}_{\B}+\frac{2(1+\g)}{l+1}\left( \dot{\cal M}^{kL}_\A{\cal M}^{N}_\B v^k_{\A\B}-\frac1{n+1} \dot{\cal M}^{kL}_\A\dot{\cal M}^{kN}_\B\right)\;,\\
\la{vert3s}
\alpha^{pLN}&=&\frac12\frac{2l+2n+3}{2l+2n+7}{\cal M}^{L}_\A{\cal M}^{kN}_\B v^p_\B v^k_{\B}+\frac{2(1+\g)}{n+2}\varepsilon_{pkq}{\cal M}^{L}_\A{\Sc}^{qN}_\B  v^k_{\A\B}\\\nonumber
&& +\Bigg(v^k_\A v^p_{\A\B}-\frac12 v^k_\A v^p_{\A}+\frac{2}{2l+2n+7} v^p_\B v^k_\B\Bigg){\cal M}^{kL}_\A{\cal M}^{N}_\B \\\nonumber
&&+\frac{2(1+\g)}{(l+1)(n+2)}\varepsilon_{pkq}\dot{\cal M}^{kL}_\A{\Sc}^{qN}_{\B}-\frac{2(1+\g)}{l+2}\varepsilon_{pkq}{\Sc}^{kL}_\A\Bigg({\cal M}^{N}_\B v^q_{\A\B}  -\frac1{n+1}\dot{\cal M}^{qN}_\B\Bigg)\;,\\
\alpha^{pqLN}&=&-\frac1{2l+2n+9}{\cal M}^{kL}_\A{\cal M}^{kN}_\B v^p_\B v^q_{\B}-\frac{2(1+\g)}{(n+2)(l+2)}{\Sc}^{pL}_\A{\Sc}^{qN}_\B\;,\\
%%%%%%%%%%%%%%%%%%%%%%%%%%%%%%%%%%%%%%%%%%%%
\la{bet1111}
\beta^{iLN}&=&\left[\Big(2+2\g-\frac1{2l+2n+3}\Big)a^i_{\B}-\Big(l+2+2\g\Big)a^i_\A\right]{\cal M}^L_\A{\cal M}^N_\B\;,\\
%%%%%%%%%%%%%%%%%%%%%%%%%%%%%
\la{bet2222}
\beta^{ipLN}&=&\left[\left(l+1-\frac1{2l+2n+5}\right)a^p_{\B}-la^p_\A\right]{\cal M}^{iL}_\A{\cal M}^{N}_{\B}+\frac{1}{2l+2n+5}a^p_\B{\cal M}^{L}_\A{\cal M}^{iN}_\B\;,\\
%%%%%%%%%%%%%%%%%%%%%%%
\la{btg42x}
\beta^{LN}&=&-(l+1){\cal M}^{kL}_\A{\cal M}^{N}_\B a^k_{\A}-(n+1){\cal M}^{L}_\A{\cal M}^{kN}_\B a^k_{\B}-\frac1{2l+2n+5}\Big({\cal M}^{kL}_\A{\cal M}^{N}_{\B}-{\cal M}^{L}_\A{\cal M}^{kN}_\B\Big)a^k_\B\;,\\
\la{nv2c6a}
\beta^{pLN}&=&\frac1{2l+2n+7}{\cal M}^{kL}_\A{\cal M}^{kN}_\B  a^p_\B \;,\\\label{uu33vv11}
\gamma^{LN}&=&-\g\Bigg[l\bar U({\bm x}_\A)+(n+1)\bar U({\bm x}_\B)\Bigg]{\cal M}^L_\A{\cal M}^N_\B\;,\\
%%%%%%%%%%%%%%%%%%%%%%%
\label{zw3k4b}
\gamma^{pLN}&=&F^{kp}_\A{\cal M}^{kL}_\A{\cal M}^{N}_\B-F^{kp}_\B{\cal M}^L_\A{\cal M}^{kN}_{\B}\;.
%%%%%%%%%%%%%%%%%%%%%%%%%%%%%%%%%%%
\ea
The PN coefficients \eqref{oq7v3d}--\eqref{zw3k4b} depend on the {\it active} mass and spin multipoles of the bodies of the $N$-body system and their time derivatives taken in the local frame adapted to the corresponding body. They also depend on velocities of the centers of mass of the bodies and their accelerations defined with respect to the global coordinates. Coefficient \eqref{zw3k4b} describes dependence of the force on the matrix $F^{ij}_\A$ of relativistic precession which obeys an ordinary differential equation \citep[Eq. (151)]{k2019PRD}
\ba
  \label{5.18}
  \dot{F}^{ij}_\A&=&2(1+\gamma)\pd^{[i}\bar{U}^{j]}({\bm x}_\A)+(1+2\gamma)v_{\A}^{[i}\pd^{j]}_{\phantom A}\bar{U}({\bm x}_\A)+v_{\A}^{[i}{\cal Q}^{j]}_{\phantom A}\;.
\ea
The matrix of the relativistic precession consists of the contribution of the Lense-Thirring, de Sitter and Thomas precessions which correspond to the first, second and third terms in the right hand side of \eqref{5.18} respectively.

The mass and spin multipoles of each body appearing in the equations of motion \eqref{MBaB1}--\eqref{eee3s} describe the internal degrees of freedom of motion of matter of the corresponding body. They are treated in the equations of motion as purely spatial tensors residing in the tangent spacetime manifold being orthogonal to the world line of the origin of the local coordinates adapted to the body. The time dependence of the multipoles is not constrained by the condition of rigid rotation. We assume the most general case that the multipoles evolves over time in accordance with their own equations of motion derived from the microscopic equations of motion of matter inside the corresponding body. In the present paper we need three equations of motion of the multipoles in the Newtonian approximation. One of them is the equation of time evolution of the mass quadrupole moment which is derived in Appendix \ref{app1a}. Second equation is the equation of evolution of the spin dipole of body A. It reads \citep{dsx2,Racine_2006CQG,k2019PRD} 
\ba\label{vw291z3}
\dot{\Sc}^i_\A&=&\varepsilon_{ipq}\sum_{\B\neq\A}\sum_{l=0}^\infty\sum_{n=0}^\infty\frac{1}{l!n!}\M_\A^{pL}\M^N_\B\pd_{qLN}R^{-1}_{\A\B}\;,
\ea
where $\varepsilon_{ipq}$ is fully-antisymmetric symbol of Levi-Civita. The third equation is the equation of time evolution of the inertial mass \eqref{x1674}. 

Post-Newtonian equations of motion \eqref{MBaB1}--\eqref{eee3s} of the $N$-body problem with all multipoles taken into account represent a system of ordinary differential equations of the forth order. Indeed, the reader can observe that the last line in equation \eqref{eee3s} explicitly contains the first and second time derivatives of the acceleration of the center of mass of body A coupled with the spin-dipole and mass quadrupole of body A. This fact has been noticed by \citet{dsx2}.  Equation of motion depending on the higher-order time derivatives of the center of mass of moving bodies may lead to the appearance of spurious, self-accelerating solutions \citep{chicone_2001PhLA} which make no physical sense. A standard way to eliminate the self-accelerated modes is to replace the terms with higher-order time derivatives with the equations of motion of lower approximations which reduces the order of the time derivatives \citep{Landau1975,Damour_1983grr}. This procedure is applied for calculating the gravitational force in inspiralling compact binaries in the second, third and forth PN approximations  \citep{Blanchet_2002LRR,Blanchet_2003,blanchet_2005PhRvD7,schaefer_2011mmgr,Schaefer2018} where the terms with the higher order time derivatives appear due to the finite speed of gravity and retardation effects \citep{Damour_1983grr,Damour_1987book,Damour_1989MG5,gk86}. It should be noticed, however, that the nature of the terms depending on the higher-order time derivatives in the post-Newtonian force \eqref{eee3s} is not associated with the retardation of gravity but rather with the accelerated motion of the center of mass of body A and non-linear (quadratic) dependence of the transformations between global and local coordinates adapted to the body, on spatial coordinates.

Finally, we notice that the post-Newtonian equations of translational motion derived by \citet{racine_2005PhRvD,racine2013PhRvD} in general relativity are in complete agreement \footnote{For complete comparison see \citep[Appendix B]{k2019PRD}} with our equations \eqref{MBaB1}--\eqref{vw291z3} in the limiting case of $\b=\g=1$. We draw attention of the reader to the fact that it is important to know the dependence of body's multipoles on the external variables for derivation of the Lagrangian of the $N$-body problem. Therefore, we have computed the multipole moments as functions of the internal and external variables and shown their explicit form in section \ref{j2x10t} of the present paper. This part of the computation is omitted in the paper by \citet{racine_2005PhRvD}. 

\subsection{Lagrangian and Variational Derivative} 
     
The central task of the present paper is to solve the inverse problem of the Lagrangian mechanics \citep{Santilli_1978,Saunders_2010,Zenkov_2015} that is to find out the Lagrangian ${\cal L}$ corresponding to the translational equations of motion \eqref{MBaB1}--\eqref{eee3s} with the PN coefficients given in \eqref{oq7v3d}--\eqref{zw3k4b}. The post-Newtonian Lagrangian ${\cal L}$ of translational motion of an $N$-body system with arbitrary structured bodies possessing the mass and spin multipoles depends on the external dynamic variables which are coordinates and velocities of the centers of mass of the bodies. It can also depend on accelerations of the bodies as was shown in case of pole-dipole particles with spin by  \citet{damour_1982CR} and Barker and O'Connell \citep{Barker1986_acceleration,Barker_Oconnell1987JMP} (see also \citep{will_spin_PRD93,Mikoczi_2017PhRvD}). The Lagrangian of point-like massive particles depends on accelerations of the bodies non-linearly starting from the second PN approximations \citep{Damour_1985GReGr,gk86,Damour_1989MG5}. We have observed that the post-Newtonian force \eqref{eee3s} depends on the second order time derivatives of the accelerations. Hence, it suggests that the post-Newtonian Lagrangian of $N$ bodies having quadrupole and higher-order multipole moments should depend on accelerations as well. 

We consider the principle of the least action for the external problem of motion of $N$ bodies in the configuration space of 3N external dynamic variables which are the global coordinates $x^i_\A\equiv x^i_\A(t)$, velocity $v^i_\A\equiv v^i_\A(t)$ and acceleration $a^i_\A\equiv a^i_\A(t)$ of the center of mass of body A$\in\{1,2,...,{\rm N}\}$. The internal dynamic variable is any integral quantity taken over volume of body A from the interior distribution of mass density, velocity, etc. Such integrals do not depend explicitly on the external variables. The internal dynamic variables are identified with the Blanchet-Damour mass multipoles \eqref{wvzx35x}, the {\it non-canonical} multipoles \eqref{NL15a}, the spin multipoles \eqref{1.32} and a number of other integrals characterizing the internal structure of bodies like those entering the virial theorems -- see appendix \ref{app1a}. 

The primary internal dynamic variables of the $N$-body system are the mass and spin multipoles -- $\I^L_\A$ and $\Sc^L_\A$. We notice that the multipole $\I^L_\A$ makes up the basic (Newtonian) part of the mass multipole moment $\M^L_\A$ of body A given in \eqref{klz2a1} which tells us that $\M^L_\A$ is a composite function of both internal and external variables. Nonetheless, $\M^L_\A$ is convenient for more economic formulation of the equations of motion \eqref{MBaB1}--\eqref{eee3s}. Definition \eqref{1.32} of spin multipoles $\Sc^L_\A$ is sufficient in description of translational equations of motion in the post-Newtonian approximation and does not depend on the external variables explicitly. 

The internal dynamic variables are related to the external dynamic variables through the differential equations of motion of the internal problem. These equations are required for finding the Lagrangian of translational motion of the $N$-body system. In this paper we need the Newtonian equations of motion for the mass-monopole, spin-dipole and mass-quadrupole of body A. The equations of motion for mass-monopole and mass-quadrupole are derived in Appendix \ref{app1a}. The equation for spin-dipole is given in \eqref{vw291z3}. These equations are employed in section \ref{rsimt129} for derivation of the Lagrangian of translational motion. 

We denote the Lagrangian of the external problem as ${\cal L}\equiv{\cal L}_{\rm ext}$. We postulate that in the global coordinates the Lagrangian is given by a scalar function ${\cal L}\equiv{\cal L}\big(x_\A,v_\A,a_\A;Q_A,\dot Q_\A,\Lambda_\A\big)$, where $x_\A$, $v_\A$, $a_\A$ are coordinates, velocity and acceleration of the center of mass of the bodies, $Q_\A=\{M^L_\A,S^L_\A \}$ denotes a set of the internal dynamic variables which are the mass $M^L_\A$ and spin $S^L_\A$ multipoles, and $\Lambda_\A$ is a set of the internal integral variables like those entering the virial theorems explained in appendix \ref{app1a}. We denote the components of the multipoles in the global coordinates as $M^L_\A$ and $S^L_\A$ to distinguish from their components $\M^L_\A$ and $\Sc^L_\A$ computed in the local frame of body A. Both external and internal dynamic variables are functions of global coordinate time $t$ but the internal variables enter the Lagrangian of the external problem as time-dependent parameters. Equations of motion for the internal variables are derived not from ${\cal L}\equiv{\cal L}_{\rm ext}$ but the Lagrangian of the internal problem ${\cal L}_{\rm int}$ which we don't discuss in the present paper. 

Translational motion of bodies of an $N$-body system is determined by the action
\ba \label{lagr892}
S&=&\int\limits_{t_0}^{t_1}{\cal L}\big(x_\A,v_\A,a_\A;Q_A,\dot Q_\A,\Lambda_\A\big)dt\;.
\ea
Variation of the action leads to the variation of the Lagrangian
\ba\label{onxye648}
\delta S&=&\int\limits_{t_0}^{t_1}\delta{\cal L}dt=\int\limits_{t_0}^{t_1}\frac{\delta {\cal L}}{\delta x^i_\A}\delta x^i_\A dt;,
\ea
where the variation of coordinates, $\delta x^i_\A=x'^i_\A(t)-x^i_\A(t)$, is taken on a hypersurface of constant coordinate time $t$, and 
\ba\label{j8c5a2}
\frac{\delta {\cal L}}{\delta x^i_\A}&\equiv&-\frac{\pd {\cal L}}{\pd x^i_\A}+\frac{d}{dt}\frac{\pd {\cal L}}{\pd v^i_\A} -\frac{d^2}{dt^2}\frac{\pd {\cal L}}{\pd a^i_\A}\;,
\ea
denotes $\delta$-variational derivative of the Lagrangian in the extended configuration space of the external dynamic variables, \citep{Damour_1985GReGr,Petrov_2017book} taken under condition that all partial derivatives are computed with keeping the internal variables $\{Q_\A,\dot Q_\A,\Lambda_\A\}$ constant. The least action principle states that the corresponding variation of the action \eqref{lagr892} vanishes on-shell, $\delta S=0$, that yields the Euler-Lagrange equations of motion of the centers of mass of bodies in terms of the variational derivatives, 
\ba\label{j8c5a2qq}
\frac{\delta {\cal L}}{\delta x^i_\A}&=&0\;.
\ea
The $\delta$-variational derivative represents a covariant variational derivative expressed in a single global coordinate chart. The internal dynamic variables $Q_\A,\dot Q_\A,\Lambda_\A$ in this representation are expressed in the same coordinates as the external dynamic variables. It is the $\delta$-variational derivative which is used in covariant calculations of equations of motion in the effective field theories \citep{goldberger_PRD,Obukhov_Puetzfeld2014,Steinhoff_2015}.

Nonetheless, physically measurable are the components of multipoles expressed in the local frame adapted to each body and these, locally-frame multipoles appear in translational equations of motion \eqref{MBaB1}--\eqref{eee3s} of extended bodies in the global coordinates. Therefore, it is conceivable         
to look for the Lagrangian of the external problem considered as a scalar function $L\equiv L\big(x_\A,v_\A,a_\A;q_A,\dot q_\A,\lambda_\A\big)$ with the internal dynamic variables $q_\A=\{\M^L_\A,\Sc^L_\A\}$ and $\lambda_\A$ computed in the local frame of body A. To make the Lagrangians ${\cal L}$ and $L$ consistent, the variables $\{q_\A,\lambda_\A\}$ must be related to $\{Q_\A,\Lambda\}$ by the frame and/or gauge transformations of the internal variables, let say $q_\A=q_\A\le(x_\A,v_\A,Q_\A\ri)$ and $\lambda_\A=\lambda_\A\le(x_\A,v_\A,\Lambda_\A\ri)$ -- see sections \ref{ii58} and \ref{746352424} for specific details. It is clear that the concept of the variational derivative must be modified to accommodate the new set of variables. 

Let us introduce a new symbol $\textswab{d}$ for the variation of the action $S$ and corresponding dynamic variables which respects the variables $q_\A,\lambda_\A$ defined in the local frame of body A, 
\ba\label{omwy634241}
\textswab{d} S&=&\int\limits_{t_0}^{t_1}\textswab{d}L\big(x_\A,v_\A,a_\A;q_A,\dot q_\A,\lambda_\A\big)dt=\int\limits_{t_0}^{t_1}\frac{\textswab{d}L}{\textswab{d}x^i_\A}\textswab{d}x^i_\A dt\;.
\ea
The $\textswab{d}$-variation of the external variables in \eqref{omwy634241} is now performed with keeping constant values of the internal variables $\{q_\A,\dot{q}_\A,\lambda_\A\}$. 
The $\textswab{d}$-variational derivative is defined by formula 
\ba\label{ybzc35s80}
\frac{\textswab{d} L}{\textswab{d} x^i_\A}&\equiv& -\frac{\pd L}{\pd x^i_\A} +\frac{d}{dt}\frac{\pd L}{\pd v^i_\A}-\frac{d^2}{dt^2}\frac{\pd L}{\pd a^i_\A}=0\;,
\ea
where, by the new convention, the partial derivatives from the Lagrangian $L$ are computed under the condition that the internal variables $\{q_\A,\dot{q}_\A,\lambda_\A\}$ in the Lagrangian $L$ are kept constant.  
The least action principle, $\textswab{d} S=0$, applied to the Lagrangian $L$ yields translational equations of motion    
\ba\label{ybc1s80}
\frac{\textswab{d} L}{\textswab{d} x^i_\A}&=& 0\;,
\ea
which should be equivalent to the equations of motion \eqref{j8c5a2qq} after accounting for the law of transformation of the internal variables.

\subsection{The inverse problem of the Lagrangian mechanics}\label{hh3545ax2}

The primary goal of the present paper is to find a Lagrangian such that its variational derivative would be equal to the equations of motion \eqref{MBaB1}. Two possible approaches are conceivable based on two types of the Lagrangians and the internal dynamic variables explained in the previous section. We have 
\ba\label{po75zc2}
\frac{\delta {\cal L}}{\delta x^i_\A}={m}^{\phantom{i}}_{\A}a^i_{\A} -{\dutchcal F}^i_{\rm N} - {\dutchcal F}^i_{\rm pN}-\ddot{I}^i_\A\;,
\ea
or
\ba\label{i92bx64z}
\frac{\textswab{d}L}{\textswab{d}x^i_\A}={m}^{\phantom{i}}_{\A}a^i_{\A} -{\dutchcal F}^i_{\rm N} - {\dutchcal F}^i_{\rm pN}-\ddot{I}^i_\A\;,
\ea
The problem of derivation of the Lagrangian by solving either \eqref{po75zc2} for ${\cal L}$ of \eqref{i92bx64z} for $L$ is known as the inverse problem of the Lagrangian mechanics. It has been thoroughly analyzed by a number of previous researchers and the most notable results are summarized in  \citep{Santilli_1978,Saunders_2010,Zenkov_2015}. A general method developed for solving the inverse problem of the Lagrangian mechanics relies upon mapping of equations of motion to an adjoint system of partial differential equations for a scalar function associated with the Lagrangian if solution of the adjoint system exists. Finding solution of the adjoint system is not an easy problem of the theory of differential equations in partial derivatives and will not be implemented in the present paper due to the complexity of post-Newtonian equations of motion with all multipoles. 

In fact, there is no need to solve equations \eqref{po75zc2} or \eqref{i92bx64z} separately as they suggest that the two variational derivatives must be equal,
\ba\label{9h36xca}
\frac{\delta {\cal L}}{\delta x^i_\A}=\frac{\textswab{d}L}{\textswab{d}x^i_\A}\;.
\ea
which means that the Lagrangians ${\cal L}$ and $L$ are to be interrelated by some transformation which can be found by solving \eqref{9h36xca}. Assuming that the transformation between the internal variables is of the post-Newtonian order $q_\A=Q_\A+f_\A(x_\A,v_\A,Q_\A)$ and the dependence of the Lagrangians on $\dot q_\A$ and $\lambda_\A$ variables is only at the post-Newtonian terms, equation \eqref{9h36xca} is reduced to
\ba\label{yuwt34za12}
\frac{\delta}{\delta x^i_\A}\Bigl({\cal L}-L\Bigr)&=&\frac{\pd{\cal L}}{\pd Q_\A}\frac{\pd f_\A}{\pd x^i_\A}-\frac{d}{dt}\le[\frac{\pd{\cal L}}{\pd Q_\A}\frac{\pd f_\A}{\pd v^i_\A}\ri]\;,
\ea
which suggests that $L$ can be obtained from ${\cal L}$ by doing some transformation of the internal variables which is not obvious in the most general case of the bodies depending on all multipoles.
In what follows, we shall focus on solving equation \eqref{po75zc2} and finding the Lagrangian ${\cal L}$. Its counterpart, the Lagrangian $L$, which is a solution of \eqref{i92bx64z} will be found in section \ref{746352424} by applying a series of transformations of the internal variables and the virial theorems from appendix \ref{app1a} which is equivalent to solving equation \eqref{yuwt34za12}. The reason for this approach to the problem is simply due to the fact that we could solve the inverse problem by integrating \eqref{po75zc2} and to find the Lagrangian ${\cal L}$ while solution of the inverse problem by integrating \eqref{i92bx64z} turned out to be less obvious to tackle.  

The procedure that shall be used in the present paper to find out the Lagrangian ${\cal L}$ is based on the observation that the right hand side of \eqref{po75zc2} is represented in the form of the differential operator from coordinates of the center of mass of the bodies and depend on velocities and accelerations algebraically. The Lagrangian must have a similar algebraic-differential structure and can be found by a series of transformations bringing the terms with the second time derivatives of the multipole moments and accelerations of the bodies in the post-Newtonian force \eqref{eee3s}, to the form of the variational derivative \eqref{j8c5a2} accompanied with additional terms preserving the identity of the transformation. Some of the additional terms are, then, canceled out with similar terms in the post-Newtonian force. Repeating this procedure several times reveals that all terms in the right hand side of \eqref{po75zc2} are unambiguously reduced to the form of the variational derivative \eqref{j8c5a2} taken from a linear combination of scalar functions which, thus, represents the major part of the Lagrangian that we are looking for. Nonetheless, there remain a few troublesome terms in the right hand side of \eqref{po75zc2} which cannot be directly reduced to the variational derivative by applying the above procedure. These terms are associated with the implicit coupling of the internal and external degrees of freedom, and their reduction to the variational derivative requires utilization of the equations of motion of the internal problem for spin-dipole and mass-quadrupole moments. These equations map the troublesome terms in the translational equations of motion to a number of the integrals and their time derivatives which are the intrinsic variables. Rather remarkably, these intrinsic variables couple to the velocity of body A in such a way that they can be reduced to the variational derivative. It completes derivation of the Lagrangian.  

It should be clarified that solution of the inverse problem of the Lagrangian mechanics is not unique. Indeed, the $\delta$-variational derivative \eqref{j8c5a2} taken from a total time derivative of an arbitrary function $F(t,x_\A,v_\A;Q_\A,\dot{Q}_\A,\Lambda_\A)$ depending on coordinates and velocities of the bodies vanishes identically. Therefore, the Lagrangian ${\cal L}={\cal L}(t,x_\A,v_\A,a_\A;Q_\A,\dot{Q}_\A,\Lambda_\A)$ and 
\ba\label{imx5ea1}
\tilde{\cal L}\le(t,x_\A,v_\A,a_\A;Q_\A,\dot{Q}_\A\ri)&=&{\cal L}\le(t,x_\A,v_\A,a_\A;Q_\A,\dot{Q}_\A,\Lambda_\A\ri)+\frac{d}{dt}F\le(t,x_\A,v_\A;Q_\A,\dot{Q}_\A,\Lambda_\A\ri)\;,
\ea
are equivalent in the sense that they yield identical equations of motion. However, it may be necessary to take into account that adding a total time derivative to the Lagrangian must be accompanied with a corresponding change of the initial and/or boundary conditions for the $N$-body system. Naturally, the total time derivative added to the Lagrangian also contributes to the laws of conservation of the $N$-body system \citep{Petrov_2017book}. We shall discuss these subtleties somewhere else.  

Besides the Lagrangian transformations mentioned above, there are three auxiliary transformations which we have to perform before looking for the solution of the inverse problem. First, equations of motion \eqref{MBaB1}--\eqref{eee3s} have been derived and are given in terms of the STF partial derivatives taken from the coordinate distance $R_{\A\B}=|{\bm x}_\A-{\bm x}_\B|$ between each couple of bodies in the $N$-body system. However, the variational derivative \eqref{j8c5a2} has a free spatial index which is clearly split from the STF projection. It points out that we are to convert all STF partial derivatives to a symmetric form to peel off the index of the variational derivative. Second, the procedure of the matched asymptotic expansions used for derivation of the translational equations of motion makes the multipole moment of each body $B\neq A$ of the $N$-body system a function of the local coordinate time $u_\B$ which is simultaneous with the time argument $u_\A$ of multipoles of body A. However, the least action principle applied to the action \eqref{lagr892} suggests that the variation of the positions of the centers of mass of all bodies are taken on the hypersurface of constant time $t$ of the global coordinates. It means that all multipole moments in the equations of motion \eqref{MBaB1}--\eqref{eee3s} should be transported to the hypersurface of constant time $t$ before solving the inverse problem. Third, the components of multipole moments, $\M^L_A$ and $\Sc^L_\A$, entering the equations of motion \eqref{MBaB1}--\eqref{eee3s} refer to the local coordinates adapted to body A. Similarly, the components of multipole moments, $\M^L_B$ and $\Sc^L_\B$, refer to the local coordinates adapted to body B. However, the coordinates of the center of mass of the bodies, their velocities and accelerations refer to the global coordinates. We are working with the variational principle in the form of equation \eqref{onxye648} which assumes that components of the multipoles $M^L_A$ and $S^L_\A$ expressed in the global frame do not change under variation of the external dynamic variables. On the other hand, the variational derivative \eqref{j8c5a2} taken from the components of the multipole moments $\M^L_A$ and $\Sc^L_\A$ expressed in the local coordinates does not vanish as the local coordinates relate to the global ones by means of relativistic transformation which depends on the external variables. To eliminate this coordinate effect in our computation, we convert the components of multipole moments from the local to global coordinates. Next section discusses details of these three transformations.  

\section{Preparatory transformations of gravitational force}\label{mx5ql1}
\subsection{Frame conversion of multipoles}\label{ii58}

Gravitational force in the right hand side of equations of motion \eqref{MBaB1} is a function of body's internal multipoles which are considered as tensors in the tangent spacetime attached at each instant of time to the center of mass of the body \citep{th_1985,Dixon2015,kop_2019EPJP}. We introduce two reference frames in the tangent spacetime corresponding to the local and global coordinates. The local frame (L-frame) consists of four coordinate basis vectors ${\bf e}_\a$ being tangent to the axes of the local coordinates adapted to the body and passing through the center of mass of the body: ${\bf e}_\a=\le[\pd/\pd w^\a\ri]_{w^i=0}$. Notice that vectors of the coordinate basis are not normalized. By construction, the timelike vector ${\bf e}_0$ of the local frame defines direction of the four-velocity of the body's center of mass. Three other vectors ${\bf e}_i$ of the L-frame are spatial and orthogonal to each other as well as to ${\bf e}_0$. The global frame (G-frame) consists of four coordinate basis vectors ${\bf E}_\a$ being tangent to the axes of the global coordinates passing at each instant of time through the center of mass of the body: ${\bf E}_\a=\le[\pd/\pd x^\a\ri]_{x^i=x^i_\A}$. The coordinate basis vectors ${\bf E}_\a$ are neither mutually orthogonal in the tangent spacetime nor they are unit vectors. 

Transformation between the basis vectors on a tangent manifold affiliated with body A$\in\{1,2,...,{\rm N}\}$ reads 
\be \label{ii4s21}
{\bf E}_\a=\Lambda^\b{}_\a {\bf e}_\b\;,
\ee
where the components of the matrix of transformation $\Lambda^{\a}{}_{\b}$ are given by the law of transformation of the coordinate bases of the local and global coordinates taken at the origin of the local coordinates \citep[Eqs. 480--483]{k2019PRD}.
\ba
\label{1as1a}
\Lambda^0{}_0&=&1+\frac12 v^2_\A-\bar U({\bm x}_\A)+{\cal O}(4)\;,\\
\label{1as1b}
\Lambda^0{}_i&=&-v^i_{\A}(1+\frac12v^2_{\A})+2(1+\g)\bar U^i({\bm x}_\A)-(1+2\g)v^i_{\A}\bar U({\bm x}_\A)+{\cal O}(5)\;,\\
\label{1as1c}
\Lambda^i{}_0&=&-v^i_{\A}\left[1+\frac12v^2_{\A}+\g\bar U({\bm x}_\A)\right]-F^{ij}_\A v^j_{\A}+{\cal O}(5)\;,\\
\label{1as1d}
\Lambda^i{}_j&=&\d^{ij}\left[1+\g\bar U({\bm x}_\A)\right]+\frac12v^i_{\A}v^j_{\A}+F^{ij}_\A+{\cal O}(4)\;,
\ea
and $F^{ij}_\A$ is the skew-symmetric matrix of relativistic precession of the spatial axes of the L-frame with respect to the G-frame defined by solution of equation \eqref{5.18}. 

We have used calligraphic letters to denote four-dimensional tensor components of the mass, $\M^{\a_1...\a_l}_\A$, and spin, ${\Sc}^{\a_1...\a_l}_\A$, multipoles of body A in the L-frame. Components of the same tensors taken with respect to the G-frame will be denoted with roman letters, $M^{\a_1...\a_l}_\A$ and $S^{\a_1...\a_l}_\A$, respectively \footnote{Of course, the tensor indices of the components $\M^{\a_1...\a_l}_\A$, ${\Sc}^{\a_1...\a_l}_\A$ and $M^{\a_1...\a_l}_\A$, $S^{\a_1...\a_l}_\A$ belong correspondingly to the L and G frames.}. From the point of view of the Lagrangian formalism in the configuration space of the $N$-body system, the multipole moments of body A are the internal dynamic variables whose variation is independent of the variation of the external dynamic variable -- the global coordinates $x^i_\A$ of the center of mass of body A. We consider the variation of the coordinates of the center of mass of body A on the hypersurface of constant coordinate time $t$ of the global coordinates. It does not change the basis vectors of G-frame of body A. Therefore, the Lagrangian variation of the G-frame components $M^{\a_1...\a_l}_\A$ and $S^{\a_1...\a_l}_\A$ of the multipoles is nil. On the other hand, transformation \eqref{ii4s21} demonstrates that the variation of the basis vectors of the L-frame will not vanish because the components of the matrix $\Lambda^\b{}_\a$ depend on the coordinates and velocity of the center of mass of body A. Therefore, the L-frame components of multipoles $\M^L_\A$ and $\Sc^L_\A$ are functions of the external variables through the components of the matrix of transformation $\Lambda^\b{}_\a$. It means that the variational derivative from the L-frame components of the multipoles $\M^{\a_1...\a_l}\A$ and $\Sc^{\a_1...\a_l}\A$ do not vanish and must be computed explicitly. For this purpose it is convenient to transform the L-frame components of multipoles to the G-frame to make the dependence of the L-frame components on the external variables explicit.

To this end, we recollect that, by definition, the multipoles of body A are tensors which have only spatial components in the L-frame of the body. It means that the multipoles of body A are orthogonal to four-velocity of the world line of its center of mass.  We denote the four-velocity as ${\cal U}^\a$ in the L-frame, and as $U^\a$ in the G-frame. The condition of orthogonality of the multipole moments of body A to the four-velocity of its center of mass reads,
\be \label{qx410}
{\cal U}_\b \M^{\a_1...\a_{l-1}\b}_\A=U_\b M^{\a_1...\a_{l-1}\b}_\A\equiv 0\;,\qquad\qquad {\cal U}_\b {\Sc}^{\a_1...\a_{l-1}\b}_\A=U_\b S^{\a_1...\a_{l-1}\b}_\A\equiv 0\;.
\ee
In the L-frame of body A we have, ${\cal U}^0=1$ and ${\cal U}^i=0$, so that the condition \eqref{qx410} agrees with the condition that in the L-frame multipoles have only the spatial components $\M^L_\A$ and ${\Sc}^L_\A$. This property does not hold in the G-frame that is the time-like components of $M^{\a_1...\a_{l-1}\b}_\A$ and $S^{\a_1...\a_{l-1}\b}_\A$ do not vanish. However, all time components of the G-frame multipoles like $M^{0i_1..i_l}$, etc., can be expressed solely in terms of their spatial components by making use of \eqref{qx410}.    

The components of the multipoles in the L-frame are defined by volume integrals \eqref{act29}, \eqref{1.32}, where the local coordinates $w^\a=(u,w^i)$ should be treated as the affine coordinates of the L-frame. The affine coordinates of the G-frame in the tangent manifold are denoted as $X^\a=(T,X^i)$ where $T$ coincide with the global coordinate time, $T=t$, and $X^i=x^i-x^i_\A(t)$. Transformation between the affine coordinates of the tangent spacetime is equivalent to the transformation of 1-forms on the tangent manifold which is inverse to the transformation \eqref{ii4s21} of the basis vectors of the two frames,
\ba\label{berxc24d}
w^\a=\Lambda^\a{}_\b X^\b\;.
\ea
The frame transformation of the mass multipoles of body A should be performed with the post-Newtonian accuracy as the mass multipoles enter the Newtonian force in equations of motion \eqref{MBaB1}--\eqref{eee3s}. At the same time, the spin multipoles enter only the post-Newtonian force \eqref{eee3s} and their frame transformation is sufficient with the Newtonian accuracy. Thus, for the spin multipoles we simply have
\ba\label{nb4cz} 
\Sc^L_\A=S^L_\A\;,\qquad \Sc^N_\B=S^N_\B\;.
\ea

The L-frame tensor components, $\M^L_\A$, of the mass multipoles are transformed to the G-frame components, $M^L_\A$,
with the help of the coordinate transformation \eqref{berxc24d}. It yields
\be\label{hy2x}
\M^{i_1...i_l}_\A=\Lambda^{i_1}{}_{\a_1}...\Lambda^{i_l}{}_{\a_l}M^{\a_1...\a_l}_\A=l\Lambda^{i_1}{}_{j_1}...\Lambda^{i_{l-1}}{}_{j_{1-1}}\Lambda^{i_l}{}_{0}M^{j_1...j_{l-1}0}_\A+\Lambda^{i_1}{}_{j_1}...\Lambda^{i_l}{}_{j_l}M^{j_1...j_l}_\A+{\cal O}(4)\;.
\ee
This transformation includes the time component, $M^{\a_1...\a_{l-1}0}_\A$, of the mass multipole in the G-frame but it can be found from the condition \eqref{qx410} of orthogonality of the multipoles to the four-velocity of body A,
\be\label{lkn5d}
U_\a M^{\a_1...\a_{l-1}\a}_\A\equiv U_0 M^{\a_1...\a_{l-1}0}_\A+U_j M^{\a_1...\a_{l-1}j}_\A=0\;.
\ee
Four-velocity of body A in the G-frame is given in the Newtonian approximation by $U_0=-1$ and $U_i=v^i_\A$. Substituting this to \eqref{lkn5d} yields 
\be\label{oev52}
M^{j_1...j_{l-1}0}_\A=v_\A^p M^{j_1...j_{l-1}p}_\A\;,
\ee
that allows us to compute all time components of the multipoles in the G-frame from the corresponding spatial components of the multipole.
Hence, the frame transformation \eqref{hy2x} of the mass multipole components of body A can be written down solely in terms of the spatial components, 
\ba\label{nx5w}
\M^L_\A&=&
 M^L_\A-\frac{l}{2}v_{\A}^kv_{\A}^{<i_l}M_\A^{L-1>k}-lF_{\A}^{k<i_l}M_\A^{L-1>k}+ l\g\bar U({\bm x}_\A)M_\A^L-\frac{l(l-1)}{2(2l-1)} v_{\A}^{<i_l}v_{\A}^{i_{l-1}}{\aleph}_\A^{L-2>}\;,
  \ea
and similarly for body B,
\ba\label{gcvt5}
\M^N_\B
%&=&M^N_\B-\frac{n}{2}v_{\B}^kv_{\B}^{<i_n}M_\B^{N-1>k}-nF_{\B}^{k<i_n}M_\B^{N-1>k}+ nD_{\B}^{k<i_n}M_\B^{N-1>k}\\\nonumber
 &=&M^N_\B-\frac{n}{2}v_{\B}^kv_{\B}^{<i_n}M_\B^{N-1>k}-nF_{\B}^{k<i_n}M_\B^{N-1>k}+ n\g\bar U({\bm x}_\B)M_\B^N-\frac{n(n-1)}{2(2n-1)} v_{\B}^{<i_n}v_{\B}^{i_{n-1}}{\aleph}_\B^{N-2>}\;,
\ea
where the G-frame multipoles $M^L_\A$, $M^N_\B$ are explained below in \eqref{klz2b1}, \eqref{klzsd2}, and
\ba\label{aleph7354}
\aleph^L_\A = \int\limits_{{\cal V}_{\A}}\rho^*(T_\A,{\bm X})|{\bm X}|^2X^Ld^3X\;,\quad\quad
\aleph^N_\B = \int\limits_{{\cal V}_{\B}}\rho^*(T_\B,{\bm X})|{\bm X}|^2X^Nd^3X\;,
\ea
are the G-frame components of the {\it non-canonical} multipoles defined earlier in the L-frames of the corresponding bodies as ${\cal N}^L_\A$ and ${\cal N}^N_\B$. Time $T_\A$ in the G-frame of body A corresponds to the coordinate time $u_A$ in the L-frame of this body in accordance with \eqref{berxc24d}. 

The {\it active} mass multipoles $M^L_\A$ of body A take in the G-frame on the following form 
\ba\label{klz2b1}
 M_\A^L &=&\mM_\A^{L}+a_\A^p M^{pL}_\A+\gamma\bar U({\bm x}_\A)M_\A^L+(1-2\b)\sum_{k=0}^{\infty}\frac{1}{k!}\pd_K\bar U({\bm x}_\A){\gimel}_\A^{KL}+\Delta M^L_\A+\frac{l}{2l+1}a^{<i_l}{\aleph}_\A^{L-1>}\;,
\end{eqnarray}
where the Blanchet-Damour mass multipoles
\ba\label{wvzx35y}
\mM_\A^{L}&=&\int\limits_{{\cal V}_{\A}}\tilde{\sigma}(T_\A,\bm{X})X^{<L>} d^3X+\frac{1}{(2l+3)}\left[\frac12\ddot{\aleph}_\A^{L}-2(1+\gamma)\frac{2l+1}{l+1} \dot{\Re}_\A^{L}\right]\;,
\ea
and
\ba
\label{un141d}
  \Re^L_\A=\int\limits_{{\cal V}_{\A}}\rho^*(T_\A,{\bm X})\nu^p(T_\A,{\bm X})X^{<pL>}d^3X\;,
 \ea 
are the {\it non-canonical} multipoles computed in the G-frame, and the density $\tilde{\sigma}(T_\A,\bm{X})$ is exactly the same as defined above in \eqref{vv44hh66aa} after replacing coordinates $u\rightarrow T_\A$, ${\bm w}\rightarrow\bm{X}$. 
Integration in \eqref{wvzx35y}, \eqref{un141d} is performed over the hyperplane of constant time $T_\A$ of body A with making use of 
the post-Newtonian law of transformation the invariant mass density $\rho^*=\rho\sqrt{-g}u^0$  (see, e.g.,  \citep[Eq.4.81]{willbook})
\ba\label{jj77zz33aa}
\rho^*(T,{\bm X})d^3X&=&\rho^*(u,{\bm w})d^3w\;.
\ea

Similar expression defines the {\it active} mass multipoles of body B
\ba\label{klzsd2}
 M_\B^N &=&\mM_\B^{N}+a_\B^p M^{pN}_\B+\gamma\bar U({\bm x}_\B)M_\B^N+ (1-2\b)\sum_{k=0}^{\infty}\frac{1}{k!}\pd_K\bar U({\bm x}_\B){\gimel}_\B^{KN}+\Delta M_\B^N+\frac{n}{2n+1}a^{<i_n}{\aleph}_\B^{N-1>} \;,
 \ea 
 where
 \ba\label{e5z310}
\mM_\B^{N}&=&\int\limits_{{\cal V}_{\B}}\tilde{\sigma}(T_\B,\bm{X})X^{<N>} d^3X+\frac{1}{(2n+3)}\left[\frac12\ddot{\aleph}_\B^{N}-2(1+\gamma)\frac{2n+1}{n+1} \dot{\Re}_\B^{N}\right]\;, 
\end{eqnarray}
and
\ba\label{un1wd}
  \Re^N_\B&=&\int\limits_{{\cal V}_{\B}}\rho^*(T_\B,{\bm X})\nu^p(T_\B,{\bm X})X^{<pN>}d^3X\;.
 \ea 
Integration in \eqref{e5z310}, \eqref{un1wd} is performed over the hyperplane of a constant time $T_\B$ of body B. Time $T_\B$ in the G-frame of body B corresponds to the coordinate time $u_\B$ in the L-frame of body B in accordance with \eqref{berxc24d}. Correspondence between the times $T_\A$ and $T_\B$ will be explained in the next section \ref{ii57}. 

Post-Newtonian frame transformations \eqref{nx5w} and \eqref{gcvt5} do not affect the multipoles in the post-Newtonian force \eqref{eee3s} but bring about additional post-Newtonian corrections to the Newtonian force \eqref{w1q5} which now takes on the following form,
\ba\label{nn33ss66}
{\dutchcal F}^i_{\rm N}&=&
\sum_{\B\neq \A}\sum_{l=0}^{\infty}\sum_{n=0}^{\infty}\frac{(-1)^n}{l!n!}M_{\A}^{L}(T_\A)M_\B^{N}(T_\B)\pd_{iLN}R_{\A\B}^{-1}\\\nonumber
&&-\frac12\sum_{\B\neq \A}\sum_{l=0}^{\infty}\sum_{n=0}^{\infty}\frac{(-1)^n}{l!n!}\le(v^k_\A v^p_\A M_{\A}^{kL}M_\B^{N}- v^k_\B v^p_\B M_{\A}^{L}M_\B^{kN}\ri)\pd_{ipLN}R_{\A\B}^{-1}\\\nonumber
&&-\frac12\sum_{\B\neq \A}\sum_{l=0}^{\infty}\sum_{n=0}^{\infty}\frac{(-1)^n}{l!n!}\le(\frac{v^k_\A v^p_\A {\aleph}_{\A}^{L}M_\B^{N}}{2l+3}+\frac{v^k_\B v^p_\B M_{\A}^{L}{\aleph}_\B^{N}}{2n+3}\ri)\pd_{ikpLN}R_{\A\B}^{-1}\\\nonumber
&&-\sum_{\B\neq \A}\sum_{l=0}^{\infty}\sum_{n=0}^{\infty}\frac{(-1)^n}{l!n!}\le(F_{\A}^{kp} M_{\A}^{kL}M_\B^{N}-F_{\B}^{kp}  M_{\A}^{L}M_\B^{kN}\ri)\pd_{ipLN}R_{\A\B}^{-1}\\\nonumber
&&+\g\sum_{\B\neq \A}\sum_{l=0}^{\infty}\sum_{n=0}^{\infty}\frac{(-1)^n}{l!n!}\le[l\bar U({\bm x}_\A)M_\A^L M_\B^{N}+n\bar U({\bm x}_\B)M^L_\A M_\B^N\ri]\pd_{iLN}R_{\A\B}^{-1}\;,
\ea
where we have explicitly shown the time dependence of the mass multipoles in the first, purely Newtonian term in the right hand side of \eqref{nn33ss66}. 
Times $u_\A$ and $u_\B$ are simultaneous in the L-frame of body B but they are not simultaneous in the G-frame of this body \citep{k2019PRD}. Since times $T_\A$ and $T_\B$ are related to the local times $u_\A$ and $u_\B$ by linear transformation \eqref{berxc24d} the times $T_\A$ and $T_\B$ are not simultaneous either. We consider the time $T_A$ as a reference time for all bodies of the $N$-body system and take $T_A=t$. The time delay between $T_\A$ and $T_\B=T_A+\Delta_{\A\B}$ makes the multipoles of body B dependent on the external variables like the coordinates and velocities of the centers of mass of the bodies from the $N$-body system through the time delay which depends on coordinates and velocities of the center of mass of body B, $\Delta_{\A\B}=\Delta(v_\B,R_{\A\B})$. This dependence of the time arguments of multipoles on the external variables should be made explicit in order to compute the variational derivative from the multipoles. This is done in the next section.  

\subsection{Time transformation of the Newtonian force}\label{ii57}

Coordinates of the center of mass of each body of the $N$-body system are computed in the equations of motion \eqref{MBaB1}--\eqref{eee3s} at the same instant of global coordinate time $t$, that is $x^i_{\A}\equiv x^i_{\A}(t)$ and $x^i_{\B}\equiv x^i_{\B}(t)$. On the other hand, the mass multipole moments in the Newtonian force \eqref{nn33ss66} are taken at different times $T_A$ and $T_\B\neq T_\A$. The time difference is a consequence of the application of the matching procedure to establish a correspondence between the local and global coordinates which yields \citep[Section IX.B]{k2019PRD} 
\ba\label{nvz5sa}
T_\A=t\;,\qquad\qquad T_\B=t-v^p_\B R^p_{\A\B}\;,
\ea 
where $t$ is the value of the global coordinate time. 
The mass multipoles of body A are computed at the time $t$, that is $M^L_\A(T_\A)=M^L_\A(t)$. The mass multipoles $M^N_\B(T_\B)$ are expanded in the Taylor series around the instant of time $t$ by making use of \eqref{nvz5sa},
\ba\label{jn24z}
M^N_\B(T_\B)&\equiv& M^N_\B(t)-\dot{M}^N_\B v^p_\B R^p_{\A\B}+{\cal O}(4)\;.
\ea
From now on, we adopt the following notations for the G-frame components of the mass and spin multipoles taken on the hypersurface of constant time $t$, 
\ba\label{koe41xb}
M^L_\A\equiv M^L_\A(t)\;,\qquad\qquad M^N_\B\equiv M^N_\B(t)\;,\qquad\qquad S^L_\A\equiv S^L_\A(t)\;,\qquad\qquad S^N_\B\equiv S^N_\B(t)\;.
\ea 

Our next step is to
substitute \eqref{jn24z} to the first term in the right hand side of the Newtonian force \eqref{nn33ss66} and discard all residual terms of the order of ${\cal O}(4)$ which are redundant. It yields, 
\ba\label{ked9v2x0}
\sum_{{\B}\not=\A}\sum_{l=0}^\infty\sum_{n=0}^\infty\frac{(-1)^n}{l!n!}M^L_\A(T_\A){M}^N_{\B}(T_\B)\pd_{iLN}R^{-1}_{\A\B}
&=&\sum_{{\B}\not=\A}\sum_{l=0}^\infty\sum_{n=0}^\infty\frac{(-1)^n}{l!n!}\le[M^L_\A{M}^N_{\B}-{M}^L_\A\dot{M}^N_{{\B}}v^p_{{\B}}R^p_{\A{\B}}\ri]\pd_{iLN}R^{-1}_{\A\B}\;,
\ea
which can be recast to STF form with the help of Eq. (B15) from paper \citep{k2019PRD},
\ba\label{un3f}
&&\sum_{l=0}^\infty\sum_{n=0}^\infty\frac{(-1)^n}{l!n!}{M}^L_\A\dot{ M}^N_{{\B}}v^p_{{\B}}R^p_{{\A\B}}\pd_{iLN}R^{-1}_{\A\B} 
%\\\nonumber
=
\sum_{l=0}^\infty\sum_{n=0}^\infty\frac{(-1)^n}{l!n!}{M}^L_\A\left[\dot{ M}^N_{{\B}}v^p_{{\B}}\pd_{<iL>pN}R_{{\A\B}}
+v^p_{{\B}}\dot{M}^{pN}_{{\B}}\pd_{iLN}R^{-1}_{\A\B}\right]\;.
\ea
The term with the STF derivative in \eqref{un3f} should be transformed to a symmetric derivative by applying the peeling off decomposition of indices \citep[Eq. (A1)]{k2019PRD},  
\ba\label{ybw32}
M^L_\A\pd_{<iL>}R_{\A\B}
%&=&\frac1{l+1}M^L_\A\le[\pd_i\pd_{<L>}+l\pd_{i<L-1}\pd_{i_l>}-\frac{2l}{2l+1}\pd_k\pd_{<k<L-1}>\d_{i_l>i}\ri]R_{\A\B}
%\\\nonumber
&=&M^L_\A\pd_{iL}R_{\A\B}-\frac{2l}{2l+1}M^{iL-1}_\A\pd_{L-1}R^{-1}_{\A\B}\;.
\ea
It allows us to bring \eqref{ked9v2x0} to the following form,
\ba\label{nb3cx}
\sum_{{\B}\not=\A}\sum_{l=0}^\infty\sum_{n=0}^\infty\frac{(-1)^n}{l!n!}M^L_\A(T_\A){M}^N_{\B}(T_\B)\pd_{iLN}R^{-1}_{\A\B}
&=&\sum_{\B\neq \A}\sum_{l=0}^{\infty}\sum_{n=0}^{\infty}\frac{(-1)^n}{l!n!}{M}_{\A}^{L}{M}_\B^{N}\pd_{iLN}R_{\A\B}^{-1}\\\nonumber
&+&\sum_{\B\neq \A}\sum_{l=0}^{\infty}
\sum_{n=0}^\infty\frac{(-1)^n}{l!n!}\frac{2}{2l+3}M^{iL}_\A\dot{M}^N_\B v^p_{{\B}}\pd_{pLN}R^{-1}_{\A\B}\\\nonumber
&-&\sum_{\B\neq \A}\sum_{l=0}^{\infty}\sum_{n=0}^\infty\frac{(-1)^n}{l!n!}{M}^L_\A\left[\dot{M}^N_{{\B}}v^p_{{\B}}\pd_{ipLN}R_{\A\B}+v^p_{{\B}}\dot{M}^{pN}_{{\B}}\pd_{iLN}R^{-1}_{\A\B}\right]\;,
\ea
which should be substituted to the very first (Newtonian) term in the right hand side of \eqref{nn33ss66}.

\subsection{Transformation of the STF partial derivatives}\label{ii56}

Computation of the Lagrangian by solving the inverse problem equation \eqref{po75zc2}, is significantly simplified if all STF partial derivatives in the post-Newtonian gravity force ${\dutchcal F}^i_{\rm pN}$  are replaced with the symmetric partial derivatives entering definition of the STF projection \eqref{stfformula}. To this end, we notice that the STF derivatives from $R^{-1}_{\A\B}$ are already equivalent to the symmetric derivatives from $R^{-1}_{\A\B}$, in the sense that
\be
\pd_{<L>}R^{-1}_{\A\B}=\pd_L R^{-1}_{\A\B}\;.
\ee
This property is a direct consequence of the Laplace equation $\Delta R^{-1}_{\A\B}\equiv \pd^i\pd_i R^{-1}_\B|_{{\bm x}={\bm x}_\A}=0$ for function $R^{-1}_{\B}=|{\bm x}-{\bm x}_\B|^{-1}$ which nullifies all traces in $\pd_{<L>}R^{-1}_{\A\B}$. However, the post-Newtonian force ${\dutchcal F}^i_{\rm pN}$ contains STF derivatives from $R_{\A\B}$ which are not equivalent to the symmetric derivatives, that is $\pd_{<L>}R_{\A\B}\neq\pd_L R_{\A\B}$. Terms with the STF derivatives from $R_{\A\B}$ appear in the third line of \eqref{eee3s} and we are to bring all of them back to the symmetric form. 

For converting the term depending on the second time derivative from multipole ${{\cal M}}_\B^N$ we employ \citep[Eq. (271)]{k2019PRD} which yields,
\ba\la{yuv3}
\sum_{\B\not=\A}\sum_{l=0}^\infty\sum_{n=0}^{\infty}\frac{(-1)^n}{l!n!}{\cal M}_\A^L\ddot{{\cal M}}_\B^N\pd_{<iLN>}R_{\A\B}
&=&\sum_{\B\not=\A}\sum_{l=0}^\infty\sum_{n=0}^{\infty}\frac{(-1)^n}{l!n!}{M}_\A^L\ddot{{M}}_\B^N\pd_{iLN}R_{\A\B}
\\\nonumber
&+&2\sum_{\B\not=\A}\sum_{l=0}^\infty\sum_{n=0}^{\infty}\frac{(-1)^n}{l!n!}\frac{{M}_\A^{L}\ddot{M}_\B^{iN}-{M}_\A^{iL}\ddot{M}_\B^N}{2l+2n+3}\pd_{LN}R^{-1}_{\A\B}
\\\nonumber
&+&2\sum_{\B\not=\A}\sum_{l=0}^\infty\sum_{n=0}^{\infty}\frac{(-1)^n}{l!n!}\frac{{M}_\A^{kL}\ddot{{M}}_\B^{kN}}{2l+2n+5}\pd_{iLN}R^{-1}_{\A\B}\;,
\ea
where we have also utilized the frame transformation of the multipoles which is reduced in the post-Newtonian terms to a trivial replacement of all multipoles defined in the L-frame of a body to its G-frame.
%%%%%%%%%%%%%%%%%%%%%%%%%%%%%%%%%%%%%%%%%%%%

To convert the term depending on the acceleration $a_\B^p$ of the center of mass of body B in the third line of \eqref{eee3s} we resort to \citep[Eq. (273)]{k2019PRD}. It results in
\ba\label{nj8e3}
\sum_{\B\not=\A}\sum_{l=0}^\infty\sum_{n=0}^{\infty}\frac{(-1)^n}{l!n!}{\cal M}_\A^L{{\cal M}}_\B^Na_\B^p\pd_{<ipLN>}R_{\A\B}
&=&\sum_{\B\not=\A}\sum_{l=0}^\infty\sum_{n=0}^{\infty}\frac{(-1)^n}{l!n!}{M}_\A^L{{M}}_\B^Na_\B^p\pd_{ipLN}R_{\A\B}
\\\nonumber
&&-2\sum_{\B\not=\A}\sum_{l=0}^\infty\sum_{n=0}^{\infty}\frac{(-1)^n}{l!n!}\frac{{M}_\A^{L}{M}_\B^N}{2l+2n+3}\;a^i_\B\pd_{LN}R^{-1}_{\A\B}
\\\nonumber
&&+2\sum_{\B\not=\A}\sum_{l=0}^\infty\sum_{n=0}^{\infty}\frac{(-1)^n}{l!n!}\frac{{M}_\A^{L}{M}_\B^{iN}-{M}_\A^{iL}{M}_\B^N}{2l+2n+5}\;a^p_\B\pd_{pLN}R^{-1}_{\A\B}
\\\nonumber
&&+2\sum_{\B\not=\A}\sum_{l=0}^\infty\sum_{n=0}^{\infty}\frac{(-1)^n}{l!n!}\frac{{M}_\A^{L}{M}_\B^{pN}-{M}_\A^{pL}{M}_\B^N}{2l+2n+5}\;a^p_\B\pd_{iLN}R^{-1}_{\A\B}
\\\nonumber
&&+2\sum_{\B\not=\A}\sum_{l=0}^\infty\sum_{n=0}^{\infty}\frac{(-1)^n}{l!n!}\frac{{M}_\A^{kL}{M}_\B^{kN}}{2l+2n+7}\;a^p_\B\pd_{ipLN}R^{-1}_{\A\B}\;,
\ea
where again we have replaced in the right side of \eqref{nj8e3} the mass multipoles defined in the L-frame to those defined in the G-frame.
 
For transforming the term being quadratic with respect to velocities of the center of mass of the bodies in the third line of \eqref{eee3s} we employ \citep[Eq. (272)]{k2019PRD} along with \eqref{ybw32} of the present paper. It brings forth the following identity, 
\ba
\la{vet5v7}
\sum_{\B\not=\A}\sum_{l=0}^\infty\sum_{n=0}^{\infty}\frac{(-1)^n}{l!n!}{\cal M}^L_{\A}{\cal M}^N_{\B}v^p_{\B}v^q_{\B}\pd_{<ipqLN>}R_{\A\B}
&=&\sum_{\B\not=\A}\sum_{l=0}^\infty\sum_{n=0}^{\infty}\frac{(-1)^n}{l!n!}{M}^L_{\A}{M}^N_{\B}v^p_{\B}v^q_{\B}\pd_{ipqLN}R_{\B}
\\\nonumber
&&-\sum_{\B\not=\A}\sum_{l=0}^\infty\sum_{n=0}^{\infty}\frac{(-1)^n}{l!n!}\frac{2}{2l+2n+5}{M}^{L}_{\A}{M}^N_{\B}v^2_{\B}\pd_{iLN}R^{-1}_{\A\B}
\\\nonumber
&&-\sum_{\B\not=\A}\sum_{l=0}^\infty\sum_{n=0}^{\infty}\frac{(-1)^n}{l!n!}\frac{4}{2l+2n+5}{M}^{L}_{\A}{M}^{N}_{\B}v^i_{\B}v^p_{\B}\pd_{pLN}R^{-1}_{\A\B}
\\\nonumber
&&-\sum_{\B\not=\A}\sum_{l=0}^\infty\sum_{n=0}^{\infty}\frac{(-1)^n}{l!n!}\frac{2}{2l+2n+7}{M}^{iL}_{\A}{M}^N_{\B}v^p_{\B}v^q_{\B}\pd_{pqLN}R^{-1}_{\A\B}
\\\nonumber
&&+\sum_{\B\not=\A}\sum_{l=0}^\infty\sum_{n=0}^{\infty}\frac{(-1)^n}{l!n!}\frac{2}{2l+2n+7}{M}^{L}_{\A}{M}^{iN}_{\B}v^p_{\B}v^q_{\B}\pd_{pqLN}R^{-1}_{\A\B}
\\\nonumber
&&-\sum_{\B\not=\A}\sum_{l=0}^\infty\sum_{n=0}^{\infty}\frac{(-1)^n}{l!n!}\frac{4}{2l+2n+7}{M}^{qL}_{\A}{M}^N_{\B}v^p_{\B}v^q_{\B}\pd_{ipLN}R^{-1}_{\A\B}
\\\nonumber
&&+\sum_{\B\not=\A}\sum_{l=0}^\infty\sum_{n=0}^{\infty}\frac{(-1)^n}{l!n!}\frac{4}{2l+2n+7}{M}^{qN}_{\B}{M}^L_{\A}v^p_{\B}v^q_{\B}\pd_{ipLN}R^{-1}_{\A\B}
\\\nonumber
&&+\sum_{\B\not=\A}\sum_{l=0}^\infty\sum_{n=0}^{\infty}\frac{(-1)^n}{l!n!}
\frac{2}{2l+2n+9}{M}^{kL}_{\A}{M}^{kN}_{\B}v^p_{\B}v^q_{\B}\pd_{ipqLN}R^{-1}_{\A\B}\;.
\ea
This completes the requested transformation of the STF partial derivatives to a standard symmetric derivatives. 

\subsection{Reduced form of the equations of motion}

Now, we summarize the results of the preceding transformations from sections \ref{ii58}--\ref{ii57} by substituting them to equations \eqref{MBaB1}--\eqref{eee3s} and contracting similar terms. It yields the equations of motion \eqref{MBaB1}-\eqref{eee3s} in a reduced form that is suitable for construction of the Lagrangian by solving equation \eqref{po75zc2} of the inverse problem. It is convenient to split various terms in the reduced equations of motion in several distinctive groups depending separately on the mass and spin multipoles. as they can be treated independently. More explicitly, the reduced form of the equations of motion \eqref{MBaB1}--\eqref{eee3s} reads 
\ba\label{hywc5p}
m_\A a^i_\A&=&F^i_{\nn}+F^i_{\mm}+F^i_{\ss}\\\nonumber
&+&\frac12\sum_{\B\not=\A}\sum_{l=0}^\infty\sum_{n=0}^{\infty}\frac{(-1)^n}{l!n!}{M}_\A^L\Big[\ddot{{M}}_\B^N\pd_{iLN}- {{M}}_\B^Na_\B^p\pd_{ipLN}-2\dot{M}^N_{{\B}}v^p_{{\B}}\pd_{ipLN}+{{M}}_\B^Nv_\B^pv_\B^q\pd_{ipqLN}\Big]R_{\A\B}\\\nonumber
&+&3\left(a_\A^k\ddot{M}_{\A}^{ik}+2\dot{a}_\A^k\dot{M}_{\A}^{ik}+\ddot{a}_\A^k{ M}_{\A}^{ik}\right)-\varepsilon^{ikq}\left(2a^k_\A\dot{S}_\A^q+\dot{a}^k_\A{S}_\A^q\right)+\ddot{I}^i_\A\;,
\ea
where $m_\A$ is the inertial mass of body A defined above in \eqref{confmass}, $F^i_{\nn}$ is the Newtonian force, $F^i_{\mm}$ and $F^i_{\ss}$ are two components of the post-Newtonian force depending respectively on the mass and spin multipole components,
\ba\label{in2cz}
F^i_\nn&=&\sum_{\B\neq \A}\sum_{l=0}^{\infty}\sum_{n=0}^{\infty}\frac{(-1)^n}{l!n!}{M}_{\A}^{L}{M}_\B^{N}\pd_{iLN}R_{\A\B}^{-1}\\
\label{jne6x3}
F^i_{\mm}&=&
\sum_{\B\not=\A}\sum_{l=0}^{\infty}\sum_{n=0}^{\infty}\frac{(-1)^n}{l!n!}\Big[\Big(\a_\mm^{iLN}+\beta_\mm^{iLN}\Big)\pd_{LN}+\Big(\a_\mm^{ipLN}+\beta_\mm^{ipLN}\Big)\pd_{pLN}+\a_\mm^{ipqLN}\pd_{pqLN}\\\nonumber
&&\phantom{------------}+\Big(\a_\mm^{LN}+\beta_\mm^{LN}+\gamma_\mm^{LN}\Big)\pd_{iLN}+\a_\mm^{pLN}\pd_{ipLN}+\a_\mm^{pqLN}\pd_{ipqLN}\Big]R_{\A\B}^{-1}\\
\label{uev20v}
F^i_{\ss}&=&\sum_{\B\not=\A}\sum_{l=0}^{\infty}\sum_{n=0}^{\infty}\frac{(-1)^n}{l!n!}\Big[{\a}_\ss^{ipLN}\pd_{pLN}+\a_\ss^{ipqLN}\pd_{pqLN}
+\a_\ss^{pLN}\pd_{ipLN}+\a_\ss^{pqLN}\pd_{ipqLN}\Big]R_{\A\B}^{-1}
\;.
\ea
We emphasize that all spatial derivatives in \eqref{hywc5p}--\eqref{uev20v} are now the symmetric partial derivatives, and the differentiation with respect to spatial coordinates of the bodies is understood in the sense of equations \eqref{acser243}, \eqref{nrvx31w}. The PN coefficients of the differential operator in \eqref{jne6x3}, \eqref{uev20v} depend on time and are divided in two clusters depending on the mass and spin multipoles. These coefficients are explained below in full detail.
 
\subsubsection{Mass multipole coefficients}\label{b3eaz1}
The mass multipole coeffcients form three groups.
The first group of mass multipole coefficients includes those which depend solely on the mass multipoles, their time derivatives and/or velocities of the bodies. They are:
\ba\label{oq71}
\a_\mm^{iLN}&=&\Big[v^i_{\A}-2(1+\g)v^i_{\A\B}\Big]{M}^{L}_\A\dot{M}^{N}_{\B}-2(1+\g)\dot{M}^{L}_\A{M}^N_\B v^i_{\A\B}\\\nonumber
&&+\frac{2(1+\g)}{n+1}\le({M}^{L}_\A\ddot{M}^{iN}_\B+\dot{M}^{L}_\A\dot{M}^{iN}_\B\ri)-\frac{2(1+\g)}{l+1} \le(\ddot{M}^{iL}_\A{M}^N_\B-\dot{M}^{iL}_\A\dot{M}^N_\B\ri)\\\nonumber
&&-(l+1){M}^{iL}_\A\ddot{M}^N_\B -l \ddot{M}^{iL}_\A{M}^N_\B-(2l+1)\dot{M}^{iL}_\A\dot{M}^N_\B\;,\\
\la{xu3d2}
\a_\mm^{ipLN}&=&-\left[2(1+\g)v^i_{\A\B} v^p_{\A\B}+v^i_\A v^p_{\B}\right]{M}^{L}_\A{M}^{N}_\B-2(l+1){M}^{iL}_\A\dot{M}^{N}_\B v^p_{\A\B}-(2l+1)\dot{M}^{iL}_\A{M}^{N}_\B v^p_{\A\B}\\\nonumber
&&+\frac{2(1+\g)}{n+1}{M}^{L}_\A\dot{M}^{iN}_\B v^p_{\A\B}-\frac{2(1+\g)}{l+1}\dot{M}^{iL}_\A{M}^{N}_\B v^p_{\A\B}\;,\\
\la{pax4z8}
\a_\mm^{ipqLN}&=&-(l+1){M}^{iL}_\A{M}^N_\B v^p_{\A\B} v^q_{\A\B}\;,\\
\la{yes23}
\alpha_\mm^{LN}&=&\left[(1+\g)v^2_{\A\B}-\frac12v^2_{\B}-v^2_\A\right]{M}^L_\A{M}^N_{\B}+{M}^{kL}_\A\dot{M}^N_\B v^k_{\A}
-{M}^{L}_\A\dot{M}^{kN}_\B v^k_{\B}\\\nonumber
&&+\frac{2(1+\g)}{l+1}\dot{M}^{kL}_\A{M}^{N}_\B v^k_{\A\B}-\frac{2(1+\g)}{n+1}{M}^{L}_\A\dot{M}^{kN}_\B v^k_{\A\B}-\frac{2(1+\g)}{(l+1)(n+1)}\dot{M}^{kL}_\A\dot{M}^{kN}_\B\;,\\
\la{v3w}
\a_\mm^{pLN}&=&v^k_\B v^p_\B{M}^{L}_\A{M}^{kN}_\B- v^k_\A v^p_{\B}{M}^{kL}_\A{M}^{N}_\B  \;,\\\label{aaaxxx}
\a_\mm^{pqLN}&=&-\frac12\le(\frac{v^k_\A v^p_\A {\aleph}_{\A}^{L}M_\B^{N}}{2l+3}+\frac{v^k_\B v^p_\B M_{\A}^{L}{\aleph}_\B^{N}}{2n+3}\ri)\;,
\ea
%%%%%%%%%%%%%%%%%%%%%%%%%%%%%%%%%%
%%%%%%%%%%%%%%%%%%%%%%%%%%%%%%%%%
The second group of the mass multipole coefficients depend on the mass multipoles and accelerations of the center of mass of the bodies: 
\ba
\la{bet22}
\beta_\mm^{iLN}&=&\left[2\Big(1+\g\Big)a^i_{\B}-\Big(l+2+2\g\Big)a^i_\A\right]{M}^L_\A{M}^N_\B\;,
\\
\la{beta23z}
\beta_\mm^{ipLN}&=&\left[\Big(l+1\Big)a^p_{\B}-la^p_\A\right]{M}^{iL}_\A{M}^{N}_\B\;,
\\
\la{btg21q}
\beta_\mm^{LN}&=&-(l+1){M}^{kL}_\A{M}^{N}_\B a^k_{\A}-(n+1){M}^{L}_\A{M}^{kN}_\B a^k_\B\;,
\ea
%%%%%%%%%%%%%%%%%%%%%%%%%%
%%%%%%%%%%%%%%%%%%%%%%%%%%
The third group consists of just a single coefficient
\ba
\label{034z}
\gamma_\mm^{LN}&=&-\g\bar U({\bm x}_\B){M}^L_\A{M}^N_\B\;,
%\gamma_\mm^{pLN}&=&F^{kp}_\A{M}^{kL}_\A{M}^{N}_{\B}-F^{kp}_{\B}{M}^L_\A{M}^{kN}_{\B}\;,
\ea
describing gravitational coupling of the mass multipoles with the gravitational potential of the bodies being external to body B in the $N$-body system.
%%%%%%%%%%
\subsubsection{Spin multipole coefficients}
The spin multipole coefficients are split in two groups. The first group depends on the products of spin and mass multipoles and either their time derivatives or velocities of the bodies. They are:
\ba
\label{juz5q0}
\a_\ss^{ipLN}&=&\frac{2(1+\g)}{n+2}\varepsilon_{ipq}\left({M}^L_\A\dot{S}^{qN}_{\B}+\dot{M}^L_\A{S}^{qN}_\B\right)+\frac{2(1+\g)}{l+2}\varepsilon_{ipq}\left(\dot{S}^{qL}_\A{M}^N_{\B}+{S}^{qL}_\A\dot{M}^N_\B\right)
\;,\\
\label{hb4cz0}
\a_\ss^{ipqLN}&=&\frac{2(1+\g)}{l+2}\varepsilon_{ipk}{S}^{kL}_\A{M}^N_\B v^q_{\A\B}+\frac{2(1+\g)}{n+2}\varepsilon_{ipk}{M}^L_\A{S}^{kN}_\B v^q_{\A\B}\;,\\
\label{x6a01}
\a_\ss^{pLN}&=&\frac{2(1+\g)}{n+2}\varepsilon_{pkq}{M}^{L}_\A{S}^{qN}_\B  v^k_{\A\B}-\frac{2(1+\g)}{l+2}\varepsilon_{pkq}{S}^{kL}_\A{M}^{N}_\B v^q_{\A\B}\\\nonumber
&&+\frac{2(1+\g)}{(l+2)(n+1)}\varepsilon_{pkq}{S}^{kL}_\A\dot{M}^{qN}_\B+\frac{2(1+\g)}{(l+1)(n+2)}\varepsilon_{pkq}\dot{M}^{kL}_\A{S}^{qN}_{\B}\;,
\ea
The second group includes only one coefficient describing mutual gravitational coupling of the spin multipoles of bodies from the $N$-body system:
\ba
\la{bey6c4}
\a_\ss^{pqLN}&=&-\frac{2(1+\g)}{(l+2)(n+2)}{S}^{pL}_\A{S}^{qN}_\B\;.
\ea

Equations of motion \eqref{hywc5p}--\eqref{uev20v} replace the right hand side of \eqref{po75zc2} and will be used below for solving the inverse problem of the Lagrangian mechanics in the form
\ba\label{zeq48n}
\frac{\delta {\cal L}}{\delta x^i_\A}&=&{m}^{\phantom{i}}_{\A}a^i_{\A} -F^i_{\nn}-F^i_{\mm}-F^i_{\ss}\\\nonumber
&-&\frac12\sum_{\B\not=\A}\sum_{l=0}^\infty\sum_{n=0}^{\infty}\frac{(-1)^n}{l!n!}{M}_\A^L\Big[\ddot{{M}}_\B^N\pd_{iLN}- {{M}}_\B^Na_\B^p\pd_{ipLN}-2\dot{M}^N_{{\B}}v^p_{{\B}}\pd_{ipLN}+{{M}}_\B^Nv_\B^pv_\B^q\pd_{ipqLN}\Big]R_{\A\B}\\\nonumber
&-&3\left(a_\A^k\ddot{M}_{\A}^{ik}+2\dot{a}_\A^k\dot{M}_{\A}^{ik}+\ddot{a}_\A^k{ M}_{\A}^{ik}\right)+\varepsilon^{ikq}\left(2a^k_\A\dot{S}_\A^q+\dot{a}^k_\A{S}_\A^q\right)-\ddot{I}^i_\A\;.
\ea
We shall find the Lagrangian ${\cal L}$ by consecutively picking up the coefficients in the right hand side of \eqref{zeq48n} depending on the second time derivatives and bringing them to the form of the variational derivative \eqref{j8c5a2} with simultaneous reduction and simplification of the remaining terms. After converting all those terms to the variational derivatives and mutual cancellation of all similar terms we shall be left with a few residual terms which are not directly reducible to the variational derivative. Nonetheless, the residual terms can be converted to the variational derivative after transforming them to the time derivatives from the internal dynamic variables by applying the equations of motion \eqref{vw291z3} for spin dipole $S^i_\A$  and the virial theorem \eqref{jex61az8} for the mass symmetric moment ${\gimel}^{ij}_\A$. 
 
\section{The Lagrangian form of the Newtonian force} \label{z2a8gc}

We start solution of the inverse problem of translational motion of $N$ bodies from transformation of the first two terms in the right hand side of \eqref{zeq48n} which constitute the Newtonian second law of classic mechanics. The inertial term, ${m}^{\phantom{i}}_{\A}a^i_{\A}$, corresponds to the kinetic energy of translational motion of the bodies and the Newtonian force, $F^i_{\nn}$, complies with the potential energy of their Newtonian gravitational attraction.  

\subsection{Kinetic energy}\label{iop23az}

The kinetic energy of body A depends on its inertial mass that was defined in \eqref{confmass}. The inertial mass contains the intrinsic part $\mathfrak{m}_\A$ which does not depend explicitly on the external variables of the configuration space of the $N$-body system. However, it also contains the external multipoles moments ${\cal Q}_L$ and ${\cal P}$ which are the explicit functions of the coordinates of the centers of mass of the external bodies in accordance with definitions \eqref{jje8c} and \eqref{ju3cz41}. Due to the fact that {\it active} dipole moments of body A have a post-Newtonian order of magnitude, $\M^i_\A={\cal O}(2)$, its exact structure in the post-Newtonian terms is not important as the difference goes over to the second post-Newtonian order. To make equations look more homogeneous we replace the dipole term in the last (post-Newtonian) term of \eqref{confmass} $Q_i\M^i\rightarrow {\cal P}_i\M^i$. We can also neglect in this term the difference between the multipole moments defined in the local and global frames and equate, $\M^L_\A=M^L_\A$. We also use in the post-Newtonian terms notation $M_\A$ for gravitational mass. Under these conditions the inertial mass of body A \eqref{confmass} can be written down in the following form
\ba\la{rtg2}
m_\A&=&
{\mathfrak m}_\A+\g M_\A\sum_{\B\not=\A}\sum_{n=0}^\infty\frac{(-1)^n}{n!}{M}^N_\B\pd_N R^{-1}_{\A\B}-\sum_{\B\not=\A}\sum_{l=0}^\infty\sum_{n=0}^\infty\frac{(-1)^n}{l!n!}(l+1)M^L_\A{M}^N_\B\pd_{LN} R^{-1}_{\A\B}\;.
\ea
%Similarly, the inertial mass of body C is
%\ba
%\la{xndere3}
%m_\B&=&{\mathfrak m}_\B+\g \M_\B\bar U({\bm x}_\B)-\sum_{l=0}^\infty\frac{l+1}{l!}\pd_L \bar U({\bm x}_\B)\M^L_\B\\\nonumber
%&=&m_\B+\g \M_\B\sum_{\C\not=\B}\sum_{n=0}^\infty\frac{(-1)^n}{n!}{\M}^N_\C\pd_N R^{-1}_{\B\C}-\sum_{\C\not=\B}\sum_{l=0}^\infty\sum_{n=0}^\infty\frac{(-1)^n}{l!n!}(l+1)\M^L_\B{\M}^N_\C\pd_{LN} R^{-1}_{\B\C}\;,
%\ea

The inertial mass $m_\A$ is not constant. The rate of change of the inertial mass has been computed in \citep{dsx2,kovl_2004,Racine_2006CQG} and can be determined directly by taking a time derivative from both sides of \eqref{rtg2} with accounting for the time derivative of the mass $\mathfrak{m}_\A$ -- see \citep[Eq. 163]{k2019PRD} and \eqref{aa55zz77}. It yields
\ba \label{x1674} 
\dot{m}_\A&=&\g {M}_\A\dot{\bar U}({\bm x}_\A)-\sum_{l=0}^\infty\frac{1}{l!}\le[l\pd_L \bar U({\bm x}_\A)\dot{ M}^L_\A+(l+1)\pd_L \dot{\bar U}({\bm x}_\A){M}^L_\A\ri]\\\nonumber
&=&\g {M}_\A\sum_{\B\not=\A}\sum_{n=0}^\infty\frac{(-1)^n}{n!}\le[\dot{M}^N_\B\pd_{N} R^{-1}_{\A\B} +{M}^N_\B  v^p_{\A\B}\pd_{pN} R^{-1}_{\A\B}\ri]\\\nonumber
&-&\sum_{\B\not=\A}\sum_{l=0}^\infty\sum_{n=0}^\infty\frac{(-1)^n}{l!n!}\le[l\dot{M}^L_\A{M}^N_\B \pd_{LN} R^{-1}_{\A\B}+(l+1){M}^L_\A\dot{M}^N_\B \pd_{LN} R^{-1}_{\A\B}+(l+1){M}^L_\A{M}^N_\B v^p_{\A\B}\pd_{pLN} R^{-1}_{\A\B}\ri]\;.
\ea

Let us define the Newtonian kinetic energy $K_{\rm N}$ of translational motion of all bodies in the $N$-body system by a standard formula of classic mechanics 
\ba\label{yy34ccqq}
K_{\rm N}=\frac12\sum_\A{m}_\A v^2_\A\;.
\ea
The variational derivative \eqref{j8c5a2} from the kinetic energy $K_{\rm N}$ is computed with taking into account that the inertial mass $m_\A$ of body A depends explicitly on the coordinates of the external bodies of the $N$-body system. We have
\ba\label{m9v2c}
\frac{\delta K_{\rm N}}{\delta x^i_\A}&=&m_\A a^i_\A+\dot{m}_{\A}v^i_\A-\frac12v^2_\A \frac{\pd{m}_\A}{\pd x^i_\A}-\frac12\sum_{\B\not=\A} v^2_\B \frac{\pd{m}_\B}{\pd x^i_\A}\\\nonumber
%&=&M_\A a^i_\A+\g {M}_\A\dot{\bar U}({\bm x}_\A)v^i_\A-\sum_{l=0}^\infty\frac{1}{l!}\le[l{\cal Q}_L\dot{M}^L_\A+(l+1)\dot{\cal Q}_L{M}^L_\A\ri]v^i_\A\\\nonumber
%&&-\frac12v^2_\A\le[\g{M}_\A\pd_i{\bar U}({\bm x}_\A)- \sum_{l=0}^\infty\frac{l+1}{l!}{\cal Q}_{iL}{M}^L\ri] -\frac12\sum_{\A\not=\A} v^2_\A \frac{\pd M_\A}{\pd x^i_\A}    \\\nonumber
%
&=&{m}_\A a^i_\A+\g M_\A v^i_\A\sum_{\B\not=\A}\sum_{n=0}^\infty\frac{(-1)^n}{n!}\le[\dot{M}^N_\B\pd_{N} R^{-1}_{\A\B} +{M}^N_\B  v^p_{\A\B}\pd_{pN} R^{-1}_{\A\B}\ri]\\\nonumber
&-&\sum_{\B\not=\A}\sum_{l=0}^\infty\sum_{n=0}^\infty\frac{(-1)^n}{l!n!}\le[l\dot{ M}^L_\A{ M}^N_\B \pd_{LN} R^{-1}_{\A\B}+(l+1){ M}^L_\A\dot{ M}^N_\B \pd_{LN} R^{-1}_{\A\B}+(l+1){ M}^L_\A{ M}^N_\B v^p_{\A\B}\pd_{pLN} R^{-1}_{\A\B}\ri]v^i_\A\\\nonumber
&-&\frac12\g \sum_{\B\not=\A}\sum_{n=0}^\infty\frac{(-1)^n}{n!}v^2_\A{M}_\A{M}^N_\B\pd_{iN} R^{-1}_{\A\B}+\frac12 \sum_{\B\not=\A}\sum_{l=0}^\infty\sum_{n=0}^\infty\frac{(-1)^n}{l!n!}(l+1)v^2_\A{M}^L_\A{ M}^N_\B \pd_{iLN} R^{-1}_{\A\B}     \\\nonumber
&+&\frac12\g \sum_{\B\not=\A}\sum_{n=0}^\infty\frac{(-1)^n}{n!}v^2_\B{ M}_\B{ M}^N_\A\pd_{iN} R^{-1}_{\B\A}+\frac12 \sum_{\B\not=\A}\sum_{l=0}^\infty\sum_{n=0}^\infty\frac{(-1)^n}{l!n!}(n+1)v^2_\B{ M}^L_\A{ M}^N_\B \pd_{iLN} R^{-1}_{\A\B}\;,
\ea
where in the first term of the very last line the partial derivatives are taken with respect to coordinates of body B, that is $\pd_{iN} R^{-1}_{\B\A}\equiv\pd_{iN} R^{-1}_{\A}|_{{\bm x}={\bm x}_\B}$, where $R_\A=|{\bm x}-{\bm x}_\A|$. 

Some terms in the right hand side of \eqref{m9v2c} can be grouped together to form a variational derivative which can be taken to the left hand side of \eqref{m9v2c} to combine with the variational derivative from $K_{\rm N}$. More specifically, we notice that
\ba\la{sehy}
&&\frac{\delta }{\delta x^i_\A}\sum_{A}\le[\frac12 M_\A\bar U({\bm x}_\A)v^2_\A\ri]\;=\;\frac{\delta }{\delta x^i_\A}\le[\frac12 M_\A\bar U({\bm x}_\A)v^2_\A\ri]+\frac{\delta }{\delta x^i_\A}\sum_{\B\neq\A}\le[\frac12 M_\B\bar U({\bm x}_\B)v^2_\B\ri]\\\nonumber
&&\hspace{1cm}=M_\A \bar U({\bm x}_\A) a^i_\A-\frac12 \sum_{\B\not=\A}\sum_{n=0}^\infty\frac{(-1)^n}{n!}v^2_\A{M}_\A{M}^N_\B\pd_{iN} R^{-1}_{\A\B}+\frac12 \sum_{\B\not=\A}\sum_{n=0}^\infty\frac{(-1)^n}{n!}v^2_\B{ M}_\B{ M}^N_\A\pd_{iN} R^{-1}_{\B\A}\\\nonumber
&&\hspace{1cm}+M_\A v^i_\A\sum_{\B\not=\A}\sum_{n=0}^\infty\frac{(-1)^n}{n!}\le[\dot{M}^N_\B\pd_{N} R^{-1}_{\A\B} +{M}^N_\B  v^p_{\A\B}\pd_{pN} R^{-1}_{\A\B}\ri]\;.
\ea
Moreover, it is straightforward to show that in the first PN approximation the following identity is valid, 
\ba \label{he52cz5}
\frac{\delta}{\delta x^i_\A}\sum_\A\le(\frac18 {m}_\A v^4_\A\ri)&=&m_\A\le(\frac12 v^2_\A a^i_\A+v^i_{\A}v^p_{\A}a^p_\A\ri)\\\nonumber
&=&\sum_{{\B}\not=\A}\sum_{l=0}^{\infty}\sum_{n=0}^{\infty}\frac{(-1)^n}{l!n!}\le[\frac12 v^2_\A M^L_\A M^N_\B\pd_{iLN}R^{-1}_{\A\B}+v^i_\A v^p_\A M^L_\A M^N_\B\pd_{pLN}R^{-1}_{\A\B}\ri]\;.
\ea
This term represents a post-Newtonian correction to the kinetic energy \eqref{yy34ccqq} and it is reasonable to add it to the variational derivative \eqref{m9v2c} of the kinetic energy along with the term \eqref{sehy}.

Subtracting \eqref{sehy} from \eqref{m9v2c} and adding \eqref{he52cz5}, we notice that some similar terms in the so-obtained expression cancel out and the inertial, acceleration-dependent term in the right hand side of equation \eqref{zeq48n} takes on the form of the variational derivative plus some other terms,
\ba\label{uv3cz5}
{m}_\A a^i_\A&=&\frac{\delta {\cal L}_1}{\delta x^i_\A}+\g M_\A \bar U({\bm x}_\A) a^i_\A-M_\A v^i_{\A}v^p_{\A}a^p_\A-\frac12 M_\A v^2_\A a^i_\A\\\nonumber
&+&\sum_{\B\not=\A}\sum_{l=0}^\infty\sum_{n=0}^\infty\frac{(-1)^n}{l!n!}\le[l\dot{ M}^L_\A{ M}^N_\B \pd_{LN} R^{-1}_{\A\B}+(l+1){ M}^L_\A\dot{ M}^N_\B \pd_{LN} R^{-1}_{\A\B}+(l+1){ M}^L_\A{ M}^N_\B v^p_{\A\B}\pd_{pLN} R^{-1}_{\A\B}\ri]v^i_\A\\\nonumber
&-&\frac12\sum_{\B\not=\A}\sum_{l=0}^\infty\sum_{n=0}^\infty\frac{(-1)^n}{l!n!}(l+1)v^2_\A{M}^L_\A{ M}^N_\B \pd_{iLN} R^{-1}_{\A\B} -\frac12 \sum_{\B\not=\A}\sum_{l=0}^\infty\sum_{n=0}^\infty\frac{(-1)^n}{l!n!}(n+1)v^2_\B{ M}^L_\A{ M}^N_\B \pd_{iLN} R^{-1}_{\A\B}\;,
\ea 
where we have denoted
\ba \label{oo99mm33}
{\cal L}_1&=&\frac12\sum_\A {m}_\A v^2_\A\Bigl[1+\frac14 v^2_\A-\g\bar U({\bm x}_\A)\Bigr]\;.
\ea  
We shall use \eqref{uv3cz5} in section \ref{rsimt129} to combine it with other terms in the right hand side of \eqref{zeq48n} to derive the Lagrangian of translational motion of the $N$-body problem.

\subsection{Potential energy}\label{uq376xv2}
The Newtonian potential energy of gravitational interactions in the $N$-body system is associated with the gravitational force $F^i_\nn$ presented by \eqref{in2cz}. It depends on the active mass multipoles \eqref{klz2b1}, \eqref{klzsd2} which are functions the external dynamic variables. In what follows, we shall put the external variables apart in the gravitational force $F^i_\nn$ which allows to track them down explicitly and facilitate the derivation of the Lagrangian of the $N$-body system. Replacing the mass multipoles $M^L_\A$ and $M^N_\B$ with their definitions \eqref{klz2b1}, \eqref{klzsd2} in the Newtonian gravitational force $F^i_\nn$, we get a fairly long formula,
\ba \label{uev26xc}
F^i_\nn&=&\sum_{\B\not=\A}\sum_{l=0}^\infty\sum_{n=0}^\infty\frac{(-1)^n}{l!n!}{\mM}^L_\A{\mM}^N_\B\pd_{iLN}R^{-1}_{\A\B}\\\nonumber
&+&\g\sum_{\B\not=\A}\sum_{l=0}^\infty\sum_{n=0}^\infty\frac{(-1)^n}{l!n!}\Biggl[\bar U({\bm x}_\A)+\bar U({\bm x}_\B)\Biggr]{M}^L_\A M^N_\B\pd_{iLN}R^{-1}_{\A\B}\\\nonumber
&+&\sum_{\B\not=\A}\sum_{l=0}^\infty\sum_{n=0}^\infty\frac{(-1)^n}{l!n!}\le(a^p_\A M^{pL}_\A{M}^N_\B+a^p_\B M^{L}_\A{ M}^{pN}_\B\ri)\pd_{iLN}R^{-1}_{\A\B}\\\nonumber
&+&\sum_{\B\not=\A}\sum_{l=0}^\infty\sum_{n=0}^\infty\frac{(-1)^n}{l!n!}\le[\frac{la^{<i_l}_\A {\aleph}^{L-1>}_\A{ M}^N_\B}{2l+1}+\frac{na^{<i_n}_\B {\aleph}^{N-1>}_\B M^{L}_\A}{2n+1}\ri]\pd_{iLN}R^{-1}_{\A\B}\\\nonumber
&+&(1-2\b)\sum_{\B\not=\A}\sum_{l=0}^\infty\sum_{n=0}^\infty\sum_{k=0}^{\infty}\frac{(-1)^n}{l!n!k!}\le[\pd_K\bar U({\bm x}_\A){\gimel}_\A^{KL}M_\B^N+\pd_K\bar U({\bm x}_\B){\gimel}_\B^{KL}M^N_\A\ri]\pd_{iLN}R^{-1}_{\A\B}\;,
\ea
where the mass multipoles $\mM^L_\A\equiv \mM^L_\A(t)$ and $\mM^N_\B\equiv \mM^N_\B(t)$ are defined respectively in \eqref{wvzx35y}, \eqref{e5z310}, and taken at the same instant of time $t$ as $M^L_\A$ and $M^N_\B$ -- see \eqref{koe41xb}. We have also introduced new notation for the symmetric moments in the G-frame,
\be\label{symmom9374}
 \gimel^{L}_\A=\int\limits_{{\cal V}_{\A}}\rho^*(T_\A,{\bm X})X^{(i_1...i_l)}d^3X\;,\quad\quad \gimel^{L}_\B=\int\limits_{{\cal V}_{\B}}\rho^*(T_\B,{\bm X})X^{(i_1...i_l)}d^3X\;.
\ee
It is worth noticing that in the case $l=0$ the monopole symmetric moment coincides with the mass of body A, that is $\gimel_\A=M_\A$. We are also allowed to consider the dipole $\gimel^i_\A=M^i_\A=0$ in any post-Newtonian term. 

Almost each term in the decomposition \eqref{uev26xc} of the Newtonian force can be separately converted to the variational derivative either exactly or up to a term being equal to the ordinary, second-order time derivative from a linear combination of some functions which will be specified below. Indeed, it is straightforward to establish that the very first term in the right hand side of \eqref{uev26xc} is equivalent to the variational derivative  
\ba\label{rwteas}
\sum_{\B\not=\A}\sum_{l=0}^\infty\sum_{n=0}^\infty\frac{(-1)^n}{l!n!}{\mM}^L_\A{\mM}^N_\B\pd_{iLN}R^{-1}_{\A\B}&=&-\frac12\frac{\delta }{\delta x^i_\A}\biggl[\sum_\A\sum_{\B\not=\A}\sum_{l=0}^\infty\sum_{n=0}^\infty\frac{(-1)^n}{l!n!}{\mM}^L_\A{\mM}^N_\B\pd_{LN}R^{-1}_{\A\B}\biggr]\;,
\ea
where we have taken into account that the variational derivatives from the internal variables vanish: $\delta \mM^L_\A/\delta x^i_\A=0$, $\delta \mM^N_\B/\delta x^i_\A=0$.

The terms with the accelerations in the third and forth lines of the right hand side of \eqref{uev26xc} are converted to 
\ba \label{jj87}
&&\sum_{\B\not=\A}\sum_{l=0}^\infty\sum_{n=0}^\infty\frac{(-1)^n}{l!n!}\le(a^p_\A M^{pL}_\A{M}^N_\B+a^p_\B M^{L}_\A{ M}^{pN}_\B\ri)\pd_{iLN}R^{-1}_{\A\B}=\\\nonumber
&&\hspace{2cm} -\frac12\frac{\d}{\delta x^i_\A}\sum_\A\sum_{\B\not=\A}\sum_{l=0}^\infty\sum_{n=0}^\infty\frac{(-1)^n}{l!n!}\le(a^p_\A M^{pL}_\A{M}^N_\B+a^p_\B M^{L}_\A{ M}^{pN}_\B\ri)\pd_{LN}R^{-1}_{\A\B}\\\nonumber
&&\hspace{2cm}\phantom{=}-\frac{d^2}{dt^2}\sum_{\B\not=\A}\sum_{l=0}^\infty\sum_{n=0}^\infty\frac{(-1)^n}{l!n!}M^{iL}_\A{M}^N_\B\pd_{LN}R^{-1}_{\A\B}\;,
\ea
and
\ba
\label{hyw39z1}
&&\sum_{\B\not=\A}\sum_{l=0}^\infty\sum_{n=0}^\infty\frac{(-1)^n}{l!n!}\le[\frac{la^{<i_l}_\A {\aleph}^{L-1>}_\A{ M}^N_\B}{2l+1}+\frac{na^{<i_n}_\B {\aleph}^{N-1>}_\B M^{L}_\A}{2n+1}\ri]\pd_{iLN}R^{-1}_{\A\B}=\\\nonumber
&&\sum_{\B\not=\A}\sum_{l=0}^\infty\sum_{n=0}^\infty\frac{(-1)^n}{l!n!}\le[\frac{a^p_\A {\aleph}^{L}_\A{ M}^N_\B}{2l+3}-\frac{a^p_\B M^{L}_\A{\aleph}^{N}_\B}{2n+3}\ri]\pd_{ipLN}R^{-1}_{\A\B}=\\\nonumber
&&\hspace{2cm}-\frac12\frac{\d}{\delta x^i_\A}\sum_\A\sum_{\B\not=\A}\sum_{l=0}^\infty\sum_{n=0}^\infty\frac{(-1)^n}{l!n!}\le[\frac{a^p_\A {\aleph}^{L}_\A{ M}^N_\B}{2l+3}-\frac{a^p_\B M^{L}_\A{\aleph}^{N}_\B}{2n+3}\ri]\pd_{pLN}R^{-1}_{\A\B}\\\nonumber
&&\hspace{2cm}\phantom{=}-\frac{d^2}{dt^2}\sum_{\B\not=\A}\sum_{l=0}^\infty\sum_{n=0}^\infty\frac{(-1)^n}{l!n!}\frac{{\aleph}^{L}_\A{ M}^N_\B}{2l+3}\pd_{iLN}R^{-1}_{\A\B}\;.
\ea
The terms with the second-order time derivatives in the right hand side of \eqref{jj87}, \eqref{hyw39z1} can be removed from the equations of motion by appropriately choosing the shift function $I^i_\A$ in \eqref{zeq48n}. It corresponds to making translation of the coordinates of the center of mass of body A from the origin of its local coordinates, point $w^i=0$, to another point $w^i_\A$ defined by equation \eqref{yxc36cf} under the condition of vanishing conformal dipole moment ${\cal J}^i_\A=0$. More exactly, we choose
\ba\label{op1qz4}
I^i_\A&=&\sum_{\B\not=\A}\sum_{l=0}^\infty\sum_{n=0}^\infty\frac{(-1)^n}{l!n!}M^{iL}_\A{M}^N_\B\pd_{LN}R^{-1}_{\A\B}+\sum_{\B\not=\A}\sum_{l=0}^\infty\sum_{n=0}^\infty\frac{(-1)^n}{l!n!}\frac{{\aleph}^{L}_\A{ M}^N_\B}{2l+3}\pd_{iLN}R^{-1}_{\A\B}\;,
\ea
which means that the local coordinates of the center of mass of body A are now 
\ba \label{vv88zz22}
w^i_{\A}=-{m}^{-1}_\A I^i_\A\;.
\ea
It should be emphasized that the {\it active} mass and spin multipole moments $M^L_\A$ and $S^L_\A$ with $l\ge 1$ are still computed with respect to the origin of the local coordinates of the corresponding body, that is with respect to the point $w^i=0$ which is defined by the condition ${\cal J}^i_\A=0$. The shifted values of the mass multipoles, $\hat{M}^L_\A$ are related to the mass multipoles defined with respect to the origin of the local frame, by transformations
\ba
\hat{M}^L_\A=M^L_\A-lM^{<L-1}_\A w^{i_l>}_\A+{\cal O}(4)\;,\qquad&&\qquad \hat{M}^N_\B=M^N_\B-nM^{<N-1}_\B w^{i_l>}_\B+{\cal O}(4)\;,
\ea
where we have taken into account that the translation $w^i_\A$ has a post-Newtonian order of magnitude according to \eqref{op1qz4}, \eqref{vv88zz22}. 
Because the equations of motion depend on the mass and spin multipoles, $M^L_\A$ and $S^L_\A$, defined with respect to the origin of the local coordinates, we continue to use these multipoles in the rest of the present paper. 

The terms in the very last line of \eqref{uev26xc} represent the most laborious part in the computation of the variational derivative of the gravitational potential energy of $N$ bodies. These terms describe two-point and three-point non-linear interactions of the mass multipoles.  We follow \citet[\S 77]{fockbook} who did similar computations but only for monopole terms (the indices $l=k=0$). Extending Fock's calculation for all other indices $l\ge 0, k\ge 0$, we have
\ba
(1-2\b)\sum_{\B\not=\A}\sum_{l=0}^\infty\sum_{n=0}^\infty\sum_{k=0}^{\infty}\frac{(-1)^n}{l!n!k!}\Bigl[\pd_K \bar U({\bm x}_\A){\gimel}_\A^{KL}M_\B^N+\pd_K \bar U({\bm x}_\B){\gimel}_\B^{KL}M^N_\A\Bigr]\pd_{iLN}R^{-1}_{\A\B}
&=&-\frac{\d}{\delta x^i_\A}\Bigl({\cal L}_{ab}+{\cal L}_{abc}\Bigr)\;,
\ea
where
\ba\label{om02x1z}
{\cal L}_{ab}=\le(\frac12-\b\ri)\sum_\A\sum_{\B\neq\A}\Psi_{\A\B}\;,\qquad\qquad {\cal L}_{abc}=\le(\frac12-\b\ri)\sum_\A\sum_{\B\neq\A}\sum_{\C\neq\A,\B}\Psi_{\A\B\C}\;,
\ea
are defined in terms of fully-symmetric (with respect to the summation indices A,B,C) functions $\Psi_{\A\B}=\Psi_{(\A\B)}$, $\Psi_{\A\B\C}=\Psi_{(\A\B\C)}$. Function
\mathleft
\begin{flalign}
\label{ow5x41q}
\Psi_{\A\B}
&=\frac12\sum_{l=0}^{\infty}\sum_{n=0}^{\infty}\frac{1}{l!n!}\biggl[\gimel_\A^{LN}\pd_{L} \bar U_\B({\bm x}_\A)\pd_N \bar U_\B({\bm x}_\A)+\gimel_\B^{LN}\pd_L  \bar U_\A({\bm x}_\B)\pd_{N} \bar U_\A({\bm x}_\B)\biggr]\\\nonumber
%&=\frac12\sum_{l=0}^{\infty}\sum_{n=0}^{\infty}\sum_{k=0}^{\infty}\sum_{s=0}^{\infty}\frac{(-1)^{k+s}}{l!n!k!s!}\biggl[\gimel_\A^{LN}M^K_\B M_\B^S
%\le(\pd_{LK}R^{-1}_{\A\B}\ri)\le(\pd_{NS}R^{-1}_{\A\B}\ri) +\gimel_\B^{LN}M^K_\A M_\A^S\le(\pd_{LK}R^{-1}_{\B\A}\ri)\le(\pd_{NS}R^{-1}_{\B\A}\ri)\biggr]\\\nonumber
&=\frac12\sum_{l=0}^{\infty}\sum_{n=0}^{\infty}\sum_{k=0}^{\infty}\sum_{s=0}^{\infty}\frac{(-1)^{k+s}}{l!n!k!s!}\biggl[\gimel_\A^{LN}M^K_\B M_\B^S
 +\gimel_\B^{KS}M^L_\A M_\A^N\biggr]\le(\pd_{LK}R^{-1}_{\A\B}\ri)\le(\pd_{NS}R^{-1}_{\A\B}\ri)
\;, 
\end{flalign}
describes the two-point gravitational interaction, and function
\mathleft
\begin{flalign}
\label{iu3s5az}
\Psi_{\A\B\C}&=\frac13\sum_{l=0}^{\infty}\sum_{n=0}^{\infty}\frac{1}{l!n!}\biggl[\gimel_\A^{LN}\pd_{L} \bar U_\B({\bm x}_\A)\pd_N \bar U_\C({\bm x}_\A) +\gimel_\B^{LN}\pd_L \bar U_\C({\bm x}_\B)\pd_{N}\bar U_\A({\bm x}_\B)+\gimel_\C^{LN}\pd_L \bar U_\A({\bm x}_\C)\pd_{N} \bar U_\B({\bm x}_\C)\biggr]\\\nonumber
&=\frac13\sum_{l=0}^{\infty}\sum_{n=0}^{\infty}\sum_{k=0}^{\infty}\sum_{s=0}^{\infty}\frac{(-1)^{k+s}}{l!n!k!s!}\biggl[
\gimel_\A^{LN}M^K_\B M^S_\C\le(\pd_{LK}R^{-1}_{\A\B}\ri)\le(\pd_{NS}R^{-1}_{\A\C}\ri)
\\\nonumber
&\hspace{4.3cm}+
\gimel_\B^{LN}M^K_\C M^S_\A\le(\pd_{LK}R^{-1}_{\B\C}\ri)\le(\pd_{NS}R^{-1}_{\B\A}\ri)+
\gimel_\C^{LN}M^K_\A M^S_\B\le(\pd_{LK}R^{-1}_{\C\A}\ri)\le(\pd_{NS}R^{-1}_{\C\B}\ri)\biggr]
%\\\nonumber
%&=\frac13\sum_{l=0}^{\infty}\sum_{n=0}^{\infty}\sum_{k=0}^{\infty}\sum_{s=0}^{\infty}\frac{(-1)^{l+n}}{l!n!k!s!}\biggl[
%\gimel_\A^{LN}M^K_\B M^S_\C\pd^{LK}_\B\pd^{NS}_\C \frac{1}{ R_{\A\B}R_{\A\C}}\\\nonumber
%&\hspace{4.3cm}+
%\gimel_\B^{LN}M^K_\C M^S_\A\pd^{LK}_\C \pd^{NS}_\A \frac{1}{R_{\B\C} R_{\B\A}}+
%\gimel_\C^{LN}M^K_\A M^S_\B\pd^{LK}_\A\pd^{NS}_\B\frac{1}{ R_{\C\A} R_{\C\B}}\biggr]
%\\\nonumber
%&=\frac13\sum_{l=0}^{\infty}\sum_{n=0}^{\infty}\sum_{k=0}^{\infty}\sum_{s=0}^{\infty}\frac{(-1)^{l+n}}{l!n!k!s!}\biggl[
%\gimel_\A^{LN}M^K_\B M^S_\C\pd^{LK}_\B\pd^{NS}_\C\Delta_\A \\\nonumber
%&\hspace{4.3cm}+
%\gimel_\B^{LN}M^K_\C M^S_\A\pd^{LK}_\C \pd^{NS}_\A\Delta_\B +
%\gimel_\C^{LN}M^K_\A M^S_\B\pd^{LK}_\A\pd^{NS}_\B\Delta_\C \biggr]\log\le(R_{\A\B}+R_{\A\C}+R_{\B\C} \ri)
\;,
\end{flalign}
describes the three-point interaction. Notice that the {\it non-canonical} symmetric multipole moments $\gimel_\A^{LN}$ that appear in \eqref{ow5x41q}, \eqref{iu3s5az}, have been introduced in \eqref{symmom9374}. They can be replaced with their STF projections ${\cal I}_{\A(p)}^{IJ}\equiv{\cal I}_{\A(p)}^{<IJ>}$ by making use of formula (A22b) from paper by \citet{Blanchet_1989AIHPA}. For example, 
\ba\label{qq66tt22}
\gimel_\A^{LN}\le(\pd_{LK}R^{-1}_{\A\B}\ri)\le(\pd_{NS}R^{-1}_{\A\C}\ri)&=&\sum_{p=0}^{{\rm min}(l,n)}\frac{l!n!}{p!(l-p)!(n-p)!}\frac{(2l+2n-4p+1)!!}{(2l+2n-2p+1)!!}{\cal I}_{\A(p)}^{IJ}\le(\pd_{IPK}R^{-1}_{\A\B}\ri)\le(\pd_{JPS}R^{-1}_{\A\C}\ri)\;,
\ea
where
${\rm min}(l,n)$ denotes the minimum of two numbers $l$ and $n$, the indices $I\equiv L-P$, $J\equiv N-P$, 
\ba
{\cal I}_{\A(p)}^{IJ}=\int\limits_{{\cal V}_{\A}}\sigma(T_\A,{\bm X})|{\bm X}|^{2p}X^{<IJ>}d^3X\;,
\ea
and we have taken into account that the symmetric derivative from $R^{-1}_{\A\B}$ is equivalent to its STF derivative \footnote{Notice that the STF multipoles ${\cal I}_{\A(0)}^{IJ}=M_\A^{<IJ>}$ for $p=0$, and ${\cal I}_{\A(1)}^{IJ}={\aleph}_\A^{<IJ>}$ when $p=1$.}. Similar STF-decomposition equations can be written for other terms in \eqref{ow5x41q} and \eqref{iu3s5az}.

Accounting for the results obtained in this section, we can express the Newtonian force \eqref{uev26xc} in the form of the variational derivative supplemented with terms depending on the external gravitational potentials $\bar U({\bm x}_\A)$, $\bar U({\bm x}_\B)$ for body A and B respectively, and a second time derivative of vector function \eqref{op1qz4} which cancel out the corresponding shift function in the right hand side of the inverse-problem equation \eqref{zeq48n},
\ba\label{rt45xc12}
F^i_\nn&=&-\frac{\delta }{\delta x^i_\A}\Bigl({\cal L}_2+{\cal L}_{ab}+{\cal L}_{abc}\Bigr)
+\g\sum_{\B\not=\A}\sum_{l=0}^\infty\sum_{n=0}^\infty\frac{(-1)^n}{l!n!}\Bigl[\bar U({\bm x}_\A)+\bar U({\bm x}_\B)\Bigr]{M}^L_\A M^N_\B\pd_{iLN}R^{-1}_{\A\B}-\ddot{I}^i_\A\;,
\ea
where 
\ba\label{iuw53cz}
{\cal L}_2&=&\frac12\sum_\A\sum_{\B\not=\A}\sum_{l=0}^\infty\sum_{n=0}^\infty\frac{(-1)^n}{l!n!}{\mM}^L_\A{\mM}^N_\B\pd_{LN}R^{-1}_{\A\B}\\\nonumber
&+&\frac12\sum_\A\sum_{\B\not=\A}\sum_{l=0}^\infty\sum_{n=0}^\infty\frac{(-1)^n}{l!n!}\le(a^p_\A M^{pL}_\A{M}^N_\B+a^p_\B M^{L}_\A{ M}^{pN}_\B\ri)\pd_{LN}R^{-1}_{\A\B}\\\nonumber
&+&\frac12\sum_\A\sum_{\B\not=\A}\sum_{l=0}^\infty\sum_{n=0}^\infty\frac{(-1)^n}{l!n!}\le[\frac{a^p_\A {\aleph}^{L}_\A{ M}^N_\B}{2l+3}-\frac{a^p_\B M^{L}_\A{\aleph}^{N}_\B}{2n+3}\ri]\pd_{pLN}R^{-1}_{\A\B}\;,
\ea
is the Newtonian-like potential energy of gravitational field  
where the mass multipole $\mM^L_\A$ and $\mM^N_\B$ belong to the G-frame of the bodies and have been defined in \eqref{wvzx35y}, \eqref{e5z310}.

Equation \eqref{rt45xc12} can be further simplified to 
\ba \label{jwq62v8}
F^i_\nn&=&-\frac{\delta }{\delta x^i_\A}\Bigl({\cal L}_2+{\cal L}_{ab}+{\cal L}_{abc}\Bigr)+\g M_\A\bar U({\bm x}_\A)a^i_\A+
\g\sum_{\B\not=\A}\sum_{l=0}^\infty\sum_{n=0}^\infty\frac{(-1)^n}{l!n!}\bar U({\bm x}_\B){M}^L_\A M^N_\B\pd_{iLN}R^{-1}_{\A\B}-\ddot{I}^i_\A\;,
\ea
where we have used the Newtonian equations of motion, $M_\B a^i_\A=F^i_\nn$, to reduce the term depending on the gravitational potential $\bar U({\bm x}_\A)$ of the bodies being external to body A. This replacement elucidates that the term depending on $\bar U({\bm x}_\A)$ in \eqref{jwq62v8} is exactly equal to a similar term in the right hand side of \eqref{uv3cz5} so that the two terms will mutually cancel out in the final reduction of similar terms in the right hand side of \eqref{zeq48n}.

\section{The Lagrangian form of the post-Newtonian force}\label{in2q9ze}

In this section we consider transformation of the post-Newtonian terms in the right hand side of \eqref{zeq48n} which contain time derivatives of the first, second and higher order, to the variational derivative. The terms with the time derivatives consist of four groups. The first group includes terms being proportional to the products of the partial derivatives from the coordinate distance $R_{\A\B}$ between the bodies and the time derivatives of the mass multipoles or acceleration of the bodies. The second group consists of terms being proportional to the products of the partial derivatives from the inverse of the coordinate distance $R^{-1}_{\A\B}$ with time derivatives of the mass multipoles or accelerations of the bodies which appear in force $F^i_\mm$. The third group consists of terms being proportional to the products of the partial derivatives from the inverse of the coordinate distance $R^{-1}_{\A\B}$ and the first time derivatives of the spin multipoles which constitute the force $F^i_\ss$. The forth group consists of the residual terms depending on the higher-order time derivatives of the multipoles and the external variables like velocity and acceleration of body A and their time derivatives. Below we consider these groups of terms one by one.

\subsection{Transformation of the higher-order derivative terms coupled to distance \texorpdfstring{$R_{\A\B}$}{RBC}}

Terms being proportional to the symmetric partial derivatives from $R_{\A\B}$ appear in the second line of equation \eqref{zeq48n}.
First, we transform to the variational derivative the term which is proportional to the second time derivative,
\ba\label{bc2xz7}
&&\sum_{\B\not=\A}\sum_{l=0}^\infty\sum_{n=0}^\infty\frac{(-1)^n}{l!n!}{M}_\A^L\ddot{{M}}_{\B}^N\pd_{iLN}R_{\A\B}
\;=\;\frac{\delta }{\delta x^i_\A}\sum_{\A}\sum_{\B\not=\A}\sum_{l=0}^\infty\sum_{n=0}^\infty\frac{(-1)^n}{l!n!}{M}_\A^L\dot{{M}}_{\B}^Nv^p_\A\pd_{pLN}R_{\A\B}\\\nonumber
&&\hspace{4cm}+\sum_{\B\not=\A}\sum_{l=0}^\infty\sum_{n=0}^\infty\frac{(-1)^n}{l!n!}\le[\le({M}_\A^L\dot{{M}}_{\B}^N-\dot{M}_\A^L{M}_{\B}^N\ri)v^p_\B\pd_{ipLN}R_{\A\B}
-\dot{M}_\A^L\dot{{M}}_{\B}^N\pd_{iLN}R_{\A\B}\ri]\;.
\ea
Next term is proportional to the acceleration. It is transformed as follows,
\ba\label{nrv3z}
&&\sum_{\B\not=\A}\sum_{l=0}^\infty\sum_{n=0}^\infty\frac{(-1)^n}{l!n!}{M}_\A^L{M}_{\B}^Na^p_\B\pd_{ipLN}R_{\A\B}
\;=\;\frac12\frac{\delta }{\delta x^i_\A}\sum_{\A}\sum_{\B\not=\A}\sum_{l=0}^\infty\sum_{n=0}^\infty\frac{(-1)^n}{l!n!}{M}_\A^L{M}_{\B}^Nv^p_{\A}v^q_{\B}\pd_{pqLN}R_{\A\B}\\\nonumber
&&+\sum_{\B\not=\A}\sum_{l=0}^\infty\sum_{n=0}^\infty\frac{(-1)^n}{l!n!}\le[{M}_\A^L{M}_{\B}^Nv^p_\B v^q_{\B}\pd_{ipqLN}R_{\A\B}-\dot{M}_\A^L{M}_{\B}^Nv^p_\B\pd_{ipLN}R_{\A\B}
-{M}_\A^L\dot{M}_{\B}^Nv^p_\B\pd_{ipLN}R_{\A\B}\ri]\;.
\ea
Making use of the above formulas we get that the combination of terms in the second line of \eqref{zeq48n} is completely reduced to a variational derivative without residual terms,
\ba\label{jn3vs0} 
\frac12\sum_{\B\not=\A}\sum_{l=0}^\infty\sum_{n=0}^{\infty}\frac{(-1)^n}{l!n!}{M}_\A^L\Big[\ddot{{M}}_\B^N\pd_{iLN}- {{M}}_\B^Na_\B^p\pd_{ipLN}-2\dot{M}^N_{{\B}}v^p_{{\B}}\pd_{ipLN}+{{M}}_\B^Nv_\B^pv_\B^q\pd_{ipqLN}\Big]R_{\A\B}
&=&-\frac{\delta {\cal L}_3}{\delta x^i_\A}\;.
\ea
where
\ba\label{pob36xr}
{\cal L}_3&=&\frac14\sum_{\A}\sum_{\B\not=\A}\sum_{l=0}^\infty\sum_{n=0}^\infty\frac{(-1)^n}{l!n!}\le[{M}_\A^L{M}_{\B}^Nv^p_{\A}v^q_{\B}\pd_{pqLN}-2{M}_\A^L\dot{{M}}_{\B}^Nv^p_\A\pd_{pLN}-\dot{M}_\A^L\dot{M}_{\B}^N\pd_{LN}\ri]R_{\A\B}\;.
\ea

\subsection{Transformation of the higher-order derivative terms in the force $F^i_\mm$}

The force $F^i_\mm$ is given in \eqref{jne6x3} with the PN coefficients listed in section \ref{b3eaz1}.  The higher-order derivative terms include the second time derivatives of the mass multipole moments and accelerations of the center of mass of the bodies of the $N$-body system. 

\subsubsection{Terms with the second time derivatives of the mass multipoles}

The second time derivatives of mass multipole moments appear only in the mass multipole coefficient $\a^{iLN}_\mm$ shown in \eqref{oq71}. The transformation equations for the four different terms with the second time derivatives, which appear in this coefficient, are as follows:
\ba\label{pev3z}
&&\sum_{\B\not=\A}\sum_{l=0}^\infty\sum_{n=0}^\infty\frac{(-1)^n}{l!n!}\frac{{M}_\A^L\ddot{{M}}_{\B}^{iN}}{n+1}\pd_{LN}R_{\A\B}^{-1}
\;=\;\frac{\delta }{\delta x^i_\A}\sum_{\A}\sum_{\B\not=\A}\sum_{l=0}^\infty\sum_{n=0}^\infty\frac{(-1)^n}{l!n!}\frac{{M}_\A^L\dot{{M}}_{\B}^{pN}v^p_\A}{n+1}\pd_{LN}R_{\A\B}^{-1}\\\nonumber
&+&\sum_{\B\not=\A}\sum_{l=0}^\infty\sum_{n=0}^\infty\frac{(-1)^n}{l!n!}\le[\le(\frac{\dot{{M}}_{\A}^{pL}{M}_\B^N v^p_\B}{l+1}+\frac{{M}_\A^L\dot{{M}}_{\B}^{pN}v^p_\A}{n+1}\ri)\pd_{iLN}R_{\A\B}^{-1}
-\frac{\dot{M}_\A^L\dot{{M}}_{\B}^{iN}}{n+1}\pd_{LN}R_{\A\B}^{-1}-\frac{{M}_\A^L\dot{{M}}_{\B}^{iN}v^p_{\A\B}}{n+1}\pd_{pLN}R_{\A\B}^{-1}\ri]\;,
\ea
\newline
\ba\label{peaz01}
&&\sum_{\B\not=\A}\sum_{l=0}^\infty\sum_{n=0}^\infty\frac{(-1)^n}{l!n!}\frac{\ddot{M}_\A^{iL}{{M}}_{\B}^{N}}{l+1}\pd_{LN}R_{\A\B}^{-1}
\;=\;\frac{\delta }{\delta x^i_\A}\sum_{\A}\sum_{\B\not=\A}\sum_{l=0}^\infty\sum_{n=0}^\infty\frac{(-1)^n}{l!n!}\frac{\dot{M}_\A^{pL}{M}_{\B}^{N}v^p_\A}{l+1}\pd_{LN}R_{\A\B}^{-1}\\\nonumber
&+&\sum_{\B\not=\A}\sum_{l=0}^\infty\sum_{n=0}^\infty\frac{(-1)^n}{l!n!}\le[\le(\frac{{{M}}_{\A}^{L}\dot{M}_\B^{pN} v^p_\B}{n+1}+\frac{\dot{M}_\A^{pL}{{M}}_{\B}^{N}v^p_\A}{l+1}\ri)\pd_{iLN}R_{\A\B}^{-1}
-\frac{\dot{M}_\A^{iL}\dot{{M}}_{\B}^{N}}{l+1}\pd_{LN}R_{\A\B}^{-1}-\frac{\dot{M}_\A^{iL}{{M}}_{\B}^{N}v^p_{\A\B}}{l+1}\pd_{pLN}R_{\A\B}^{-1}\ri]\;,
\ea
\newline
\ba\label{peaz02}
&&\sum_{\B\not=\A}\sum_{l=0}^\infty\sum_{n=0}^\infty\frac{(-1)^n}{l!n!}l\ddot{M}_\A^{iL}{{M}}_{\B}^{N}\pd_{LN}R_{\A\B}^{-1}
\;=\;\frac{\delta }{\delta x^i_\A}\sum_{\A}\sum_{\B\not=\A}\sum_{l=0}^\infty\sum_{n=0}^\infty\frac{(-1)^n}{l!n!}l\dot{M}_\A^{pL}{M}_{\B}^{N}v^p_\A\pd_{LN}R_{\A\B}^{-1}\\\nonumber
&+&\sum_{\B\not=\A}\sum_{l=0}^\infty\sum_{n=0}^\infty\frac{(-1)^n}{l!n!}\le[\le(n{{M}}_{\A}^{L}\dot{M}_\B^{pN} v^p_\B+l\dot{M}_\A^{pL}{{M}}_{\B}^{N}v^p_\A\ri)\pd_{iLN}R_{\A\B}^{-1}
-l\dot{M}_\A^{iL}\dot{{M}}_{\B}^{N}\pd_{LN}R_{\A\B}^{-1}-l\dot{M}_\A^{iL}{{M}}_{\B}^{N}v^p_{\A\B}\pd_{pLN}R_{\A\B}^{-1}\ri]\;,
\ea
\newline
\ba\label{peaz07}
&&\sum_{\B\not=\A}\sum_{l=0}^\infty\sum_{n=0}^\infty\frac{(-1)^n}{l!n!}(l+1){M}_\A^{iL}\ddot{{M}}_{\B}^{N}\pd_{LN}R_{\A\B}^{-1}
\;=\;\frac{\delta }{\delta x^i_\A}\sum_{\A}\sum_{\B\not=\A}\sum_{l=0}^\infty\sum_{n=0}^\infty\frac{(-1)^n}{l!n!}(l+1){M}_\A^{pL}\dot{M}_{\B}^{N}v^p_\A\pd_{LN}R_{\A\B}^{-1}\\\nonumber
&+&\sum_{\B\not=\A}\sum_{l=0}^\infty\sum_{n=0}^\infty\frac{(-1)^n}{l!n!}\le[(n+1){\dot{M}}_{\A}^{L}{M}_\B^{pN} v^p_\B+(l+1){M}_\A^{pL} \dot{M}_{\B}^{N}v^p_\A\ri]\pd_{iLN}R_{\A\B}^{-1}
\\\nonumber
&-&\sum_{\B\not=\A}\sum_{l=0}^\infty\sum_{n=0}^\infty\frac{(-1)^n}{l!n!}\le[
(l+1)\dot{M}_\A^{iL}\dot{{M}}_{\B}^{N}\pd_{LN}R_{\A\B}^{-1}+(l+1){M}_\A^{iL}\dot{{M}}_{\B}^{N}v^p_{\A\B}\pd_{pLN}R_{\A\B}^{-1}\ri]\;.
\ea

Transformations \eqref{pev3z}--\eqref{peaz07} convert all terms with the second time derivatives of the mass multipoles in the force $F^i_\mm$ to the variational derivative and a number of other terms which bring additional contributions to other mass multipole coefficients entering \eqref{jne6x3}. After reduction of similar terms the post-Newtonian force \eqref{jne6x3} is modified and takes on the following form
\ba\label{jk3z7}
F^i_\mm &=&-\frac{\delta {\cal L}_4}{\delta x^i_\A}+\sum_{\B\not=\A}\sum_{l=0}^{\infty}\sum_{n=0}^{\infty}\frac{(-1)^n}{l!n!}\Bigg[\le(\hat\a_\mm^{iLN}+\beta_\mm^{iLN}\ri)\pd_{LN}+\le(\hat\a_\mm^{ipLN}+\beta_\mm^{ipLN}\ri)\pd_{pLN}+\hat\a_\mm^{ipqLN}\pd_{pqLN}\\\nonumber
&&\hspace{3cm}+\le(\hat\a_\mm^{LN}+\beta_\mm^{LN}+\gamma_\mm^{LN}\ri)\pd_{iLN}+\a_\mm^{pLN}\pd_{ipLN}+\a_\mm^{pqLN}\pd_{ipqLN}\Bigg]R_{\A\B}^{-1}\;,
\ea
where
\ba\label{iyevd09}
{\cal L}_4&=&2(1+\g)
\sum_{\A}\sum_{\B\not=\A}\sum_{l=0}^\infty\sum_{n=0}^\infty\frac{(-1)^n}{l!n!}\le[\le(\frac{\dot{M}_\A^{pL}{M}_{\B}^{N}}{l+1}-\frac{M_\A^L\dot{{M}}_{\B}^{pN}}{n+1}\ri)v^p_{\A}- \frac{\dot{M}^{kL}_\A\dot{M}^{kN}_\B}{2(l+1)(n+1)} \ri]\pd_{LN}R_{\A\B}^{-1}\\\nonumber
&&\hspace{1cm}+\sum_{\A}\sum_{\B\not=\A}\sum_{l=0}^\infty\sum_{n=0}^\infty\frac{(-1)^n}{l!n!}\le[(l+1){M}_\A^{pL}\dot{M}_{\B}^{N}+l\dot{M}_\A^{pL}{M}_{\B}^{N}\ri]v^p_\A\pd_{LN}R_{\A\B}^{-1}\;,
\ea
and the coefficients in the force $F^i_\mm$, which have been modified by the above transformations, are marked with hat. They are 
\ba\label{oq12}
\hat\a_\mm^{iLN}&=&{M}^{L}_\A\dot{M}^{N}_{\B}v^i_{\A}-2(1+\g)\le({M}^{L}_\A\dot{M}^{N}_{\B}+\dot{M}^{L}_\A{M}^N_\B\ri) v^i_{\A\B}\;,
\\\la{xu3nq}
\hat\a_\mm^{ipLN}&=&-\left[2(1+\g)v^i_{\A\B} v^p_{\A\B}+v^i_\A v^p_{\B}\right]{M}^{L}_\A{M}^{N}_\B-(l+1)\le({M}^{iL}_\A\dot{M}^{N}_\B+\dot{M}^{iL}_\A{M}^{N}_\B\ri) v^p_{\A\B}\;,
\\
\la{px41zc}
\hat\a_\mm^{ipqLN}&=&-(l+1){M}^{iL}_\A{M}^N_\B v^p_{\A\B} v^q_{\A\B}\;,\\
\la{jevx23}
\hat\alpha_\mm^{LN}&=&\left[(1+\g)v^2_{\A\B}-\frac12v^2_{\B}-v^2_\A\right]{M}^L_\A{M}^N_{\B}\\\nonumber
&&-l\,\le({M}^{kL}_\A\dot{M}^N_\B+\dot{M}^{kL}_\A{M}^N_\B\ri) v^k_{\A}-(n+1)\le({M}^{L}_\A\dot{M}^{kN}_\B +\dot{M}^{L}_\A{M}^{kN}_\B\ri) v^k_{\B}\;.
\ea
The remaining coefficients in \eqref{jk3z7} (without the hat) are not affected by the transformations \eqref{pev3z}--\eqref{peaz07} and remain the same as shown in \eqref{v3w}--\eqref{034z}. Now, we are to single out the variational derivatives from acceleration-dependent terms in \eqref{jk3z7}. 

\subsubsection{Terms with the accelerations}

Equation for the force \eqref{jk3z7} includes four terms depending explicitly on accelerations of the center of mass of the bodies of the $N$-body system. These terms appear in the PN coefficients $\beta^{iLN}_\mm$ and $\beta^{ipLN}_\mm$. We transform these terms to the form of the variational derivative and combine the rest of the terms, obtained after these transformation, with similar terms in the remaining PN coefficients of the force \eqref{jk3z7}.

Two terms depend on the acceleration of body A. Their corresponding transformations are:
\ba\label{yyxx41a}
&&\sum_{\B\not=\A}\sum_{l=0}^\infty\sum_{n=0}^\infty\frac{(-1)^n}{l!n!}(l+2+2\g)a^i_\A M_\A^{L}M_{\B}^{N}\pd_{LN}R_{\A\B}^{-1}=\\\nonumber
&&\frac{\delta }{\delta x^i_\A}\sum_{\A}\sum_{\B\not=\A}\sum_{l=0}^\infty\sum_{n=0}^\infty\frac{(-1)^n}{l!n!}\le(\frac{l}2+1+\g\ri)v^2_\A M_\A^{L}M_{\B}^{N}\pd_{LN}R_{\A\B}^{-1}
\\\nonumber
&&\quad+\sum_{\B\not=\A}\sum_{l=0}^\infty\sum_{n=0}^\infty\frac{(-1)^n}{l!n!}\le[\le(\frac{l}2+1+\g\ri)v^2_\A+\le(\frac{n}2+1+\g\ri)v^2_\B\ri]M_\A^{L}M_{\B}^{N}\pd_{iLN}R_{\A\B}^{-1}\\\nonumber
&&\quad-\sum_{\B\not=\A}\sum_{l=0}^\infty\sum_{n=0}^\infty\frac{(-1)^n}{l!n!}\le(l+2+2\g\ri)\le[
\le(\dot{M}_\A^{L}M_{\B}^{N}+M_\A^{L}\dot{M}_{\B}^{N}\ri)v^i_\A\pd_{LN}R_{\A\B}^{-1}+ M_\A^{L}M_{\B}^{N}v^i_\A v^p_{\A\B}\pd_{pLN}R_{\A\B}^{-1}\ri]\;,
\ea
and
\ba
&&\sum_{\B\not=\A}\sum_{l=0}^\infty\sum_{n=0}^\infty\frac{(-1)^n}{l!n!}la^p_\A{M}_\A^{iL}{M}_{\B}^{N}\pd_{pLN}R_{\A\B}^{-1}=
\\\nonumber
&&\frac{\delta }{\delta x^i_\A}\sum_{\A}\sum_{\B\not=\A}\sum_{l=0}^\infty\sum_{n=0}^\infty\frac{(-1)^n}{l!n!}lv^p_\A v^q_A{M}_\A^{qL}{M}_{\B}^{N}\pd_{pLN}R_{\A\B}^{-1}
\\\nonumber
&&-\sum_{\B\not=\A}\sum_{l=0}^\infty\sum_{n=0}^\infty\frac{(-1)^n}{l!n!}la^p_\A{M}_\A^{pL}{M}_{\B}^{N}\pd_{iLN}R_{\A\B}^{-1}
-\sum_{\B\not=\A}\sum_{l=0}^\infty\sum_{n=0}^\infty\frac{(-1)^n}{l!n!}lv^p_\A {M}_\A^{iL}{M}_{\B}^{N}v^q_{\A\B}\pd_{pqLN}R_{\A\B}^{-1}
\\\nonumber
&&+\sum_{\B\not=\A}\sum_{l=0}^\infty\sum_{n=0}^\infty\frac{(-1)^n}{l!n!}\le[\le(lv^p_\B v^q_B{M}_\A^{qL}{M}_{\B}^{N}-nv^p_\B v^q_\B{M}_{\A}^{L}{M}_\B^{qN}\ri)\pd_{ipLN}R_{\A\B}^{-1}
-lv^p_\A\le( \dot{M}_\A^{iL}{M}_{\B}^{N}
+{M}_\A^{iL}\dot{M}_{\B}^{N}\ri)\pd_{pLN}R_{\A\B}^{-1}\ri]
\\\nonumber
&&-\sum_{\B\not=\A}\sum_{l=0}^\infty\sum_{n=0}^\infty\frac{(-1)^n}{l!n!}lv^p_\A\le(\dot{M}_\A^{pL}{M}_{\B}^{N}
+{M}_\A^{pL}\dot{M}_{\B}^{N}\ri)\pd_{iLN}R_{\A\B}^{-1}\;.
\ea
There are also two terms which depend on the acceleration of body B. Their transformations read:
\ba
&&\sum_{\B\not=\A}\sum_{l=0}^\infty\sum_{n=0}^\infty\frac{(-1)^n}{l!n!}a^i_\B{M}_\A^{L}{M}_{\B}^{N}\pd_{LN}R_{\A\B}^{-1}=
\\\nonumber
&&\frac{\delta }{\delta x^i_\A}\sum_{\A}\sum_{\B\not=\A}\sum_{l=0}^\infty\sum_{n=0}^\infty\frac{(-1)^n}{l!n!}\le[\frac12 v^p_\A v^p_\B{M}_\A^{L}{M}_{\B}^{N}\pd_{LN}R_{\A\B}^{-1}\ri]
\\\nonumber
&&\quad+\sum_{\B\not=\A}\sum_{l=0}^\infty\sum_{n=0}^\infty\frac{(-1)^n}{l!n!}{M}_\A^{L}{M}_{\B}^{N}v^p_\A v^p_\B\pd_{iLN}R_{\A\B}^{-1}\\\nonumber
&&\quad
-\sum_{\B\not=\A}\sum_{l=0}^\infty\sum_{n=0}^\infty\frac{(-1)^n}{l!n!}\le[\le(\dot{M}_\A^{L}{M}_{\B}^{N}+{M}_\A^{L}\dot{M}_{\B}^{N}\ri)v^i_\B\pd_{LN}R_{\A\B}^{-1}+{M}_\A^{L}{M}_{\B}^{N}v^i_\B v^p_{\A\B}\pd_{pLN}R_{\A\B}^{-1}\ri]\;,
\ea
and
\ba
&&\sum_{\B\not=\A}\sum_{l=0}^\infty\sum_{n=0}^\infty\frac{(-1)^n}{l!n!}(l+1)a^p_\B{M}_\A^{iL}{M}_{\B}^{N}\pd_{pLN}R_{\A\B}^{-1}=\\\nonumber
&&\frac{\delta }{\delta x^i_\A}\sum_{\A}\sum_{\B\not=\A}\sum_{l=0}^\infty\sum_{n=0}^\infty\frac{(-1)^n}{l!n!}(l+1)v^p_\B v^q_A{M}_\A^{qL}{M}_{\B}^{N}\pd_{pLN}R_{\A\B}^{-1}
\\\nonumber 
&&\quad
+\sum_{\B\not=\A}\sum_{l=0}^\infty\sum_{n=0}^\infty\frac{(-1)^n}{l!n!}\le[(l+1)v^p_\B v^q_B{M}_\A^{qL}{M}_{\B}^{N}-(n+1)v^p_\B v^q_\B{M}_{\A}^{L}{M}_\B^{qN}\ri]\pd_{ipLN}R_{\A\B}^{-1}\\\nonumber
&&\quad
-\sum_{\B\not=\A}\sum_{l=0}^\infty\sum_{n=0}^\infty\frac{(-1)^n}{l!n!}(l+1)\le[v^p_\B \le(\dot{M}_\A^{iL}{M}_{\B}^{N}+{M}_\A^{iL}\dot{M}_{\B}^{N}\ri)\pd_{pLN}R_{\A\B}^{-1}+v^p_\B {M}_\A^{iL}{M}_{\B}^{N}v^q_{\A\B}\pd_{pqLN}R_{\A\B}^{-1}\ri]
\\\nonumber
&&\quad
+\sum_{\B\not=\A}\sum_{l=0}^\infty\sum_{n=0}^\infty\frac{(-1)^n}{l!n!}(n+1)\le[a^q_\B{M}_{\A}^{L}{M}_\B^{qN}+v^q_\B\le(\dot{M}_{\A}^{L}{M}_\B^{qN}+{M}_{\A}^{L}\dot{M}_\B^{qN}\ri)\ri]\pd_{iLN}R_{\A\B}^{-1}\;. 
\ea
 
After replacing the above transformations of the accelerations in \eqref{jk3z7}, the force $F^i_\mm$ takes on the following form:
\ba\label{koe7vq}
F^i_\mm &=&-\frac{\delta {\cal L}_4}{\delta x^i_\A}-\frac{\delta {\cal L}_5}{\delta x^i_\A}\\\nonumber   
&+&\sum_{\B\not=\A}\sum_{l=0}^{\infty}\sum_{n=0}^{\infty}\frac{(-1)^n}{l!n!}\Bigg[\tilde\a_\mm^{iLN}\pd_{LN}+\tilde\a_\mm^{ipLN}\pd_{pLN}+\tilde\a_\mm^{ipqLN}\pd_{pqLN}+\le(\tilde\a_\mm^{LN}+\tilde\beta_\mm^{LN}+\gamma_\mm^{LN}\ri)\pd_{iLN}+\a_\mm^{pqLN}\pd_{ipqLN}\Bigg]R_{\A\B}^{-1}\;,
\ea
where ${\cal L}_4$ has been derived previously in \eqref{iyevd09}. A new contribution to the variational derivative reads 
\ba\label{ivz419z}
{\cal L}_5&=&
\sum_{\A}\sum_{\B\not=\A}\sum_{l=0}^\infty\sum_{n=0}^\infty\frac{(-1)^n}{l!n!}\Biggl[\Bigl(1+\g+\frac{l}{2}\Bigr)v^2_\A-(1+\g)v^p_\A v^p_\B\Biggr]M_\A^{L}M_{\B}^{N}\pd_{LN}R_{\A\B}^{-1}  \\\nonumber
&+&\sum_{\A}\sum_{\B\not=\A}\sum_{l=0}^\infty\sum_{n=0}^\infty\frac{(-1)^n}{l!n!}\le[l v^p_\A-(l+1)v^p_\B\ri]v^k_\A M_\A^{kL}M_{\B}^{N}\pd_{pLN}R_{\A\B}^{-1}\;,
\ea  
and the PN coefficients which have been changed in $F^i_\mm$ after the reduction of similar terms, are marked with tilde and read,
\ba\label{2w12}
\tilde\a_\mm^{iLN}&=&\le[(l+1){M}^{L}_\A\dot{M}^{N}_{\B}+l\dot{M}^{L}_\A{M}^N_\B\ri]v^i_\A\;,\\\nonumber
\\\label{c4nq}
\tilde\a_\mm^{ipLN}&=&\le(l v_{\A\B}^p-v_\B^p\ri)v_\A^i{M}^{L}_\A{M}^{N}_\B- \le({M}^{iL}_\A\dot{M}^{N}_\B+\dot{M}^{iL}_\A{M}^{N}_\B\ri)v^p_\A\;,
\\\nonumber
\label{vjyrt4z8}
\tilde\a_\mm^{ipqLN}&=&-v^p_{\A} v^q_{\A\B}{M}^{iL}_\A{M}^N_\B\;,\\\nonumber
\label{ycxrz23}
\tilde\alpha_\mm^{LN}&=&-\frac12\le[v^2_\A -(l+1)v^2_\A-(n+1)v^2_\B\ri]{M}^L_\A{M}^N_{\B}\;,
\\
\label{mctw5z}
%\tilde\alpha_\mm^{pLN}&&=0\;,\\
\tilde\beta_\mm^{LN}&=&-M^{kL}_\A M^N_\B a^k_\A\;.
\ea
The coefficients $\a_\mm^{pqLN}$ and  $\gamma_\mm^{LN}$ are not affected by the above transformations and remain the same as in \eqref{aaaxxx} and \eqref{034z} respectively.

\subsection{Transformation of the spin multipole terms in the force $F^i_\ss$}

In this section we transform the spin multipole-dependent force $F^i_\ss$ given in \eqref{uev20v} to the Lagrangian form. To this end, we compute the following variational derivative,
\ba
&&\frac{\delta}{\delta x^i_\A}\sum_\A\sum_{\B\neq\A}\sum_{l=0}^\infty\sum_{n=0}^\infty\frac{(-1)^n}{l!n!}v^k_{\A\B}\varepsilon_{kpq}\frac{S_\A^{qL}M_\B^N}{l+2}\pd_{pLN}R^{-1}_{\A\B}\\\nonumber
&&\hspace{5cm}=\sum_{\B\neq\A}\sum_{l=0}^\infty\sum_{n=0}^\infty\frac{(-1)^n}{l!n!}\varepsilon_{ipq}\frac{S_\A^{qL}\dot M_\B^N+\dot S_\A^{qL}M_\B^N}{l+2}\pd_{pLN}R^{-1}_{\A\B}\\\nonumber
&&\hspace{5cm}+\sum_{\B\neq\A}\sum_{l=0}^\infty\sum_{n=0}^\infty\frac{(-1)^n}{l!n!}\varepsilon_{ipq}\frac{\dot M_\A^L S_\B^{qN}+M_\A^L\dot S_\B^{qN}}{n+2}\pd_{pLN}R^{-1}_{\A\B}\\\nonumber
&&\hspace{5cm}+\sum_{\B\neq\A}\sum_{l=0}^\infty\sum_{n=0}^\infty\frac{(-1)^n}{l!n!}\varepsilon_{ipk}\le[\frac{S_\A^{kL}M_\B^N}{l+2}+\frac{M_\A^L S_\B^{kN}}{n+2}\ri]v^q_{\A\B}\pd_{pqLN}R^{-1}_{\A\B}\\\nonumber
&&\hspace{5cm}-\sum_{\B\neq\A}\sum_{l=0}^\infty\sum_{n=0}^\infty\frac{(-1)^n}{l!n!}\varepsilon_{kpq}\le[\frac{S_\A^{qL}M_\B^N}{l+2}+\frac{M_\A^L S_\B^{qN}}{n+2}\ri]v^k_{\A\B}\pd_{ipLN}R^{-1}_{\A\B}\;.
\ea
This formula, after having been compared with expression \eqref{uev20v} for the spin multipole-dependent force $F^i_\ss$, elucidates that the force can be exactly reduced to the variational derivative,
\ba\label{nucz3ap}
\F^i_{\ss}&=&-\frac{\d{\cal L}_6}{\d x^i_\A}\;,
\ea
where
\ba\label{gtce29d}
{\cal L}_6
&=&2(1+\g)\sum_\A\sum_{\B\neq\A}\sum_{l=0}^\infty\sum_{n=0}^\infty\frac{(-1)^n}{l!n!}\varepsilon_{pkq}v^k_{\A\B}\frac{S_\A^{qL}M_\B^N }{l+2}\pd_{pLN}R^{-1}_{\A\B}\\\nonumber
&+&2(1+\g)\sum_\A\sum_{\B\neq\A}\sum_{l=0}^\infty\sum_{n=0}^\infty\frac{(-1)^n}{l!n!}\varepsilon_{pkq}\frac{\dot M_\A^{kL} S^{qN}_\B}{(l+1)(n+2)}\pd_{pLN}R^{-1}_{\A\B}\\\nonumber
&-&(1+\g)\sum_\A\sum_{\B\neq\A}\sum_{l=0}^\infty\sum_{n=0}^\infty\frac{(-1)^n}{l!n!}\frac{S_\A^{pL}S_\B^{qN}}{(l+2)(n+2)}\pd_{pqLN}R^{-1}_{\A\B}\;.
\ea

\subsection{Reduction of the residual terms. Application of the virial theorem.}\label{rsimt129}

The post-Newtonian Lagrangian is obtained by replacing \eqref{uv3cz5}, \eqref{jwq62v8}, \eqref{jn3vs0}, \eqref{koe7vq}, and \eqref{nucz3ap} in the corresponding terms of equation \eqref{zeq48n} of the inverse problem of the Lagrangian mechanics. This makes the right hand side of \eqref{zeq48n} consisting of two groups of terms. The first group is represented in the form of the variational derivative from a linear combination of scalars ${\cal L}_1$, ${\cal L}_2$,...,${\cal L}_6$. The second group includes a few residual terms which have not yet been represented in the form of the variational derivative. The goal of this subsection is to demonstrate that the residual terms are also reduced to the variational derivative after some additional transformations. 

First of all, we notice after inspection of structure of the residual terms that some of them can be directly converted to the Lagrangian form. Indeed, the terms in the last line of \eqref{zeq48n} depending on the STF quadrupole moment $M^{ij}_\A$ of the body A and its higher-order time derivatives can be written as
\ba\label{ib3x1a}
3\le(a_\A^k\ddot{M}_{\A}^{ik}+2\dot{a}_\A^k\dot{ M}_{\A}^{ik}+\ddot{a}_\A^k{M}_{\A}^{ik}\ri)&=&\frac{d^2}{dt^2}\le(3{a}_\A^k{M}_{\A}^{ik}\ri)=-\frac{\d{\cal L}_7}{\delta x^i_\A}\;, 
\ea
with 
\ba 
{\cal L}_7=\frac32\sum_\A M^{pq}_\A a^p_\A a^q_\A\;.
\ea
Alternatively, the terms depending on the STF quadrupole moment $M^{ij}_\A$ of the body A and its higher-order time derivatives can be removed from the equations of motion by making a post-Newtonian translation of the position of the center of mass of body A with respect to the origin of the local coordinates. This is what we did in our paper \citep{k2019PRD} to circumvent the appearance of possible spurious solutions of the equations of motion with higher-order time derivatives \citep{chicone_2001PhLA}. However, since the acceleration-dependent terms \eqref{ib3x1a} are fully reduced to the variational derivative from the scalar ${\cal L}_7$, we prefer in the present paper to include them to the total Lagrangian of the $N$-body problem instead of making a complementary shift in the position of the center of mass of body A in addition to that shown in \eqref{vv88zz22}. 

After putting together  different pieces of the equations of motion given by equations \eqref{uv3cz5}, \eqref{jwq62v8}, \eqref{jn3vs0}, \eqref{koe7vq}, \eqref{nucz3ap} and \eqref{ib3x1a} and canceling similar terms, the equation of the inverse problem \eqref{zeq48n} acquires the following form
\ba\label{kl9v2cz4}
\frac{\delta {\cal L}}{\delta x^i_\A}&=&\frac{\delta }{\delta x^i_\A}\Bigl({\cal L}_1+{\cal L}_2+{\cal L}_3+{\cal L}_4+{\cal L}_5+{\cal L}_6+{\cal L}_7+{\cal L}_{ab}+{\cal L}_{abc}\Bigr)\\\nonumber
&&+\sum_{{\B}\not=\A}\sum_{l=0}^{\infty}\sum_{n=0}^{\infty}\frac{(-1)^n}{l!n!}\le[\le({M}^{iL}_\A\dot{M}^{N}_\B+\dot{M}^{iL}_\A{M}^{N}_\B\ri)v^p_\A\pd_{pLN}+v^p_{\A} v^q_{\A\B}{M}^{iL}_\A{M}^N_\B\pd_{pqLN}+M^{pL}_\A M^N_\B a^p_\A\pd_{iLN}\ri]R_{\A{\B}}^{-1}\\\nonumber
&&+\frac12\sum_{{\B}\not=\A}\sum_{l=0}^{\infty}\sum_{n=0}^{\infty}\frac{(-1)^n}{l!n!}\le(\frac{v^k_\A v^p_\A {\aleph}_{\A}^{L}M_\B^{N}}{2l+3}+\frac{v^k_\B v^p_\B M_{\A}^{L}{\aleph}_\B^{N}}{2n+3}\ri)\pd_{ikpLN}R^{-1}_{\A\B}+\varepsilon_{ikp}\left(2a^k_\A\dot{S}_\A^p+\dot{a}^k_\A{S}_\A^p\right)
\;.
\ea
Despite the fact that we have moved quite far in reducing equation \eqref{zeq48n} of the inverse problem to the integrable form,  the equation \eqref{kl9v2cz4} still contains in the right hand side some residual terms that are not looking like the variational derivative. Our goal is to find out a next chain of transformations which will bring these terms to the explicit form of the variational derivative, thus, making equation \eqref{kl9v2cz4} fully-integrable. 

To this end, we notice that the residual terms in the second line of \eqref{kl9v2cz4} can be reassembled as follows,
\ba\label{h2s34}
&&\sum_{{\B}\not=\A}\sum_{l=0}^{\infty}\sum_{n=0}^{\infty}\frac{(-1)^n}{l!n!}\le[\le({M}^{iL}_\A\dot{M}^{N}_\B+\dot{M}^{iL}_\A{M}^{N}_\B\ri)v^p_\A\pd_{pLN}+v^p_{\A} v^q_{\A\B}{M}^{iL}_\A{M}^N_\B\pd_{pqLN}+M^{pL}_\A M^N_\B a^p_\A\pd_{iLN}\ri]R_{\A{\B}}^{-1}\\\nonumber
&=&\sum_{{\B}\not=\A}\sum_{l=0}^{\infty}\sum_{n=0}^{\infty}\frac{(-1)^n}{l!n!}\le[v^p_\A\frac{d}{dt}\le(M^{iL}_\A M^N_\B\pd_{pLN}R^{-1}_{\A\B}\ri)+a^p_\A M^{pL}_\A M^N_\B\pd_{iLN} R^{-1}_{\A\B}\ri]\\\nonumber
&=&2v^p_\A\frac{d}{dt}\sum_{{\B}\not=\A}\sum_{l=0}^{\infty}\sum_{n=0}^{\infty}\frac{(-1)^n}{l!n!}M^N_\B M^{L[i}_\A \pd^{p]LN}_{\phantom A} R^{-1}_{\A\B}+\frac{d}{dt}\Biggl[v^p_\A\sum_{{\B}\not=\A}\sum_{l=0}^{\infty}\sum_{n=0}^{\infty}\frac{(-1)^n}{l!n!}M^{pL}_\A M^N_\B\pd_{iLN} R^{-1}_{\A\B}\Biggr]\\\nonumber
&=&\varepsilon_{ipq}v^p_\A\ddot{S}^q_\A+\frac{d}{dt}\Biggl[v^p_\A\sum_{{\B}\not=\A}\sum_{l=0}^{\infty}\sum_{n=0}^{\infty}\frac{(-1)^n}{l!n!}M^{pL}_\A M^N_\B\pd_{iLN} R^{-1}_{\A\B}\Biggr]\;,
\ea
where the last line was obtained by making use of the rotational equations of motion \eqref{vw291z3} for the spin dipole of body A written in the reverse order,
\ba \label{op3vz6a1}
\sum_{{\B}\not=\A}\sum_{l=0}^{\infty}\sum_{n=0}^{\infty}\frac{(-1)^n}{l!n!}M^N_\B M^{L[i}_\A \pd^{p]LN}_{\phantom A} R^{-1}_{\A\B}&=&\frac12\varepsilon_{ipq}\dot{S}^q_\A\;.
\ea
Such a replacement of some terms in the post-Newtonian approximation with the help of the Newtonian equations of motion for dynamic variables is fully legitimate as this is done on-shell in the configuration space of the dynamic variables before derivation of the Lagrangian \citep{Petrov_2017book}.   

Now, we algebraically split the second term in the last line of \eqref{h2s34} in its symmetric and anti-symmetric parts, 
\ba\label{oin5da}
\sum_{{\B}\not=\A}\sum_{l=0}^{\infty}\sum_{n=0}^{\infty}\frac{(-1)^n}{l!n!}M^{pL}_\A M^N_\B\pd_{iLN} R^{-1}_{\A\B}&=&\sum_{{\B}\not=\A}\sum_{l=0}^{\infty}\sum_{n=0}^{\infty}\frac{(-1)^n}{l!n!}\Biggl(M^N_\B M^{L(p}_\A \pd^{i)LN}_{\phantom A} R^{-1}_{\A\B}+M^N_\B M^{L[p}_\A \pd^{i]LN}_{\phantom A} R^{-1}_{\A\B}\Biggr)\;,
\ea
and make use of \eqref{op3vz6a1} once again to replace the anti-symmetric piece with the time derivative from the spin ${S}^i_\A$ of body A. The symmetric piece in \eqref{oin5da} is transformed on-shell to a set of intrinsic dynamic variables by applying a tensor virial theorem \eqref{jex61az8} which is explained in Appendix \ref{app1a} in more detail. Adding the results of these on-shell transformations all together yields,
\ba\label{hui2v5}
&&\sum_{{\B}\not=\A}\sum_{l=0}^{\infty}\sum_{n=0}^{\infty}\frac{(-1)^n}{l!n!}\le[\le({M}^{iL}_\A\dot{M}^{N}_\B+\dot{M}^{iL}_\A{M}^{N}_\B\ri)v^p_\A\pd_{pLN}+v^p_{\A} v^q_{\A\B}{M}^{iL}_\A{M}^N_\B\pd_{pqLN}+M^{kL}_\A M^N_\B a^k_\A\pd_{iLN}\ri]R_{\A{\B}}^{-1}\\\nonumber
&=&\varepsilon_{ipq}v^p_\A\ddot{S}^q_\A+\frac{d}{dt}\left[v^p_\A\left(\frac12\ddot{\gimel}_\A^{ip}-2\mathfrak{T}_\A^{ip}-\mathfrak{U}_\A^{ip}-\mathfrak{S}_\A^{ip}-\sum_{\B\neq\A}\sum_{l=0}^\infty\sum_{n=0}^\infty\frac1{l!n!}\frac{\aleph^L_\A M^N_\B}{2l+3}\pd_{ipLN} R^{-1}_{\A\B}\right)-\frac12\varepsilon_{ipq}v^p_\A\dot{S}^q_\A \right]\;,
\ea
where the intrinsic dynamic variables $\gimel_\A^{ip}$, $\mathfrak{T}_\A^{ip}$, $\mathfrak{U}_\A^{ip}$, $\mathfrak{S}_\A^{ip}$ are the volume integrals performed over volume of body A as defined in \eqref{ui319}--\eqref{op25xc2} of Appendix \ref{app1a}.

This reveals that all residual terms in the right hand side of \eqref{kl9v2cz4} are fully reduced to the Lagrangian form,
\ba
&&\sum_{{\B}\not=\A}\sum_{l=0}^{\infty}\sum_{n=0}^{\infty}\frac{(-1)^n}{l!n!}\le[\le({M}^{iL}_\A\dot{M}^{N}_\B+\dot{M}^{iL}_\A{M}^{N}_\B\ri)v^p_\A\pd_{pLN}+v^p_{\A} v^q_{\A\B}{M}^{iL}_\A{M}^N_\B\pd_{pqLN}+M^{kL}_\A M^N_\B a^k_\A\pd_{iLN}\ri]R_{\A{\B}}^{-1}\\\nonumber
&&+\frac12\sum_{{\B}\not=\A}\sum_{l=0}^{\infty}\sum_{n=0}^{\infty}\frac{(-1)^n}{l!n!}\le(\frac{v^k_\A v^p_\A {\aleph}_{\A}^{L}M_\B^{N}}{2l+3}-\frac{v^k_\B v^p_\B M_{\A}^{L}{\aleph}_\B^{N}}{2n+3}\ri)\pd_{ikpLN}R^{-1}_{\A\B}+\varepsilon_{ikq}\left(2a^k_\A\dot{S}_\A^q+\dot{a}^k_\A{S}_\A^q\right)=\frac{\d{\cal L}_8}{\d x^i_\A}\;,
\ea
where
\ba\label{huy38xr}
{\cal L}_8
&=&\frac12\sum_\A\Biggl[\Bigl(\frac12\ddot{\gimel}_\A^{pq}-2\mathfrak{T}_\A^{pq}-\mathfrak{U}_\A^{pq}-\mathfrak{S}_\A^{pq}-\sum_{{\B}\not=\A}\sum_{l=0}^{\infty}\sum_{n=0}^{\infty}\frac{(-1)^n}{l!n!}\frac{{\aleph}^{L}_\A M^N_\B}{2l+3}\pd_{pqLN}R^{-1}_{\A\B}\Bigr)v^p_\A v^q_\A-\varepsilon_{kpq}a^k_\A v^p_\A S^q_\A\Biggr]\;.
\ea
It is instructive to recast ${\cal L}_8$ by making use of tensor virial theorem \eqref{jex61az8}. It yields
\ba\label{x5z8a2}
{\cal L}_8
&=&\frac12\sum_\A\le(K_\A^{pq}v^p_\A v^q_\A-\varepsilon_{kpq}a^k_\A v^p_\A S^q_\A\ri)+
\frac12\sum_{\A}\sum_{\B\not=\A}\sum_{l=0}^\infty\sum_{n=0}^\infty\frac{(-1)^n}{l!n!}v^p_\A v^k_\A M_\A^{kL}M_{\B}^{N}\pd_{pLN}R_{\A\B}^{-1}\;,
\ea
where we have introduced a shorthand notation for a virial tensor function of a single body A
\ba\label{in38cx}
K_\A^{pq}&\equiv&\frac12\ddot{\gimel}_\A^{pq}-2\mathfrak{T}_\A^{pq}-\mathfrak{U}_\A^{pq}-\mathfrak{S}_\A^{pq}\\\nonumber
&-&\sum_{{\B}\not=\A}\sum_{l=0}^{\infty}\sum_{n=0}^{\infty}\frac{(-1)^n}{l!n!}\frac{{\aleph}^{L}_\A M^N_\B}{2l+3}\pd_{pqLN}R^{-1}_{\A\B}
-\sum_{\B\neq\A}\sum_{l=0}^\infty\sum_{n=0}^\infty\frac{(-1)^n}{l!n!}M^N_\B M^{L(p}_\A\pd^{q)LN}_{\phantom{A}} R^{-1}_{\A\B}\;.
\ea
Notice that function $K_\A^{pq}$ vanishes on-shell due to the virial theorem \eqref{jex61az8} but it must be kept in the Lagrangian as the variational derivative from $K_\A^{pq}$ is not nil.

\section{Lagrangian of the $N$-body Problem}\label{ct39z6}
\subsection{Lagrangian ${\cal L}$ with the multipoles defined in the global frame}
Results of the previous section demonstrate that the right hand side of equation \eqref{po75zc2} of the inverse problem of the Lagrangian mechanics can be put to the Lagrangian form which means that \eqref{po75zc2} is reduced to a fully integrable form,
\ba\label{oo99b3vz5}
\frac{\delta {\cal L}}{\delta x^i_\A}&=&\frac{\delta }{\delta x^i_\A}\Bigl({\cal L}_{a}+{\cal L}_{ab}+{\cal L}_{abc}\Bigr)\;,
\ea
where we have denoted 
\ba\label{ee2255zz}
{\cal L}_{a}\equiv {\cal L}_1+{\cal L}_2+{\cal L}_3+{\cal L}_4+{\cal L}_5+{\cal L}_6+{\cal L}_7+{\cal L}_8\;,
\ea
and ${\cal L}_{ab}$, ${\cal L}_{abc}$ are given in \eqref{om02x1z}--\eqref{iu3s5az}. 
We can now integrate equation \eqref{oo99b3vz5} and obtain the post-Newtonian Lagrangian ${\cal L}$ of the translational equations of motion of bodies with arbitrary mass and spin multipole moments expressed in the global G-frame. The Lagrangian is a linear combination of several components,
\ba\label{uu88ff11}
{\cal L}= {\cal L}_{a}+{\cal L}_{ab}+{\cal L}_{abc}+\frac{dF}{dt}\;,
\ea
where $F\equiv\sum_\A f_\A(t,{\bm x}_\A,{\bm v}_\A)$ is an arbitrary function of coordinates and velocity of the bodies describing the gauge freedom in definition of the Lagrangian \citep{Landau1969,Petrov_2017book}. This function is omitted from subsequent discussion as it does not contribute to the equations of motion. Nonetheless, the reader should keep in mind that different choices of the gauge function $F$ yield dissimilar forms of the Lagrangian which can make more difficult the comparison with the results of other authors .    

The explicit form of ${\cal L}_{a}$ is
\ba\label{lagran23}
{\cal L}_{a}&=&\frac12\sum_\A{\mathfrak m}_\A\le(1+\frac14 v^2_\A\ri)v_\A^2+\frac12\sum_\A\Bigl(K_\A^{pq}v^p_\A v^q_\A+3M_{\A}^{pq}a_\A^pa_\A^q-\varepsilon_{kpq}a^k_\A v^p_\A S^q_\A\Bigr)\\\nonumber
&+&\frac12\sum_\A\sum_{\B\neq\A}\sum_{l=0}^\infty\sum_{n=0}^\infty\frac{(-1)^n}{l!n!}{\mathfrak M}^L_\A {\mathfrak M}^N_\B\pd_{LN}R^{-1}_{\A\B}\\\nonumber
&+&\sum_\A\sum_{\B\not=\A}\sum_{l=0}^\infty\sum_{n=0}^\infty\frac{(-1)^n}{l!n!}a^p_\A M^{pL}_\A{M}^N_\B\pd_{LN}R^{-1}_{\A\B}
+\sum_\A\sum_{\B\not=\A}\sum_{l=0}^\infty\sum_{n=0}^\infty\frac{(-1)^n}{l!n!}\frac{a^p_\A {\aleph}^{L}_\A{ M}^N_\B}{2l+3}\pd_{pLN}R^{-1}_{\A\B}\\\nonumber
&+&\frac14\sum_{\A}\sum_{\B\not=\A}\sum_{l=0}^\infty\sum_{n=0}^\infty\frac{(-1)^n}{l!n!}\le[{M}_\A^L{M}_{\B}^Nv^p_{\A}v^q_{\B}\pd_{pqLN}-2{M}_\A^L\dot{{M}}_{\B}^Nv^p_\A\pd_{pLN}-\dot{M}_\A^L\dot{M}_{\B}^N\pd_{LN}\ri]R_{\A\B}\\\nonumber
&+&\sum_{\A}\sum_{\B\not=\A}\sum_{l=0}^\infty\sum_{n=0}^\infty\frac{(-1)^n}{l!n!}\Biggl[\Bigl(\frac12+\g\Bigr)v^2_\A-(1+\g)v^p_\A v^p_\B\Biggr]M_\A^{L}M_{\B}^{N}\pd_{LN}R_{\A\B}^{-1}  \\\nonumber
&+&\sum_{\A}\sum_{\B\not=\A}\sum_{l=0}^\infty\sum_{n=0}^\infty\frac{(-1)^n}{l!n!}\le[\le(l+\frac12\ri) v^p_\A-(l+1)v^p_\B\ri]v^k_\A M_\A^{kL}M_{\B}^{N}\pd_{pLN}R_{\A\B}^{-1}\\\nonumber
&+&\sum_{\A}\sum_{\B\not=\A}\sum_{l=0}^\infty\sum_{n=0}^\infty\frac{(-1)^n}{l!n!}\le[(l+1){M}_\A^{pL}\dot{M}_{\B}^{N}+l\dot{M}_\A^{pL}{M}_{\B}^{N}\ri]v^p_\A\pd_{LN}R_{\A\B}^{-1}\\\nonumber
&+&2(1+\g)
\sum_{\A}\sum_{\B\not=\A}\sum_{l=0}^\infty\sum_{n=0}^\infty\frac{(-1)^n}{l!n!}\le[\frac{\dot{M}_\A^{pL}{M}_{\B}^{N}}{l+1}-\frac{{M}_\A^{L}\dot{M}_{\B}^{pN}}{n+1}\ri]v^p_{\A}\pd_{LN}R_{\A\B}^{-1}\\\nonumber
&-& 
(1+\g)
\sum_{\A}\sum_{\B\not=\A}\sum_{l=0}^\infty\sum_{n=0}^\infty\frac{(-1)^n}{(l+1)!(n+1)!}\dot{M}^{kL}_\A\dot{M}^{kN}_\B\pd_{LN}R_{\A\B}^{-1}\\\nonumber
&+&2(1+\g)\sum_\A\sum_{\B\neq\A}\sum_{l=0}^\infty\sum_{n=0}^\infty\frac{(-1)^n}{l!n!}\varepsilon_{pkq}\frac{v^k_{\A\B}S_\A^{qL}M_\B^N }{l+2}\pd_{pLN}R^{-1}_{\A\B}\\\nonumber
&+&2(1+\g)\sum_\A\sum_{\B\neq\A}\sum_{l=0}^\infty\sum_{n=0}^\infty\frac{(-1)^n}{l!n!}\varepsilon_{pkq}\frac{\dot M_\A^{kL} S^{qN}_\B}{(l+1)(n+2)}\pd_{pLN}R^{-1}_{\A\B}\\\nonumber
&-&(1+\g)\sum_\A\sum_{\B\neq\A}\sum_{l=0}^\infty\sum_{n=0}^\infty\frac{(-1)^n}{l!n!}\frac{S_\A^{pL}S_\B^{qN}}{(l+2)(n+2)}\pd_{pqLN}R^{-1}_{\A\B}\;,
\ea
where ${\mathfrak m}_\A$ being the post-Newtonian (Tolman) mass in general relativity defined in \eqref{grmass}, the multipole moments ${\mathfrak M}^L_\A$ and ${\mathfrak M}^L_\B$ are defined by \eqref{wvzx35y}, \eqref{e5z310} in the G-frames on the world lines of the corresponding bodies. They are Blanchet-Damour multipoles \citep{Blanchet_1989AIHPA} in scalar-tensor theory for all $l\ge 0$ including the mass monopole.

There are several points to comment on. 
\begin{enumerate}
\item The Lagrangian ${\cal L}$ is identical to the Fokker Lagrangian which would have been obtained by integrating the action of the scalar-tensor field theory. All external and internal dynamic variables of the Lagrangian ${\cal L}$ are unequivocally computed in a single global G-frame. The multipole moments entering the Lagrangian are the Blanchet-Damour multipole moments \citep{Blanchet_1989AIHPA} extended to tensor-scalar theory \citep{DamEsFar,kovl_2004}.
\item The Lagrangian ${\cal L}$ depends on accelerations of the bodies. The linear acceleration terms associated with spin of body A and its quadrupole moment have been known \citep{damour_1982CR,Barker1986_acceleration,Steinhoff_2015}. We have extended these results to the multipoles of higher order. 
\item The Lagrangian ${\cal L}$ also depends on terms being quadratic with respect to accelerations of the bodies which couple with the quadrupole moments of the bodies. The quadratic-acceleration term means that the Hessian of ${\cal L}$ computed with respect to the accelerations is non-degenerate. This brings about the Ostrogradsky instability \citep{Woodard_2015,Motohashi_PRD2015} that may be problematic in some applications. It is tempting to replace the accelerations of bodies in the Lagrangian with the equations of motion from the lower-order approximation. This is a legitimate procedure in action-based gravitational theories but it leads to an implicit (and, hence, uncontrolled) change in the original coordinates used for derivation of equations of motion and the laws of conservation \citep{SCHAFER1984,Damour_1985GReGr,Damour_1989MG5}. Correct mathematical procedure of elimination accelerations and higher-order time derivatives from the Lagrangians relies upon the method redefinition of position variables \citep{Damour_Schafer1991JMP} which is a generalization of the method of the "double zero" proposed by \citet{Barker1986_acceleration}. We notice that the quadratic acceleration terms can be also removed from the Lagrangian ${\cal L}$ by redefinition of the center of mass of body A -- see \citep[section IX.6]{k2019PRD}.   
\item The Lagrangian ${\cal L}$ explicitly depends on the {\it non-canonical} multipoles like $\aleph^L_\A$ which are multipole extension of the second moment of inertia $\aleph_\A$ of body A depending on the finite radius of the body. This fact has been noticed by \citet{nordtvedt_1994PhRvD}. 
The moment of inertia for a compact body $\aleph_\A\simeq G^2\mathfrak{m}_\A^2/c^4$, where $\mathfrak{m}_\A$ is the Tolman mass of the object. Since the moment of inertia $\aleph_\A$ appears in the post-Newtonian Lagrangian explicitly, the finite-size effects of compact bodies enter dynamics at the 3PN ($\sim c^{-6}$) order. This conclusion is valid, however, only in the scalar-tensor theory with the parameter $\beta\neq 1$. In general relativity $\beta=1$ so that all terms with the moment of inertia $\aleph_\A$ vanish in the equations of motion (see section \ref{mw71cz52} for the details) that postpones the appearance of the finite-size effects in general relativity to the 5PN ($\sim c^{-10}$) order. Previous results of some other researchers, for instance, \citep{vab,dallas_1977CeMec,spyrou_1975ApJ,spyrou_1978GReGr}, suggested that 1PN equations of motion in general relativity do depend on the moment of inertia $\aleph_\A$ of the bodies (not associated with spin). These results are based on underdeveloped mathematical technique and are unjustified as the authors of the papers \citep{vab,dallas_1977CeMec,spyrou_1975ApJ,spyrou_1978GReGr} did not take into account that physical moment of inertia $\aleph_\A$ must be calculated in the local coordinates adapted to body A but not in the global coordinates \citep{vincent_1986CeMec,kovl_2004,kopeikin_2011book}.
\item  The Lagrangian ${\cal L}_{a}$ depends on tensor function $K^{pq}_\A$ given in \eqref{in38cx} which explicitly includes the integral characteristics of the bodies depending on their specific internal structure -- the second symmetric moments $\gimel^{ij}_\A$,  the kinetic energy of the internal motion of matter $\mathfrak{T}^{ij}_\A$, the potential energy of gravitational field $\mathfrak{U}^{ij}_\A$, and the integral of stresses $\mathfrak{S}^{ij}_\A$ -- see their definitions in Appendix \ref{app1a}. It can make an impression that the Lagrangian ${\cal L}$ can be applied only to the bodies with the known internal structure. However, this is not true since, after taking the variational derivative, we can take function $K^{pq}_\A=0$ on-shell due to the virial theorem and, thus, it does not enter explicitly the results of all possible applications like computation of the equations of motion or conserved quantities. Furthermore, the explicit dependence on the integrals characterizing the internal structure of the bodies can be completely removed from the Lagrangian by switching from the internal dynamic variables (multipoles) expressed in the global G-frame to physical multipoles computed in the local L-frame along with applying the virial theorems \eqref{jex61az8}, \eqref{ec7} for transformation of gravitational masses $\mM_\A$ of the bodies. This procedure is worked out in next section.
\end{enumerate}

\subsection{Lagrangian $L$ with the multipoles defined in the local frame}\label{746352424} 
\subsubsection{Gauge transformation of the internal dynamic variables}\label{fsfw77785}
This section solves the inverse problem in the form of equation \eqref{i92bx64z} with the Lagrangian $L$ expressed in terms of the physical multipoles defined in the local L-frame of each body. First of all, we transform the monopole which is the gravitational mass ${\mathfrak M}_\A$ given by equation \eqref{wvzx35y} for index $l=0$. It reads
\ba\label{ww33oocc21}
\mM_\A&=&\mathfrak{m}_\A+\eta\mathfrak{U}_\A+\gamma\le(2\mathfrak{T}_\A+\mathfrak{U}_\A+\mathfrak{S}_\A\ri)+\frac16\ddot{\aleph}_\A-(1+\g)\frac2{3}\dot{\Re}_\A\;,
\ea
where, $\mathfrak{m}_\A$, is the Tolman mass,
a constant coefficient $\eta=4\b-\g-3$ is known as the Nordtvedt parameter \citep{willbook}, the quantities $\aleph_\A$ and $\Re_\A$ are scalar {\it non-canonical} monopoles defined in \eqref{aleph7354} and \eqref{un141d}, the integrals $\mathfrak{T}_\A$, $\mathfrak{U}_\A$, ${\aleph}_\A$, $\mathfrak{S}_\A$ characterize the internal structure of the body A as explained in equations \eqref{ui8hx}, \eqref{kilc6}, \eqref{ui320} and \eqref{ui320bbb} respectively. Accounting for the relationship, $\ddot{\aleph}_\A=2\dot{\Re}_\A$, and applying the virial theorem \eqref{ec7} we bring \eqref{ww33oocc21} to the following form
\ba\label{kk9922xx}
\mM_\A&=&\mathfrak{m}_\A+\eta\,{\mathfrak U}_A+\frac16(\g-1)\ddot{\aleph}_\A-\g\le[K_\A+\sum_{\B\not=\A}\sum_{l=1}^\infty\sum_{n=0}^{\infty}\frac{(-1)^n }{(l-1)!n!}{M}_\A^{L}M^N_\B\pd_{LN} R^{-1}_{\A\B}\ri]\;,
\ea
where the scalar function $K_\A$ is the trace of tensor function $K^{pq}_\A$,
\ba\label{pn27zda31s}
K_\A&=&K^{pp}_\A=\frac12\ddot{\aleph}_\A-2\mathfrak{T}_\A-\mathfrak{U}_\A-\mathfrak{S}_\A-\sum_{\B\not=\A}\sum_{l=1}^\infty\sum_{n=0}^{\infty}\frac{(-1)^n }{(l-1)!n!}{M}_\A^{L}M^N_\B\pd_{LN} R^{-1}_{\A\B}\;.
\ea
Function $K_\A$ vanishes on-shell but must be kept explicitly in the Lagrangian as its variational derivative does not disappear.

The terms in the square brackets of \eqref{kk9922xx} are of the post-Newtonian order of magnitude. Hence, the dipole $(l=1)$ term in the sum could, in principle, be neglected as it is of the post-Newtonian order of magnitude itself. Expression \eqref{kk9922xx} helps us to understand that in case of general relativity, when parameters $\g=\b=1$, $\eta=0$, the inertial, $\mathfrak{m}_\A$, and gravitational, $\mM_\A$, masses are identical for spherically-symmetric bodies in complete agreement with the strong principle of equivalence (SEP). In scalar-tensor theory of gravity the two masses are not identical -- mainly due to the self-gravity term $\eta{\mathfrak U}_A$ in \eqref{kk9922xx} -- and the difference between them causes violation of the SEP that was first noted by \citet{nord_PhysRev1968a,nord_PhysRev1968b,nord_1973PhRvD}. Testing this violation in the Earth-Moon system and binary pulsars is one of the most effective experimental probes for detecting the presence of a hypothetical long-range scalar fields in nature \citep{nord_1973PhRvD,nordtvedt_2001LNP,2012CQGra,LLR_2018CQGra,mitchell_2007PhRvD,damsch_1991PhRvL,wex_1997A&A,lorimer_2005ASPC,wex2014,Archibald_2018Natur}. 
Our calculations reveal that there is a non-static term contributing to the SEP violation. Indeed, according to \eqref{kk9922xx} the gravitational mass $\mM_\A$ of body A disagrees with its inertial mass $\mathfrak{m}_\A$ not only due to the static Nordtvedt effect (term with $\eta$) but also because the body's moment of inertia $\aleph_\A$ (defined in \eqref{ui320}) changes in the course of time (for example, radial pulsations of star) and/or due to a tidal gravitational interaction. 

Let us introduce new notation for the G-frame multipoles 
\ba\label{kbcgs64us}
\hat{\mM}_\A^L&\equiv&\le\{\begin{array}{ll}\mathfrak{m}_\A+\eta\,{\mathfrak U}_A+\displaystyle\frac16(\g-1)\ddot{\aleph}_\A&\;,\qquad (l=0)\\
&\\
                               \mM_\A^L&\;,\qquad (l\ge 1)
                               \end{array}\right.
\ea
which allows to decompose the Newtonian gravitational energy as follows
\ba\label{rc82cz4q}
\frac12\sum_\A\sum_{\B\neq\A}\sum_{l=0}^\infty\sum_{n=0}^\infty\frac{(-1)^n}{l!n!}{\mathfrak M}^L_\A {\mathfrak M}^N_\B\pd_{LN}R^{-1}_{\A\B}
&=&\frac12\sum_\A\sum_{\B\neq\A}\sum_{l=0}^\infty\sum_{n=0}^\infty\frac{(-1)^n}{l!n!}\hat{\mM}^L_\A \hat{\mM}^N_\B\pd_{LN}R^{-1}_{\A\B}\\\nonumber
&-&\g\sum_\A\bar{U}({\bm x}_\A)\le[K_\A+\sum_{l=1}^\infty\frac{1}{(l-1)!}{M}_\A^{L} \pd_L\bar{U}({\bm x}_\A)\ri]
\ea
where 
\ba 
\bar{U}({\bm x}_\A)&=&\sum_{\B\neq\A}\sum_{n=0}^{\infty}\frac{(-1)^n}{n!}M_\B\pd_N R^{-1}_{\A\B}\;, 
\ea
is the Newtonian potential of all bodies being external to body A.

Now, we consider a linear combination of three terms 
\ba\label{iq92vz62}
W&=&-\frac12\sum_\A\sum_{\B\neq\A}\sum_{l=0}^\infty\sum_{n=0}^\infty\frac{(-1)^n}{l!n!}\hat{\mathfrak M}^L_\A \hat{\mathfrak M}^N_\B\pd_{LN}R^{-1}_{\A\B}-\frac12 \sum_\A K_\A^{pq}v^p_\A v^q_\A+\g \sum_\A K_\A \bar{U}({\bm x}_\A)\;,
\ea
which appears in the Lagrangian ${\cal L}_a$ along with function $K_\A$ emerging after transfromation \eqref{kk9922xx}. We take
the variational $\delta$-derivative from $W$ in the sense of equation \eqref{j8c5a2}. It yields 
\ba\label{uyf35dc1}
\frac{\delta W}{\delta x^i_\A}&=&\sum_{\B\neq\A}\sum_{l=0}^\infty\sum_{n=0}^\infty\frac{(-1)^n}{l!n!}\hat{\mathfrak M}^L_\A \hat{\mathfrak M}^N_\B\pd_{iLN}R^{-1}_{\A\B}\\\nonumber
&+&\frac12 \le[\frac{\pd K_\A^{pq}}{\pd x^i_\A} v^p_\A v^q_\A+\sum_{\B\neq\A}\frac{\pd K_\B^{pq}}{\pd x^i_\A} v^p_\B v^q_\B\ri] 
-\g\le[\frac{\pd K_\A}{\pd x^i_\A} \bar{U}({\bm x}_\A)+\sum_{\B\neq\A}\frac{\pd K_\B}{\pd x^i_\A} \bar{U}({\bm x}_\B)\ri]\;,
\ea
where we have accounted for the fact that after taking the variational derivative functions $K^{pq}_\A=K_\A=0$ and $K^{pq}_\B=K_\B=0$ on-shell due to the virial theorem but their partial derivatives do not vanish. After developing the partial derivatives in \eqref{uyf35dc1} by using definitions \eqref{in38cx}, \eqref{pn27zda31s}, we get 
\ba\label{lkcb47598cb}
\frac{\delta W}{\delta x^i_\A}&=&
\sum_{\B\neq\A}\sum_{l=0}^\infty\sum_{n=0}^\infty\frac{(-1)^n}{l!n!}\hat{\mathfrak M}^L_\A \hat{\mathfrak M}^N_\B\pd_{iLN}R^{-1}_{\A\B}\\\nonumber
&-&\frac12\sum_{\B\neq \A}\sum_{l=0}^{\infty}\sum_{n=0}^{\infty}\frac{(-1)^n}{l!n!}\le(v^k_\A v^p_\A M_{\A}^{kL}M_\B^{N}- v^k_\B v^p_\B M_{\A}^{L}M_\B^{kN}\ri)\pd_{ipLN}R_{\A\B}^{-1}\\\nonumber
&-&\frac12\sum_{\B\neq \A}\sum_{l=0}^{\infty}\sum_{n=0}^{\infty}\frac{(-1)^n}{l!n!}\le(\frac{v^k_\A v^p_\A {\aleph}_{\A}^{L}M_\B^{N}}{2l+3}+\frac{v^k_\B v^p_\B M_{\A}^{L}{\aleph}_\B^{N}}{2n+3}\ri)\pd_{ikpLN}R_{\A\B}^{-1}\\\nonumber
&+&\g\sum_{\B\neq \A}\sum_{l=0}^{\infty}\sum_{n=0}^{\infty}\frac{(-1)^n}{l!n!}\le[l\bar U({\bm x}_\A)M_\A^L M_\B^{N}+n\bar U({\bm x}_\B)M^L_\A M_\B^N\ri]\pd_{iLN}R_{\A\B}^{-1}\;.
\ea
However, the right hand side of \eqref{lkcb47598cb} is equivalent (up to the relativistic precession term) to the transformation \eqref{nn33ss66} of the Newtonian gravity force due to the conversion of the multipoles from the global to local frame. Making use of the inverse transformations \eqref{nx5w}, \eqref{gcvt5} of the G-frame multipoles $\hat{\mathfrak M}^L_\A$ to L-frame multipoles 
\ba\label{ub36xc17s}
\dutchcal{M}_\A^L&\equiv&\le\{\begin{array}{ll}\mathfrak{m}_\A+\eta\,{\mathfrak U}_A+\displaystyle\frac16(\g-1)\ddot{\cal N}_\A&\;,\qquad (l=0)\\
&\\
                               {\cal I}_\A^L&\;,\qquad (l\ge 1)
                               \end{array}\right.
\ea
where ${\cal I}^L_\A$ are given by \eqref{wvzx35x}, we find out that all terms in the right hand side of \eqref{lkcb47598cb} cancel out with the corresponding terms from the multipole transformation formulas and it reduces to a simple Newtonian-like form
\ba\label{fwx35ac17}
\frac{\delta W}{\delta x^i_\A}
=\sum_{\B\neq\A}\sum_{l=0}^\infty\sum_{n=0}^\infty\frac{(-1)^n}{l!n!}\dutchcal{M}^L_\A \dutchcal{M}^N_\B\pd_{iLN}R^{-1}_{\A\B}
+\sum_{\B\neq \A}\sum_{l=0}^{\infty}\sum_{n=0}^{\infty}\frac{(-1)^n}{l!n!}\le(F_{\A}^{kp} \M_{\A}^{kL}\M_\B^{N}-F_{\B}^{kp}  \M_{\A}^{L}\M_\B^{kN}\ri)\pd_{ipLN}R_{\A\B}^{-1}\;,
\ea 
where the matrix $F_{\A}^{kp}$ of the relativistic precession is a solution of the precessional equation  \eqref{5.18}, and we replaced the G-frame multipoles $M^L_\A$ with their L-frame counterparts $\M^L_\A$. 

Now, we define function 
\ba\label{hyw23z6413}
{\cal W}&=&-\frac12\sum_\A\sum_{\B\neq\A}\sum_{l=0}^\infty\sum_{n=0}^\infty\frac{(-1)^n}{l!n!}\dutchcal{M}^L_\A \dutchcal{M}^N_\B\pd_{LN}R^{-1}_{\A\B}
-\sum_\A\sum_{\B\neq \A}\sum_{l=0}^{\infty}\sum_{n=0}^{\infty}\frac{(-1)^n}{l!n!}F_{\A}^{kp} \M_{\A}^{kL}\M_\B^{N}\pd_{pLN}R_{\A\B}^{-1}\;,
\ea
that is similar but not identical to function $W$ in which the relativistic frame transformation of multipoles \eqref{nx5w} from the global to local frame has been performed. The frame-transformed function $W$ differs from ${\cal W}$ by terms depending on the internal structure of bodies which are not present explicitly in ${\cal W}$. Applying the variational $\textswab{d}$-derivative to ${\cal W}$ in the sense of equation \eqref{ybzc35s80} (that is the moments $\dutchcal{M}_\A^L$ do not vary) we find out that
\ba\label{8464532gdte}
\frac{\delta W}{\delta x^i_\A}=\frac{\textswab{d}{\cal W}}{\textswab{d}x^i_\A}\;.
\ea
This equation allows us to transform the Lagrangian ${\cal L}$ with the mass multipoles $\mM^L_\A$ defined in the global frame to the Lagrangian $L$ with the mass multipoles $\dutchcal{M}_\A^L$ defined in the local frame. More specifically, we make use of the transformations of the Newtonian potential energy  described above and use them to replace the corresponding terms in the Lagrangian ${\cal L}$ from previous section. It yields the Lagrangian $L$ with all physical multipoles expressed in the local frame of the bodies.  

\subsubsection{Lagrangian} 

The Lagrangian of the $N$-body problem with all multipoles expressed in the local frame of each body is given by equation
\ba\label{xxxxxxxxxxxxx}
L= L_{a}+L_{ab}+L_{abc}\;,
\ea
where    
\ba\label{yyyyyyyyyy}
L_a&=&\frac12\sum_\A\le[{\mathfrak m}_\A\le(1+\frac14 v^2_\A\ri)v_\A^2+3{\cal M}_{\A}^{pq}a_\A^pa_\A^q-\varepsilon_{kpq}a^k_\A v^p_\A \Sc^q_\A\ri]\\\nonumber
&+&\frac12\sum_\A\sum_{\B\neq\A}\sum_{l=0}^\infty\sum_{n=0}^\infty\frac{(-1)^n}{l!n!}{\dutchcal{M}}^L_\A {\dutchcal{M}}^N_\B\pd_{LN}R^{-1}_{\A\B}
+\sum_\A\sum_{\B\neq \A}\sum_{l=0}^{\infty}\sum_{n=0}^{\infty}\frac{(-1)^n}{l!n!}F_{\A}^{kp} \M_{\A}^{kL}\M_\B^{N}\pd_{pLN}R_{\A\B}^{-1}\\\nonumber
&+&\sum_\A\sum_{\B\not=\A}\sum_{l=0}^\infty\sum_{n=0}^\infty\frac{(-1)^n}{l!n!}a^p_\A {\M}^{pL}_\A{\M}^N_\B\pd_{LN}R^{-1}_{\A\B}
+\sum_\A\sum_{\B\not=\A}\sum_{l=0}^\infty\sum_{n=0}^\infty\frac{(-1)^n}{l!n!}\frac{a^p_\A {\cal N}^{L}_\A{\M}^N_\B}{2l+3}\pd_{pLN}R^{-1}_{\A\B}\\\nonumber
&+&\frac14\sum_{\A}\sum_{\B\not=\A}\sum_{l=0}^\infty\sum_{n=0}^\infty\frac{(-1)^n}{l!n!}\le[{\M}_\A^L{\M}_{\B}^Nv^p_{\A}v^q_{\B}\pd_{pqLN}-2{\M}_\A^L\dot{{\M}}_{\B}^Nv^p_\A\pd_{pLN}-\dot{\M}_\A^L\dot{\M}_{\B}^N\pd_{LN}\ri]R_{\A\B}\\\nonumber
&+&\sum_{\A}\sum_{\B\not=\A}\sum_{l=0}^\infty\sum_{n=0}^\infty\frac{(-1)^n}{l!n!}\Biggl[\Bigl(\frac12+\g\Bigr)v^2_\A-(1+\g)v^p_\A v^p_\B\Biggr]\M_\A^{L}\M_{\B}^{N}\pd_{LN}R_{\A\B}^{-1}  \\\nonumber
&+&\sum_{\A}\sum_{\B\not=\A}\sum_{l=0}^\infty\sum_{n=0}^\infty\frac{(-1)^n}{l!n!}\le[\le(l+\frac12\ri) v^p_\A-(l+1)v^p_\B\ri]v^k_\A \M_\A^{kL}\M_{\B}^{N}\pd_{pLN}R_{\A\B}^{-1}\\\nonumber
&+&\sum_{\A}\sum_{\B\not=\A}\sum_{l=0}^\infty\sum_{n=0}^\infty\frac{(-1)^n}{l!n!}\le[(l+1){\M}_\A^{pL}\dot{\M}_{\B}^{N}+l\dot{\M}_\A^{pL}{\M}_{\B}^{N}\ri]v^p_\A\pd_{LN}R_{\A\B}^{-1}\\\nonumber
&+&2(1+\g)
\sum_{\A}\sum_{\B\not=\A}\sum_{l=0}^\infty\sum_{n=0}^\infty\frac{(-1)^n}{l!n!}\le[\frac{\dot{\M}_\A^{pL}{\M}_{\B}^{N}}{l+1}-\frac{{\M}_\A^{L}\dot{\M}_{\B}^{pN}}{n+1}\ri]v^p_{\A}\pd_{LN}R_{\A\B}^{-1}\\\nonumber
&-& 
(1+\g)
\sum_{\A}\sum_{\B\not=\A}\sum_{l=0}^\infty\sum_{n=0}^\infty\frac{(-1)^n}{(l+1)!(n+1)!}\dot{\M}^{kL}_\A\dot{\M}^{kN}_\B\pd_{LN}R_{\A\B}^{-1}\\\nonumber
&+&2(1+\g)\sum_\A\sum_{\B\neq\A}\sum_{l=0}^\infty\sum_{n=0}^\infty\frac{(-1)^n}{l!n!}\varepsilon_{pkq}\frac{v^k_{\A\B}{\Sc}_\A^{qL}\M_\B^N }{l+2}\pd_{pLN}R^{-1}_{\A\B}\\\nonumber
&+&2(1+\g)\sum_\A\sum_{\B\neq\A}\sum_{l=0}^\infty\sum_{n=0}^\infty\frac{(-1)^n}{l!n!}\varepsilon_{pkq}\frac{\dot \M_\A^{kL} {\Sc}^{qN}_\B}{(l+1)(n+2)}\pd_{pLN}R^{-1}_{\A\B}\\\nonumber
&-&(1+\g)\sum_\A\sum_{\B\neq\A}\sum_{l=0}^\infty\sum_{n=0}^\infty\frac{(-1)^n}{l!n!}\frac{{\Sc}_\A^{pL}{\Sc}_\B^{qN}}{(l+2)(n+2)}\pd_{pqLN}R^{-1}_{\A\B}\;,
\ea
where ${\mathfrak m}_\A$ being the post-Newtonian (Tolman) mass of body A in general relativity defined in \eqref{grmass}, the mass multipole moments ${\M}^L_\A$ and the spin multipoles $\Sc^L_\A$ are respectively defined by \eqref{klz2a1} and \eqref{1.32} in the L-frames of the body A, and the {\it non-canonical} multipoles ${\cal N}^L_\A$ are given in \eqref{NL15a}. Notice that the multipoles $\dutchcal{M}^L_\A$ are the scalar-tensor variant of the Blanchet-Damour multipoles \citep{Blanchet_1989AIHPA} only for $l\ge 1$. The monopole $\dutchcal{M}_\A$  ($l=0)$ is not the Blanchet-Damour mass but a generalization of the Tolman mass in scalar-tensor theory of gravity found by \citet{nord_PhysRev1968b}. 

The terms $L_{ab}$ and $L_{abc}$ in \eqref{xxxxxxxxxxxxx} are given by combining equation \eqref{om02x1z} along with the very last term in the right hand side of the transformation equation \eqref{rc82cz4q}. We have
\ba\label{mr536vaq}
L_{ab}&=&\sum_\A\sum_{\B\neq\A}\le[\le(1-\b\ri)\Phi_{\A\B}-\frac12\Phi_{\A\B}-\gamma \sum_{l=0}^\infty\sum_{n=0}^{\infty}\sum_{s=0}^\infty\frac{l(-1)^{k+s}}{l!k!s!}{\M}_\A^{L} \M^K_\B \M^S_\B\le(\pd_{LK} R^{-1}_{\A\B}\ri)\le(\pd_{S}R^{-1}_{\A\B}\ri)\ri]\;,
\\
\label{kxncbey364tg}
L_{abc}&=&\sum_\A\sum_{\B\neq\A}\sum_{\C\neq\A,\B}\le[\le(1-\b\ri)\Phi_{\A\B\C}-\frac12\Phi_{\A\B\C}-\gamma\sum_{l=0}^\infty\sum_{n=0}^{\infty}\sum_{s=0}^\infty\frac{l(-1)^{k+s}}{l!k!s!}{\M}_\A^{L} \M^K_\B \M^S_\C\le(\pd_{LK} R^{-1}_{\A\B}\ri)\le(\pd_{S}R^{-1}_{\A\C}\ri)\ri]\;,
\ea
where functions 
\mathleft
\ba
\label{jhfdye747}
\Phi_{\A\B}
&=&\frac12\sum_{l=0}^{\infty}\sum_{n=0}^{\infty}\sum_{k=0}^{\infty}\sum_{s=0}^{\infty}\frac{(-1)^{k+s}}{l!n!k!s!}\biggl[\mathfrak{I}_\A^{LN}\M^K_\B \M_\B^S
 +\mathfrak{I}_\B^{KS}\M^L_\A \M_\A^N\biggr]\le(\pd_{LK}R^{-1}_{\A\B}\ri)\le(\pd_{NS}R^{-1}_{\A\B}\ri)
\;, 
\ea
\mathleft
\begin{flalign}
\label{kkqa5163v}
\Phi_{\A\B\C}
&=\frac13\sum_{l=0}^{\infty}\sum_{n=0}^{\infty}\sum_{k=0}^{\infty}\sum_{s=0}^{\infty}\frac{(-1)^{k+s}}{l!n!k!s!}\biggl[
\mathfrak{I}_\A^{LN}\M^K_\B \M^S_\C\le(\pd_{LK}R^{-1}_{\A\B}\ri)\le(\pd_{NS}R^{-1}_{\A\C}\ri)
\\\nonumber
&\hspace{4.3cm}+
\mathfrak{I}_\B^{LN}\M^K_\C \M^S_\A\le(\pd_{LK}R^{-1}_{\B\C}\ri)\le(\pd_{NS}R^{-1}_{\B\A}\ri)+
\mathfrak{I}_\C^{LN}\M^K_\A \M^S_\B\le(\pd_{LK}R^{-1}_{\C\A}\ri)\le(\pd_{NS}R^{-1}_{\C\B}\ri)\biggr]
\;,
\end{flalign}
and the symmetric (not STF) moments $\mathfrak{I}^L_\A$ have been introduced in \eqref{unc62d}. We have extracted in \eqref{mr536vaq}, \eqref{kxncbey364tg} the terms being proportional to the factor $(1-\beta)$ since these terms originate from the presence of self-interaction of the scalar field and characterize its strength in scalar-tensor theories. It is these terms which lead to the appearance of dipole tidal effects in the scalar-tensor theories of gravity as has been recently pointed our by \citep{Bernard_2020PRD}. Susceptibility of each body A to the dipolar tidal field is characterized by the deformability parameter $\lambda^{\rm (s)}_\A$ introduced in \eqref{bra4xz900}. 

We emphasize that the variational derivative that must be applied to the Lagrangian $L$ in \eqref{xxxxxxxxxxxxx} is $\textswab{d}$-derivative introduced in \eqref{ybzc35s80}. The two Lagrangians ${\cal L}$ in \eqref{uu88ff11} and $L$ in \eqref{xxxxxxxxxxxxx} are equivalent in the sense of the identity \eqref{9h36xca} that is they both yield the same equations of translational motion \eqref{MBaB1}--\eqref{zw3k4b}. It is interesting to notice that the Lagrangian $L$ can be obtained from the Lagrangian ${\cal L}$ by formally replacing the G-frame multipoles $\hat{\mathfrak M}^L_\A$ to the L-frame multipoles ${\dutchcal M}^L_\A$ and using the virial theorems to nullify functions $K^{pq}_A$ and $K_\A$. Indeed, the Lagrangian ${\cal L}$ is a function of the G-frame multipoles, ${\cal L} ={\cal L}\bigl[\hat{\mathfrak M}^L_\A\bigr]$ while the Lagrangian $L$ is a function of the L-frame multipoles, $L=L\bigl[\dutchcal{M}^L_\A\bigr]$. Previous section tells us that \footnote{We ignore the matrix of the relativistic precession here.}
\ba\label{nnn564638} 
{\cal L}\bigl[\hat{\mathfrak M}^L_\A\bigr]=L\bigl[\hat{\mathfrak M}^L_\A\bigr]+\frac12 \sum_\A K_\A^{pq}v^p_\A v^q_\A-\g \sum_\A K_\A \bar{U}({\bm x}_\A)\;,
\ea
where functions $K_\A^{pq}$ and $K_\A$ vanish on-shell due to the virial theorem, which is the equation of motion for the internal variables. Hence, it is tempting to take $K_\A^{pq}=K_\A=0$ in \eqref{nnn564638}, and render a formal replacement $\hat{\mathfrak M}^L_\A\rightarrow \dutchcal{M}^L_\A$ that makes the impression that ${\cal L}=L$. However, this procedure is not a right way to establish the correspondence between the two Lagrangians. In fact, it is illegitimate since the virial theorem is valid solely on-shell and the equality $K_\A^{pq}=K_\A=0$ is not allowed in the Lagrangian before taking a variational derivative from the Lagrangian \citep{Damour_Schafer1991JMP}. Moreover, replacing the multipoles from one frame to another is not a matter of changing their labels but a corresponding mathematical transformation. The correct procedure of establishing the correspondence between the two Lagrangians must rely upon the legitimate transformations and application of the variational derivatives in the way explained in previous section \ref{fsfw77785}. The Lagrangians are equivalent in the sense of the equivalence of two variational derivatives
\ba\label{uuu4661vs}
\frac{\d}{\d x^i_\A}{\cal L}\bigl[\hat{\mathfrak M}^L_\A\bigr]= \frac{\textswab{d}}{\textswab{d} x^i_\A}L\bigl[\dutchcal{M}^L_\A\bigr]\;,
\ea
as have been discussed in section \ref{hh3545ax2}. The role of the virial theorem in derivation of the Lagrangian of the $N$-body problem with all multipoles taken into account and its importance for making a correct gauge transformation of the multipolar Lagrangian from one frame to another has not been pointed out previously in works of other researchers.

The Lagrangian $L$ with the physical multipoles ${\dutchcal M}^L_\A$ defined in the local frame of each body is more preferable in astrophysical applications as it contains the least number of terms depending on the internal structure of the bodies. The most important for gravitational wave astronomy are binary systems consisting of two objects of comparable masses. We discuss the Lagrangian $L$ of such two-body system in next section.

\section{Lagrangian of a binary system}\label{jhru747}

Two-body system is probably the most interesting case for gravitational wave astronomy. The inspiralling compact binaries are main sources of gravitational wave signal for present-day and near-future generations of gravitational-wave observatories \citep{Schutz_2018RSPTA,thorne_nobel}. Current model of inspiralling compact binary treats the bodies of the system as pole-dipole particles with constant mass and precessing spin. This model is tractable for analytic computations up to the 4PN approximation and was extensively used to construct a bank of templates of waveforms  \citep{template_PRD2014,template_PRD2019}. At the latest stages of coalescence tidal and rotational deformations of the bodies in the binary become so significant that the pole-dipole model is getting no longer accurate and should be extended to include higher-order multipoles in the Newtonian and post-Newtonian approximations \citep{Steinhoff_etal_PRD104028,Poisson_2009PhRvD,Hinderer_2008PhRvD,Nagar_2009PhRvD}. This section presents the Lagrangian $L$ for description of the orbital evolution of two-body system with all mass and spin multipole moments of the bodies taken into account.

\subsection{General Formula}
In case of two-body system, A=\{1,2\}, the post-Newtonian three-body interaction Lagrangian $L_{abc}$ vanishes, and the overall Lagrangian is simplified to a linear sum of two terms 
\ba\label{ui38x51za}
{L}={L}_{a}+{L}_{ab}\;.
\ea
The first term in \eqref{ui38x51za} reads 
\ba\label{juw71x3}
L_{a}&=&\frac12{\mathfrak m}_1\le(1+\frac14 v^2_1\ri)v^2_1+\frac12{\mathfrak m}_2\le(1+\frac14 v^2_2\ri)v^2_2\\\nonumber
&+&\frac32\Bigl({\M}_{1}^{pq}a_{1}^pa_{1}^q+{\M}_{2}^{pq}a_{2}^pa_{2}^q\Bigr)-\frac12\Bigl[{\bm a}_1\cdot\le({\bm v}_1\times{\bm \Sc}_1\ri)+{\bm a}_2\cdot\le({\bm v}_2\times{\bm \Sc}_2\ri)\Bigr]\\\nonumber
&+&\sum_{l=0}^\infty\sum_{n=0}^\infty\frac{(-1)^n}{l!n!}{\dutchcal M}^L_{1} {\dutchcal M}^N_{2}\pd_{LN}\le(\frac1R\ri)+
\sum_{l=0}^{\infty}\sum_{n=0}^{\infty}\frac{(-1)^n}{l!n!}\le[F_{1}^{kp} {\M}_{1}^{kL}{\M}_2^{N}-F_{2}^{kp}{\M}_1^{L}{\M}_{2}^{kN}\ri]\pd_{pLN}\le(\frac1R\ri)\\\nonumber
&+&\sum_{l=0}^\infty\sum_{n=0}^\infty\frac{(-1)^n}{l!n!}\le(a^p_1 {\M}^{pL}_1{{\M}}^N_2+a^p_2 {\M}^{L}_1{ {\M}}^{pN}_2\ri)\pd_{LN}\le(\frac1R\ri)\\\nonumber
&+&\sum_{l=0}^\infty\sum_{n=0}^\infty\frac{(-1)^n}{l!n!}\le[\frac{a^p_1 {{\cal N}}^{L}_1{ {\M}}^N_2}{2l+3}-\frac{a^p_2 {\M}^{L}_1{{\cal N}}^{N}_2}{2n+3}\ri]\pd_{pLN}\le(\frac1R\ri)\\\nonumber
&+&\frac12\sum_{l=0}^\infty\sum_{n=0}^\infty\frac{(-1)^n}{l!n!}\le[{{\M}}_{1}^L{{\M}}_{2}^Nv^p_{1}v^q_{2}\pd_{pqLN}+\le(\dot{{\M}}^L_1 {\M}^N_2 v^p_2      -{{\M}}_{1}^L\dot{{\M}}_{2}^Nv^p_{1}\ri)\pd_{pLN}-\dot{{\M}}_{1}^L\dot{{\M}}_{2}^N\pd_{LN}\ri]R\\\nonumber
&+&\sum_{l=0}^\infty\sum_{n=0}^\infty\frac{(-1)^n}{l!n!}\Biggl[\Bigl(\frac12+\g\Bigr)\le(v^2_{1}+v^2_{2}\ri)-2(1+\g)\le({\bm v}_1\cdot{\bm v}_2\ri)\Biggr]{\M}_{1}^{L}{\M}_{2}^{N}\pd_{LN}\le(\frac1R\ri)  \\\nonumber
&+&\sum_{l=0}^\infty\sum_{n=0}^\infty\frac{(-1)^n}{l!n!}\le[\le(l+\frac12\ri) v^p_{1}-(l+1)v^p_{2}\ri]v^k_{1} {\M}_{1}^{kL}{\M}_{2}^{N}\pd_{pLN}\le(\frac1R\ri)\\\nonumber
&+&\sum_{l=0}^\infty\sum_{n=0}^\infty\frac{(-1)^n}{l!n!}\le[(n+1)v^p_{1}-\le(n+\frac12\ri) v^p_{2}\ri]v^k_{2}{\M}_{1}^{L} {\M}_{2}^{kN}\pd_{pLN}\le(\frac1R\ri)\\\nonumber
&+&\sum_{l=0}^\infty\sum_{n=0}^\infty\frac{(-1)^n}{l!n!}\le[(l+1){{\M}}_{1}^{pL}\dot{{\M}}_{2}^{N}+l\dot{{\M}}_{1}^{pL}{{\M}}_{2}^{N}\ri]v^p_{1}\pd_{LN}\le(\frac1R\ri)\\\nonumber
&+&\sum_{l=0}^\infty\sum_{n=0}^\infty\frac{(-1)^n}{l!n!}\le[(n+1)\dot{{\M}}_{1}^{L}{{\M}}_{2}^{pN}+n{{\M}}_{1}^L\dot{{\M}}_{2}^{pN}\ri]v^p_{2}\pd_{LN}\le(\frac1R\ri)\\\nonumber
&+&2(1+\g)
\sum_{l=0}^\infty\sum_{n=0}^\infty\frac{(-1)^n}{l!n!}\le[\le(\frac{\dot{{\M}}_{1}^{pL}{{\M}}_{2}^{N}}{l+1}-\frac{{\M}_{1}^L\dot{{{\M}}}_{2}^{pN}}{n+1}\ri)\le(v_1^p-v_2^p\ri)- \frac{\dot{{\M}}^{kL}_{1}\dot{{\M}}^{kN}_{2}}{(l+1)(n+1)}\ri]\pd_{LN}\le(\frac1R\ri)\\\nonumber
&+&2(1+\g)\sum_{l=0}^\infty\sum_{n=0}^\infty\frac{(-1)^n}{l!n!}\varepsilon_{pkq}v^k\le(\frac{\Sc_{1}^{qL}{\M}_{2}^N }{l+2}+\frac{{\M}_{1}^L \Sc_{2}^{qN} }{n+2}\ri)\pd_{pLN}\le(\frac1R\ri)\\\nonumber
&+&2(1+\g)\sum_{l=0}^\infty\sum_{n=0}^\infty\frac{(-1)^n}{l!n!}\varepsilon_{pkq}\le[\frac{\dot {\M}_{1}^{kL} \Sc^{qN}_{2}}{(l+1)(n+2)}-\frac{\Sc^{qL}_{1}\dot {\M}_{2}^{kN} }{(n+1)(l+2)}\ri]\pd_{pLN}\le(\frac1R\ri)\\\nonumber
&-&2(1+\g)\sum_{l=0}^\infty\sum_{n=0}^\infty\frac{(-1)^n}{l!n!}\frac{\Sc_{1}^{pL}\Sc_{2}^{qN}}{(l+2)(n+2)}\pd_{pqLN}\le(\frac1R\ri)\;.
\ea
where the multipoles $\dutchcal{M}^L_1, \dutchcal{M}_{2}^{L}$ are given by \eqref{ub36xc17s}, $R\equiv R_{12}=|{\bm x}_1-{\bm x}_2|$, all partial derivatives from $R$ are taken with respect to ${\bm x}_1$ like $\pd_i R=\pd R_{12}/\pd x^i_1$, a dot between two spatial vectors denote their Euclidean dot product while a cross between two vectors does the Euclidean cross product. The second piece of the Lagrangian \eqref{ui38x51za} is 
\ba\label{qta523}
{L}_{ab}&=&\le(1-\b\ri)\sum_{l=0}^{\infty}\sum_{n=0}^{\infty}\frac{1}{l!n!}\biggl[{\mathfrak I}_1^{LN}{\cal P}_{1L} {\cal P}_{1N}
 +{\mathfrak I}_2^{LN}{\cal P}_{2L}{\cal P}_{2N}\biggr]\\\nonumber
&&-\frac12\sum_{l=0}^{\infty}\sum_{n=0}^{\infty}\sum_{k=0}^{\infty}\sum_{s=0}^{\infty}\frac{(-1)^{k+s}}{l!n!k!s!}\biggl[{\mathfrak I}_1^{LN}\M^K_2 \M_2^S
 +{\mathfrak I}_2^{KS}\M^L_1 \M_1^N\biggr]\pd_{LK}\le(\frac1R\ri)\pd_{NS}\le(\frac1R\ri)\\\nonumber
&&-\gamma \sum_{l=0}^\infty\sum_{n=0}^{\infty}\sum_{s=0}^\infty\frac{l}{l!k!s!}\le[(-1)^{k+s}{\M}_1^{L} \M^K_2 \M^S_2+(-1)^l\M^L_2{\M}_1^{K} \M^S_1\ri]\pd_{LK} \le(\frac1R\ri)\pd_{S}\le(\frac1R\ri)\;,
\ea
where we have singled out the piece of the Lagrangian which is generated exclusively by the self-interaction of the scalar field having external multipole moments ${\cal P}_{1L}$ and ${\cal P}_{2L}$ produced by the first and second body respectively. Definition of the external scalar multipoles has been given in \eqref{jjec54} and reads more explicitly as
\ba\label{ndhg238727} 
{\cal P}_{1L}=\sum_{k=0}^{\infty}\frac{(-1)^k}{k!}\M_2^K\pd_{LK}\le(\frac1R\ri)\qquad&,&\qquad {\cal P}_{2L}=(-1)^l\sum_{k=0}^{\infty}\frac{1}{k!}\M_1^K\pd_{LK}\le(\frac1R\ri)\;.
\ea
Formally, the structure of the term with the coefficient $1-\beta$ in \eqref{qta523} is identical with the second term in the right hand side of this equation but the physical meaning of the two terms is drastically different because the second term in \eqref{qta523} describes exclusively the effects of general relativity.

In what follows, we consider a series expansion of the Lagrangian \eqref{ui38x51za} with respect to the degrees of multipole order more explicitly, and write this series as
\ba\label{vz52x1a0}
{L}={L}_{\rm m}+{L}_{\rm d}+{L}_{\rm q}+\ldots\;,
\ea
where  ${L}_{\rm m}$ contains only monopole terms, ${L}_{\rm d}$ includes dipole terms, ${L}_{\rm q}$ takes in quadrupole terms while the terms of higher multipole order are denoted with a lower dots symbol. 

\subsection{Monopole approximation}

The monopole part ${L}_{\rm m}$ of the Lagrangian corresponds to the case of massive bodies treated as structureless point-like particles. It consists of all terms in \eqref{ui38x51za}-\eqref{qta523} constrained to the value of indices $l=n=0$. It reads, 
\ba\label{nbza12k}
{L}_{\rm m}&=&
\frac12\mathfrak{m}_{1}v^2_1\Bigl(1+\frac14 v^2_{1}\Bigr)
+\frac12\mathfrak{m}_{2}v^2_2\Bigl(1+\frac14 v^2_{2}\Bigr)+\Bigl[1+\le(\frac12-\b\ri)\frac{{\mathfrak m}_{1}+{\mathfrak m}_{2}}{r}\Bigr]\frac{{\dutchcal M}_{1} {\dutchcal M}_{2}}{R}
\\\nonumber
&&+\Bigl[(2\g+1)\le(v^2_{1}+v^2_{2}\ri)-(3+4\g)\le({\bm v}_1\cdot{\bm v}_2\ri)-({\bm n}\cdot{\bm v}_1)({\bm n}\cdot{\bm v}_2)\Bigr]\frac{{\mathfrak m}_{1} {\mathfrak m}_{2}}{2R}\;,
\ea
where the gravitational mass $\dutchcal{M}_\A$ (A=1,2) is defined in \eqref{ub36xc17s} for the value of index $l=0$. Notice that the term with the parameter $\beta$ comes from the Lagrangian $L_{ab}$ describing the nonlinear interaction of the external scalar field.   

Because $\dutchcal{M}_\A$ is not equal to the Tolman mass $\mathfrak{m}_\A$ the Newtonian dynamics is modified as pointed out by \citet{nord_PhysRev1968a,nord_PhysRev1968b}. We have neglected the difference between $\dutchcal{M}_\A$ and $\mathfrak{m}_\A$ in the post-Newtonian terms of \eqref{nbza12k} as it leads to the residual terms of the 2PN order. Notice that our treatment of the SEP accounts for the effects of non-stationarity of orbiting bodies \footnote{See the presence of the second time derivative term in definition of $\dutchcal{M}_\A$ in equation \eqref{ub36xc17s}.} which can be important in the treatment of dynamical tides in coalescing binary systems beyond the adiabatic tidal approximation in scalar-tensor theory \citep{Bernard_2018PRD,Bernard_2020PRD}.
  
The Lagrangian \eqref{nbza12k} for monopole massive particles has been derived independently many times by various researchers who have also used different mathematical techniques. Among these derivations the most notable works are by \citet{lord3} and \citet{Fichte_1950} who had derived the Lagrangian ${L}_{\rm m}$ from EIH equations of motion (see also \citep[\S 106]{Landau1975}). PPN parameter $\gamma$ has been introduced to the Lagrangian $L_{\rm m}$ by \citet{estabrook_1969ApJ}. Later on, PPN parameter $\beta$ was added to the Lagrangian of point-like bodies by \citet{willbook} and by \citet{Barker_Oconnell_1976PhRvD}. We notice that the $\beta-\gamma$ Lagrangian written by \citet[Eq. (6.80)]{willbook} treats the inertial and gravitational masses entering the Lagrangian as being equal while \citet{Barker1986_acceleration} clearly distinguish the inertial and gravitational masses to take into account the Nordtvedt effect.

\subsection{Dipole approximation}\label{nhuste432}
The dipole part ${L}_{\rm d}$ of the Lagrangian includes the active mass dipole moments ${\bm{\dutchcal M}}_{1}=\le({\dutchcal M}^i_{1}\ri)$, ${\bm{\dutchcal M}}_{2}=\le({\dutchcal M}^i_{2}\ri)$ of the bodies, the scalar tidal dipoles ${\bm{\mathcal D}}_1=\le({\cal D}^i_1\ri)$, ${\bm{\mathcal D}}_2=\le({\cal D}^i_2\ri)$ introduced in \eqref{gt3vz41}, and the body's spin ${\bm \Sc}_1=\le(\Sc^i_1\ri)$, ${\bm \Sc}_2=\le(\Sc^i_2\ri)$. It consists of two parts
\ba\label{dip9634}
{L}_{\rm d}&=&{L}_{\rm dm}+{L}_{\rm ds}\;,
\ea
where 
\ba\label{dip953432}
{L}_{\rm dm}&=&\Bigl[{\mathfrak m}_{1}\le({\bm n}\cdot{\bm{\dutchcal M}}_{2}\ri) -{\mathfrak m}_{2}\le({\bm n}\cdot{\bm{\dutchcal M}}_{1}\ri)\Bigr]\frac{1}{R^2}+\Bigl[{\mathfrak m}_{1}\le({\bm n}\cdot{\bm{\mathcal D}}_{2}\ri) -{\mathfrak m}_{2}\le({\bm n}\cdot{\bm{\mathcal D}}_{1}\ri)\Bigr]\frac{1}{2R^2}\;,
\ea
is the mass dipole part of the Lagrangian, and 
\ba\label{dip15vcx}
{L}_{\rm ds}&=&\frac12{\bm a}_1\cdot\le({\bm \Sc}_1\times{\bm v}_1\ri)+\frac12{\bm a}_2\cdot\le({\bm \Sc}_2\times{\bm v}_2\ri)+(1+\g)\le[{\bm n}\cdot\le({\bm \Sc}_1\times{\bm v}\ri)\frac{{\mathfrak m}_{2}}{R^2}
+{\bm n}\cdot\le({\bm \Sc}_2\times{\bm v}\ri)\frac{{\mathfrak m}_{1}}{R^2}\ri]\\\nonumber
&+&\frac12(1+\g)\Bigl[{\bm \Sc}_1\cdot{\bm \Sc}_2-3({\bm n}\cdot{\bm \Sc}_1)({\bm n}\cdot{\bm \Sc}_2)\Bigr]\frac1{R^3}\;,
\ea
is its spin dipole part. Here, the unit vector ${\bm n}=(n^i)$, $n^i=R^i/R$, $R^i=x^i_1-x^i_2$, the relative velocity ${\bm v}=(v^i)$, $v^i=v^i_1-v^i_2$, the dot between two spatial vectors denote their Euclidean dot product and the cross "$\times$" between vectors is the Euclidean cross product. It is worth noticing that the part of $L_{dm}$ with the scalar dipole moments ${\cal D}^i$ is taken from the first term in \eqref{qta523} which is proportional to $1-\b$. 

In general relativity the particle dipole approximation refers merely to spins of the bodies and the dipole Lagrangian of general relativity includes only $L_{\rm ds}$. The mass dipole moments of the bodies in general relativity depend merely on the displacement between the origin of body's local coordinates and its center of mass. Thus, the mass dipole is not physical and vanishes in general relativity if the origin of the local coordinates is chosen at the center of mass of the body. In scalar-tensor theory of gravity the definition of the center of mass of each body A is given in terms of the conformal dipole moment ${\cal J}^i_\A$, introduced above in \eqref{yxc36cf}, while the gravitational force in the equations of motion is governed by the {\it active} multipole moments of the bodies. The {\it active} dipole moment $\dutchcal{M}^i_\A$ of body A (A=1,2) is taken from \eqref{ub36xc17s} and \eqref{wvzx35x} for $l=1$, and reads
\ba\label{kib36dxf}
\dutchcal{M}^i_\A&=&\int_{{\cal V}_\A}\hat\sigma(u,{\bm w})w^id^3w+\eta\mathfrak{U}^i_\A+\g\le(2\mathfrak{T}^i_\A+\mathfrak{U}^i_\A+\mathfrak{S}^i_\A\ri)
+\frac1{10}\ddot{\cal N}_\A^i-\frac35(1+\g)\dot{\cal R}^i_\A\;,
\ea
where the general-relativistic mass density $\hat\sigma(u,{\bm w})$ is defined in \eqref{ik34ux5s}, and vector integrals $\mathfrak{T}^i_\A$, $\mathfrak{U}^i_\A$, $\mathfrak{S}^i_\A$ depend on the internal distribution of matter as explained in Appendix \ref{fcstrevckivt}. The dipole $\dutchcal{M}^i_\A$ does not vanish even if the {\it conformal} dipole moment ${\cal J}^i_\A=0$. Indeed, the very first term in the right hand side of \eqref{kib36dxf} can be computed from the condition ${\cal J}^i_\A=0$ by taking into account definitions 
\eqref{yxc36cf} and \eqref{op1qz4},
\ba\label{irxcw4289}
\int_{{\cal V}_\A}\hat\sigma(u,{\bm w})w^id^3w&=&\sum_{l=1}^\infty\frac{1}{(l-1)!}{\cal Q}_L\M^{iL}_\A-\frac12\sum_{l=0}^\infty\frac{1}{(2l+3)l!}{\cal Q}_{iL}{\cal N}^{L}_\A\;,
\ea
where the external gravitoelectric moments ${\cal Q}_L$ have been defined in \eqref{jje8c}.
Replacing \eqref{irxcw4289} in \eqref{kib36dxf} and making use of vector virial theorem \eqref{vect46z3a} we get
\ba\label{czd4290c}
\dutchcal{M}^i_\A&=&\eta\mathfrak{U}^i_\A+(\g-1)\le[\frac35\dot{\cal R}^i_\A-\frac1{10}\ddot{\cal N}_\A^i-\sum_{l=1}^\infty\frac{1}{(l-1)!}{\cal Q}_L\M^{iL}_\A+\frac12\sum_{l=0}^\infty\frac{1}{(2l+3)l!}{\cal Q}_{iL}{\cal N}^{L}_\A\ri]\;,
\ea
which demonstrates the presence of the Nordtvedt effect in the dipolar approximation of the scalar-tensor theory of gravity. Notice that $\dutchcal{M}^i_\A=0$ for spherically-symmetric bodies because all terms in the right hand side of \eqref{czd4290c} are integrals which angular part vanishes in case of the spherical symmetry except of ${\cal N}_\A$. However, this term couples with the local non-geodesic acceleration ${\cal Q}_i$ which is nil for spherically symmetric bodies. This consideration reveals that the {\it active} mass dipole $\dutchcal{M}^i_\A$ is sufficiently small in practical situations and can be neglected in most cases. 
  
%We emphasize once again that we do not use expression \eqref{czd4290c} to replace for $\dutchcal{M}^i_\A$ in the Lagrangian ${L}_{\rm dm}$. This is %because we have used on-shell virial vector theorem \eqref{vect46z3a} in order to get \eqref{czd4290c} that depends on the external dynamic variables 
%${\cal Q}_L$ while \eqref{kib36dxf} does not include such a dependence. Making replacement of the dipole \eqref{czd4290c} in the Lagrangian changes the %outcome of the variational derivatives and leads to the equations of motion of the bodies derived in the gauge which is different from our original %harmonic gauge \citep{SCHAFER1984,Milgrom_1994PhLA}. Nonetheless, we would like to stress that the use of \eqref{czd4290c} in the Lagrangian is admissible %but we have to track down the corresponding change in the gauge. 

Nevertheless, the scalar tidal dipole ${\cal D}^i$ does not vanish and gives contribution to $L_{\rm dm}$ which stems from the dipolar part of the term being proportional to $1-\beta$ in \eqref{qta523}. According to definition \eqref{gt3vz41}, the scalar tidal dipole is proportional to the moment of inertia ${\cal N}_\A$ coupled with the external scalar dipole ${\cal P}_i$ and reads
\ba\label{p1b5z2aq0}
{\cal D}^i_1=-\lambda^{\rm(s)}_1{\mathfrak m}_2\frac{n^i}{R^2}\qquad&,&{\cal D}^i_2=\lambda^{\rm(s)}_2{\mathfrak m}_1\frac{n^i}{R^2}
\ea
where $\lambda^{\rm(s)}_1$, $\lambda^{\rm(s)}_2$ are the scalar deformabilities of the bodies \eqref{bra4xz900}. It allows to get an explicit form of the tidal scalar dipole part of the Lagrangian, 
\ba\label{jh33z}
L_{\rm dm}^{\rm (s)}&=&\frac{{\mathfrak m}_1^2\lambda^{\rm(s)}_2+{\mathfrak m}_2^2\lambda^{\rm(s)}_1}{2R^4}\;.
\ea
It brings about the tidal dipolar effects which measurement can be used in order to put additional constraints on the parameters of the scalar-tensor theory by gravitational-wave detectors \citep{Bernard_2020PRD}.  Equations of motion with the scalar tidal dipole effects included, are obtained from the Lagrangian ${L}_{\rm dm}$ after taking the variational derivative. It yields
\ba\label{ibdfcr37a}
\mathfrak{m}_{1}a_1^i&=&-\frac{\dutchcal{M}_{1}\dutchcal{M}_{2}}{r^2}n^i-2\le[{\mathfrak m}_1^2\lambda^{\rm(s)}_2+{\mathfrak m}_2^2\lambda^{\rm(s)}_1\ri]\frac{n^i}{R^5}\;,\\
\mathfrak{m}_{2}a_2^i&=&+\frac{\dutchcal{M}_{1}\dutchcal{M}_{2}}{r^2}n^i+2\le[{\mathfrak m}_1^2\lambda^{\rm(s)}_2+{\mathfrak m}_2^2\lambda^{\rm(s)}_1\ri]\frac{n^i}{R^5}\;.
\ea
The relative acceleration $a^i=a^i_1-a^i_2$ of the two bodies is
\ba\label{nxb746dv}
a^i&=&-\frac{\alpha{\mathfrak m}n^i}{R^2}-2\le[\frac{{\mathfrak m}_2}{{\mathfrak m}_1}\lambda^{\rm(s)}_1+\frac{{\mathfrak m}_1}{{\mathfrak m}_2}\lambda^{\rm(s)}_2\ri]\frac{{\mathfrak m}n^i}{R^5}\;,
\ea
where ${\mathfrak m}={\mathfrak m}_1+{\mathfrak m}_2$ is the total gravitational mass of the system and the parameter $\alpha=\dutchcal{M}_{1}\dutchcal{M}_{2}/{\mathfrak m}_1{\mathfrak m}_2$ characterizes the Nordtvedt effect. Equation \eqref{nxb746dv} coincides in the post-Newtonian approximation with similar formula derived by \citet[eq. 7]{Bernard_2020PRD} after accounting for the definitions of the PPN parameters and the sensibility of the bodies.
 
The spin dipole part of the Lagrangian, ${L}_{\rm ds}$, has been derived by \citet{vab} in general relativity who used the harmonic gauge but replaced the body's accelerations with the Newtonian equations of motion in the Lagrangian. It was \citet{damour_1982CR} who included the acceleration-dependent part to the Lagrangian ${L}_{\rm ds}$ in general relativity while \citet{Barker1986_acceleration} extended this Lagrangian \eqref{dip15vcx} to the case of scalar-tensor theory of gravity and developed the {\it method of the double zero} to eliminate the acceleration-dependent terms from the Lagrangian without imposing an implicit gauge transformation. \citet{Barker_Oconnell1987JMP} have also analyzed the structure of the spin-dipole Lagrangian under different spin complementary conditions and in different post-Newtonian coordinates.

\subsection{Quadrupole approximation}\label{quad6353413}
The quadrupole approximation includes all terms depending on mass and spin quadrupole moments of the bodies. The quadrupole Lagrangian consists of three pieces
\ba\label{ubxyt363v}
{L}_{\rm q}&=&{L}_{\rm qm}+{L}_{\rm qs}+{L}_{\rm qa}\;.
\ea
Here, the first term in \eqref{ubxyt363v} is the mass quadrupole Lagrangian
\ba\label{rxcdit75}
{L}_{\rm qm}&=&
\frac32\Bigl[\mathfrak{m}_1\le(\M^{pq}_2+2F_{2}^{kp}\M_2^{kq}\ri)+\mathfrak{m}_2\le(\M^{pq}_1+2F_{1}^{kp}\M_1^{kq}\ri)\Bigr]\frac{n^pn^q}{R^3}\\\nonumber
&+&\frac32\Bigl(\mathfrak{m}_1\M^{pq}_2+\mathfrak{m}_2\M^{pq}_1\Bigr)\Bigl[\Bigl(\frac12+\g\Bigr)\le(v^2_{1}+v^2_2\ri)-\Bigl(\frac32+2\g\Bigr)({\bm v}_1\cdot{\bm v}_2)-\frac{5}{2}({\bm n}\cdot{\bm v}_1)({\bm n}\cdot{\bm v}_2)\Bigr]\frac{n^pn^q}{R^3}\\\nonumber
&+&\frac32\mathfrak{m}_1\M^{pq}_2\Bigl[3n^p v^q_2\le({\bm n}\cdot{\bm v}_2-{\bm n}\cdot{\bm v}_1\ri)+n^p v^q_1({\bm n}\cdot{\bm v}_2)-v^p_2 v^q_2
+v^p_1 v^q_2\Bigr]\frac{1}{R^3}\\\nonumber
&+&\frac32\mathfrak{m}_2\M^{pq}_1\Bigl[3 n^p v^q_1\le({\bm n}\cdot{\bm v}_1-{\bm n}\cdot{\bm v}_2\ri)+n^p v^q_2({\bm n}\cdot{\bm v}_1)-v^p_1 v^q_1
+v^p_1 v^q_2\Bigr]\frac{1}{R^3}\\\nonumber
&+&\mathfrak{m}_1\dot{\M}_2^{pq}\Bigl[\Bigl(2+\g\Bigr)v^p_2-\Bigl(\frac12+\g\Bigr)v^p_1\Bigr]\frac{n^q}{R^2}-\frac34\mathfrak{m}_1\dot{\M}_2^{pq}\le({\bm n}\cdot{\bm v}_1\ri)\frac{n^pn^q}{R^2}\\\nonumber
&+&\mathfrak{m}_2\dot{\M}_1^{pq}\Bigl[\Bigl(\frac12+\g\Bigr)v^p_2-\Bigl(2+\g\Bigr)v^p_1\Bigr]\frac{n^q}{R^2}+\frac34\mathfrak{m}_2\dot{\M}_1^{pq}\le({\bm n}\cdot{\bm v}_2\ri)\frac{n^pn^q}{R^2}\\\nonumber
&+&(\frac12-\b)\le[\le(3\mathfrak{m}_1+4\mathfrak{m}_2\ri)\mathfrak{m}_2\M_1^{pk}+\le(3\mathfrak{m}_2+4\mathfrak{m}_1\ri)\mathfrak{m}_1\M_2^{pk}\ri]\frac{n^pn^q}{R^4}\\\nonumber
&-&3\g\le[\mathfrak{m}_1^2\M_2^{pk}+\mathfrak{m}_2^2\M_1^{pk}\ri]\frac{n^pn^q}{R^4}
-\frac1{6}\le({\cal N}_1\mathfrak{m}_2^2+{\cal N}_2\mathfrak{m}_1^2\ri)\frac1{R^4}\;.
\ea
The second term in the right hand side of \eqref{ubxyt363v} is the spin quadrupole Lagrangian 
\ba\label{ncxge534}
{L}_{\rm qs}&=&2(1+\g)\Bigl[\mathfrak{m}_2 S^{pq}_1-\mathfrak{m}_1S^{pq}_2\Bigr]\frac{({\bm n}\times{\bm v})^pn^q}{R^3}\;.
\ea
that coincides exactly with the corresponding Lagrangian derived recently by \citet[Eq. 29]{Vines_2020PRD}.  
The third term in the right hand side of \eqref{ubxyt363v} describes the acceleration-dependent part of the mass  quadrupole Lagrangian
\ba\label{jg35xcs32}
{L}_{\rm qa}&=&\frac32\Bigl(\M_{1}^{pq}a_{1}^pa_{1}^q+\M_{2}^{pq}a_{2}^pa_{2}^q\Bigr)
+\le(\mathfrak{m}_1\M^{pq}_2a_2^p-\mathfrak{m}_2\M^{pq}_1a_1^p\ri)\frac{n^q}{R^2}+\frac13\le(\mathfrak{m}_1{\cal N}_2a_2^p-\mathfrak{m}_2{\cal N}_1a^p_1\ri)\frac{n^p}{R^2}\;.
\ea
which can be reduced to a "canonical" form depending only on the coordinates and the first time derivatives of the dynamic variables. 

More specifically, the terms depending quadratically on accelerations in the Lagrangian \eqref{jg35xcs32} can be eliminated by applying the {\it method of the double zero} having been worked out by \citet{Barker1986_acceleration}. Let us introduce two functions
\Asu\label{lom08ecvxdaw}
\ba\label{mumu8634}
Z_1^i&\equiv& a_1^i+\frac{\mathfrak{m}_2}{R^2}n^i\;,\\
Z_2^i &\equiv& a_2^i-\frac{\mathfrak{m}_1}{R^2}n^i\;,
\ea
\esu
where we have used the monopole approximation for gravitational force as it is sufficient in the quadrupole approximation. These functions vanish on-shell but they are non-vanishing if the equations of motion are not implied.

Equations \eqref{lom08ecvxdaw} allows us to recast the Lagrangian \eqref{jg35xcs32} to the following form
\ba\label{jvdt45f}
{L}_{\rm qa}&=&\frac32 \le({\M}_{1}^{pq}Z^p_1Z^q_1+{\M}_{2}^{pq}Z^p_2Z^q_2\ri)+4\le(\mathfrak{m}_1{\M}^{pq}_2a_2^p-\mathfrak{m}_2{\M}^{pq}_1a_1^p\ri)\frac{n^q}{R^2}-\frac32\le(\mathfrak{m}_1^2{\M}^{pq}_2+\mathfrak{m}_2^2{\M}^{pq}_1\ri)\frac{n^pn^q}{R^4}\\\nonumber
&+&\frac13\le(\mathfrak{m}_1{\cal N}_2a_2^p-\mathfrak{m}_2{\cal N}_1a^p_1\ri)\frac{n^q}{R^2}\;.
\ea
The products $Z^p_1Z^q_1$ and $Z^p_2Z^q_2$ are the double-zero terms \citep{Barker1986_acceleration,Damour_1989MG5} with the property that each factor in the product vanishes on-shell independently one from another.  Therefore, the double-zero terms do not contribute to the equations of motion after taking the variational derivative and can be discarded from the Lagrangian. 

Terms which are linear with respect to accelerations in the Lagrangian can be eliminated with the help of the Leibniz chain rule by interchanging time derivatives in each term of \eqref{jvdt45f} which depends on acceleration linearly. For example,
\ba\label{jvhy34}
\mathfrak{m}_2{\M}^{pq}_1a_1^p\frac{n^q}{R^2}&=&\frac{d}{dt}\le[\mathfrak{m}_2{\M}^{pq}_1v_1^p\frac{n^q}{R^2}\ri]
-\mathfrak{m}_2\dot{{\M}}^{pq}_1v_1^p\frac{n^q}{R^2}-\mathfrak{m}_2{\M}^{pq}_1v_1^p\frac{v^q}{R^3}
+3\mathfrak{m}_2{\M}^{pq}_1v_1^p\frac{({\bm n}\cdot{\bm v})n^q}{R^4}\;,
\ea
and similar procedure is applied to other terms in \eqref{jvdt45f}. After employing \eqref{jvhy34} the total time derivative appears in the Lagrangian \eqref{jvdt45f} but it has no effect on the equations of motion of the bodies and, as such, can be discarded. The remaining terms do not depend on accelerations and can be combined with similar terms in ${L}_{\rm qm}$. The elimination of the acceleration-dependent terms from the Lagrangian may be useful under certain circumstances and for comparison with the Lagrangians derived by other authors. 

\subsection{Comparison of the quadrupole Lagrangians of different authors}\label{mw71cz52}

First attempt to derive the post-Newtonian Lagrangian of an $N$-body system with accounting for the quadrupole moments of the bodies had been undertaken by \citet{vab} who conducted calculations in global coordinates. Brumberg used the Newtonian definition of the center of mass of the bodies and the multipoles defined in the global coordinates. For these reason, Brumberg's Lagrangian is expressed in terms of not physically-measurable quantities and has a lot of coordinate-dependent effects which are formidably difficult to track down. In particular, reduction of Brumberg's Lagrangian to the case of spherically-symmetric bodies lead to the appearance of the finite-size effects depending on the moments of inertia of the bodies but they can be eliminated after transformation of body's multipoles to the local coordinates as pointed out by \citet{vincent_1986CeMec} and detailed in \citep[section 6.3.3]{kopeikin_2011book}. The unphysical definitions of the center of mass and body's multipoles inhibit the range of practical applications of Brumberg's Lagrangian in astrophysics and relativistic celestial mechanics.     

More recently, \citet{xu_1997PhRvD,xu_1998ScChA} have striven to derive the post-Newtonian Lagrangian in quadrupole approximation by applying the DSX-formalism \citep{dsx1,dsx2} of the local coordinates in the $N$-body problem. However, they obtained equations of motion that significantly diverge from two subsequent and completely independent analytic derivations by \citet{racine_2005PhRvD,racine2013PhRvD} and \citet{k2019PRD} who came up to exactly the same result. We could not track down the origin of the disagreement and do not comment on the Lagrangian corresponding to the equations derived by \citet{xu_1997PhRvD,xu_1998ScChA}.

The post-Newtonian Lagrangian ${L}^{\rm VF}_{\rm qm}$ of two-body problem in the mass quadrupole approximation has been worked out by \citet{Vines_Flanagan_PRD024046} from the equations of motion having been derived previously in \citep{racine_2005PhRvD,racine2013PhRvD} for $N$ bodies possessing all mass and spin multipoles. Later on, the quadrupole Lagrangian ${L}^{\rm S}_{\rm qm}$ has been derived by \citet{Steinhoff_etal_PRD104028} in the framework of the effective field theory \citep{goldberger_PRD} of a point-particle action augmented by dynamical quadrupolar degrees of freedom. It was shown \citep{Steinhoff_etal_PRD104028} that the two Lagrangians differ by a total time derivative 
\ba\label{knxye537}
{L}^{\rm VF}&=&{L}^{\rm S}+\frac{dF}{dt}\;,
\ea
where the function $F$ is shown in \citep[Eq. 3.13]{Steinhoff_etal_PRD104028}. Since the difference between the two Lagrangians is a total time derivative, the Lagrangians ${L}^{\rm VF}$ and ${L}^{\rm S}$ are effectively the same. For computational reasons the Lagrangian ${L}^{\rm S}$ is more convenient for the comparison with our Lagrangian ${L}$ which will be done under simplifying assumption that the first body is characterized by both mass and quadrupole moment while the second body has only mass. We shall also reduce our expressions to general relativity so that the parameters $\beta=\gamma=1$ yielding $\eta=0$. 

The Lagrangian ${L}^{\rm S}$ reads
\ba\label{bxbgwge64}
{L}^{\rm S}&=&{L}_{\rm pm}-V_{\rm EQ}\;,
\ea 
where ${L}_{\rm pm}$ is the point particle (monopole) Lagrangian, the quadrupole potential $V_{\rm EQ}$ is given in \citep[Eq. 3.11]{Steinhoff_etal_PRD104028}, and we neglected for the sake of simplification the spin and quadrupole-quadrupole (tidal) terms. Following the paper \citep{Steinhoff_etal_PRD104028} we denote masses of the first and second body, $m_1$ and $m_2$, and the quadrupole moment of the first body $Q^{ij}$ \footnote{Notation $Q^{ij}$ is used by convention to denote the quadrupole induced by tides cause by the presence of a second body \citep{Vines_Flanagan_PRD024046,Steinhoff_etal_PRD104028}. It should not be confused with the gravitoelectric external multipole ${\cal Q}_{ij}$}. The second body is assumed to be point-like mass $m_2$ having no internal structure.
Then, the monopole Lagrangian
\ba\label{jvxfst364g}
{L}_{\rm pm}&=&
\frac12m_{1}v^2_1\Bigl(1+\frac14 v^2_{1}\Bigr)
+\frac12m_{2}v^2_2\Bigl(1+\frac14 v^2_{2}\Bigr)+\Bigl[3\le(v^2_{1}+v^2_{2}\ri)-7\le({\bm v}_1\cdot{\bm v}_2\ri)-({\bm n}\cdot{\bm v}_1)({\bm n}\cdot{\bm v}_2)\Bigr]\frac{m_{1} m_{2}}{2R}
\\\nonumber
&+&\Bigl(1-\frac{m_{1}+m_{2}}{2R}\Bigr)\frac{m_{1}m_{2}}R\;,
\ea
and the quadrupole Lagrangian
\ba\label{kaw8254f}
V_{\rm EQ}&=&-\frac{m_2Q^{pq}}{2R^3}\Bigl[3n^pn^q+v_1^pv^q_1-v^p_1v^q_2-3v^p_1n^q\le({\bm n}\cdot{\bm v}_1-{\bm n}\cdot{\bm v}_2\ri)+3v^p_2n^q\le({\bm n}\cdot{\bm v}_1\ri)\Bigr]\\\nonumber
&-&\frac34\frac{m_2Q^{pq}n^pn^q}{R^3}\Bigl[3v^2_1-7\le({\bm v}_1\cdot{\bm v}_2\ri)+3v^2_2-5\le({\bm n}\cdot{\bm v}_1\ri)\le({\bm n}\cdot{\bm v}_2\ri)\Bigr]+\frac{m_2Q^{pq}n^pn^q}{R^4}\Bigl(\frac32 m_1+6m_2\Bigr)\\\nonumber
&+&\frac{m_2}{R^2}\Bigl[Q^{pq}a_1^pn^q+\dot{Q}^{pq}\Bigl(v_1^pn^q-\frac32 v_2^p n^q-\frac34 n^pn^q{\bm n}\cdot{\bm v}_2\Bigr)\Bigr]\;,
\ea
where notations for coordinates, velocities and accelerations are the same as in our equation \eqref{rxcdit75}. 
 
Our Lagrangian 
\ba\label{ujdbvf98qq}
L&=&{L}_{\rm m}+{L}_{\rm q}\;.
\ea
has the monopole term 
\ba\label{mnscbvqjhjhwg}
{L}_{\rm m}&=&
\frac12\mathfrak{m}_{1}v^2_1\Bigl(1+\frac14 v^2_{1}\Bigr)
+\frac12\mathfrak{m}_{2}v^2_2\Bigl(1+\frac14 v^2_{2}\Bigr)+\Bigl[3\le(v^2_{1}+v^2_{2}\ri)-7\le({\bm v}_1\cdot{\bm v}_2\ri)-({\bm n}\cdot{\bm v}_1)({\bm n}\cdot{\bm v}_2)\Bigr]\frac{\mathfrak{m}_{1} \mathfrak{m}_{2}}{2R}
\\\nonumber
&+&\Bigl(1-\frac{\mathfrak{m}_{1}+\mathfrak{m}_{2}}{2R}\Bigr)\frac{\mathfrak{m}_{1}\mathfrak{m}_{2}}R\;.
\ea
which is identical to \eqref{jvxfst364g} under condition that masses $m_1=\mathfrak{m}_1$, $m_2=\mathfrak{m}_2$ are the Tolman masses of the bodies. The quadrupole Lagrangian in \eqref{ujdbvf98qq} reads
\ba
\label{ube9981egt}
{L}_{\rm q}&=&\frac{{\mathfrak m}_2}{2R^3}\le(\M^{pq}_1+2F_{1}^{kp}\M_1^{kq}\ri)\Bigl[3n^pn^q-3v_1^pv^q_1+3v^p_1v^q_2+9v^p_1n^q\le({\bm n}\cdot{\bm v}_1-{\bm n}\cdot{\bm v}_2\ri)+3v^p_2n^q\le({\bm n}\cdot{\bm v}_1\ri)\Bigr]\\\nonumber
&+&\frac34\frac{{\mathfrak m}_2\M_1^{pq}n^pn^q}{R^3}\Bigl[3v^2_1-7\le({\bm v}_1\cdot{\bm v}_2\ri)+3v^2_2-5\le({\bm n}\cdot{\bm v}_1\ri)\le({\bm n}\cdot{\bm v}_2\ri)\Bigr]-\frac{{\mathfrak m}_2\M_1^{pq}n^pn^q}{R^4}\Bigl(\frac32{\mathfrak m}_1+5{\mathfrak m}_2\Bigr)\\\nonumber
&-&\frac{3{\mathfrak m}_2}{R^2}\dot{\M}_1^{pq}\Bigl(v_1^pn^q-\frac12 v_2^p n^q-\frac14 n^pn^q{\bm n}\cdot{\bm v}_2\Bigr)+\frac32 \M^{pq}_1a_{1}^pa_{1}^q-\frac{\mathfrak{m}_2\M_1^{pq}}{r^2}a_1^p n^q-\frac{\mathfrak{m}_2{\cal N}_1}{3R^2}\le(a^p_1n^p+\frac{\mathfrak{m}_2}{2R^2}\ri)\;,
%\\\nonumber
%&+&\frac32 \M^{pq}_1a_{1}^pa_{1}^q-\frac{\mathfrak{m}_2\M_1^{pq}}{r^2}a_1^p n^q-\frac13\mathfrak{m}_2{\cal N}_1a^p_1\frac{n^p}{R^2}\;,
\ea
where we have neglected the acceleration of the second body.

Subtracting the Lagrangian \eqref{bxbgwge64} from our Lagrangian ${L}$ given in \eqref{ujdbvf98qq} yields
\ba\label{sjdhgcwuy364}
{L}-{L}^{\rm S}&=&-2\frac{{\mathfrak m}_2}{R^3}{\M}_1^{pq}v_1^p\Bigl[v^q-3n^q\le({\bm n}\cdot{\bm v}\ri)\Bigr]
-2\frac{{\mathfrak m}_2}{R^2}\dot{\M}_1^{pq}v_1^pn^q+\frac{{\mathfrak m}^2_2}{R^4}{\M}_1^{pq}n^pn^q+\frac32 {\M}_1^{pq}a_{1}^pa_{1}^q\\\nonumber
&&+3{\mathfrak m}_2F_{1}^{kp}\M_1^{kq}\frac{n^pn^q}{R^3}-\frac{\mathfrak{m}_2{\cal N}_1}{3R^2}\le(a^p_1n^p+\frac{\mathfrak{m}_2}{2R^2}\ri)\;,
\ea
where ${\bm v}=(v^p)$ is the relative velocity, $v^p=v^p_1-v^p_2$. The right hand side of \eqref{sjdhgcwuy364} can be rearranged to a different form depending explicitly on a {\it double zero} term and a total time derivative
\ba\label{undg3r74u}
{L}-{L}^{\rm S}&=&{\M}^{pq}_1Z^p_1Z^q_1-\frac{d}{dt}\le[2\frac{{\mathfrak m}_2}{R^2}{\M}_1^{pq}v_1^pn^q\ri]+\frac12 {\M}_1^{pq}a_{1}^pa_{1}^q+3{\mathfrak m}_2F_{1}^{kp}\M_1^{kq}\frac{n^pn^q}{R^3}-\frac{{\cal N}_1\mathfrak{m}_2}{3R^2}\le(a^p_1n^p+\frac{\mathfrak{m}_2}{2R^2}\ri)\;,
\ea
where $Z^p_1$ is a vector function given in \eqref{mumu8634} that vanishes on-shell. 
Here, the double zero term ${\M}^{pq}_1Z^p_1Z^q_1$ can be discarded as the variational derivative from this term vanishes on-shell so that it does not contribute to the equations of motion \citep{Barker1986_acceleration,Damour_Schafer1991JMP}. The term with the total time derivative can be discarded for the same reason. The matrix of relativistic precession can be eliminated by choosing the local frame of reference rotating with the angular velocity compensating the precession \citep{kopeikin_2011book}. This was the choice made in \citep{Vines_Flanagan_PRD024046,Steinhoff_etal_PRD104028}. Therefore, the only principal difference which remains between our Lagrangian ${L}$ and that derived in papers \citep{Vines_Flanagan_PRD024046,Steinhoff_etal_PRD104028} is in the terms which are proportional to the moment of inertia and the double acceleration. However, after taking the variational derivative these terms are reduced to a second time derivative in the equations of translational motion
\ba\label{uyvxv64z}
\frac{\textswab{d}}{\textswab{d}x^i}\le[\frac12{\M}_1^{pq}a_{1}^pa_{1}^q-\frac{{\cal N}_1\mathfrak{m}_2}{3R^2}\le(a^p_1n^p+\frac{\mathfrak{m}_2}{2R^2}\ri)\ri]=\frac{d^2}{dt^2}\biggl[\frac{\mathfrak{m}_2}{R^2}\mathfrak{I}_1^{ip}n^p\biggr]\;,
\ea
that is equivalent to adjusting the worldline of the center of mass of the first body, and is a purely coordinate effect. Effectively, equation \eqref{uyvxv64z} provides a proof that the finite-size effects due to the moment of inertia ${\cal N}_\A$ of the bodies, do not appear in the post-Newtonian equations of motion in general relativity.

\section{Conclusion}\label{concl73746}

We have derived the post-Newtonian Lagrangian for translational motion of $N$ bodies in an isolated astronomical system. The Lagrangian includes all mass and spin multipoles of the bodies along with their time derivatives and significantly extends the results of previous studies which were limited to the pole-dipole-quadrupole particles. The Lagrangian is given in two forms with the multipoles expressed either in the global or local coordinate frames. We have compared our Lagrangian in the pole-dipole-quadrupole approximation to the results of previous works of other researchers and observed that they are consistent up to the gauge transformation of the Lagrangian.  
The resulting Lagrangian of the present paper does not include the Lagrangian $L_{\rm int}$ for internal (rotation, pulsation, etc.) motions of matter inside the bodies. The problem of finding the Lagrangian $L_{\rm int}$ will be tackled somewhere else.  

\section*{Acknowledgments} 
I thank the anonymous referee for constructive comments and bibliographic references which helped to make it more comprehensive as well as Dr. J. Steinhoff for useful discussions. I am grateful to Dr. Yi-Zen Chu for bringing my attention to his comprehensive article \citep{Yizen_PRD2009} on the solution of an $N$-body problem in general relativity by using the technique of the Feynman diagrams. I deeply appreciate the circumstantial explanation of Dr. M. Levi on the theoretical progress achieved in the post-Newtonian solution of gravitational two-body problem within the framework of the effective field theory in general relativity \citep{Levi_2015,Levi_2015JHEP,Levi_2017CQG,Levi_2020}.  
\newpage
\begin{appendix}
\section{Virial Theorems}\label{app1a}
The virial theorems constitute a part of the post-Newtonian Lagrangian mechanics of an $N$-body system as they allow us to interconnect the external and internal degrees of freedom of a self-gravitating systems. There are global and local versions of the virial theorems. In this appendix we derive the local virial theorems which are valid for a single body A and relate its internal kinematic and structural properties to the external gravitational field.
The virial theorems are required for transformations of the post-Newtonian terms in section \ref{rsimt129} with the purpose of bringing them to a form of the variational derivative. The virial theorems are also instrumental for transforming the multipolar terms in the Lagrangians from the local to global frames as explained in section \ref{746352424}. Since the virial theorems are used in the post-Newtonian terms only it suffices to derive them in the Newtonian approximation only. In what follows, we employ the local coordinates $w^\a=(u,w^i)$ adapted to body A.  

It is important to emphasize that all virial theorems given in this appendix and applied in the different sections of the present paper, are valid solely on-shell because they intimately rely upon the use of the local equations of motion of matter. Hence, they do not work off-shell and cannot be used for simplifying some terms in the Lagrangian before taking the variational derivative. 

\subsection{Microscopic equations of motion}\label{kuku876}
The virial theorems follow directly from the microscopic equations of motion of matter of body A. There are three basic microscopic equations of motion. 
The first one is just a second Newton's law written down for an element of matter of body A. It reads,  
\ba \label{hn73ax}
\rho^*\frac{d{\nu}^i}{du}=\rho^*\pd_i U_\A-\pd_j\sigma^{ij}+\rho^*\sum_{l=0}^\infty\frac{1}{l!}{\cal Q}_{iL}w^L\;,
\ea
where $\rho^*=\rho^*(u,{\bm w})$ is the invariant mass density of matter, $\nu^i=\nu^i(u,{\bm w})$ is the spatial velocity of the element of matter, $U_\A=U_\A(u,{\bm w})$ is the Newtonian gravitational potential of body A alone, ${\cal Q}_L$ are the external gravitoelectric multipoles defined in \eqref{jje8c}, the overdot denotes a total time derivative with respect to the local coordinate time $u$ of body A, and the total time derivative in the local coordinates is
\ba 
\frac{d}{du}&\equiv&\frac{\pd}{\pd u}+\nu^k\frac{\pd}{\pd w^k}\;.
\ea

Another equation is a thermodynamic (first) law of conservation of internal energy,
\ba 
\rho^*\frac{d\Pi}{du}+\sigma^{ij}\pd_j\nu^i&=&0\;,
\ea
where $\Pi=\Pi(u,{\bm w})$ is the internal thermodynamic energy, $\sigma^{ij}=\sigma^{ij}(u,{\bm w})$ is the stress tensor.
The third equation is the equation of continuity,
\ba\label{xf3sa6}
\frac{d{\rho}^*}{du}+\rho^*\pd_i\nu^i&=&0\;,
\ea
which is exact. Equation \eqref{xf3sa6} tells us that 
\ba\label{pow37c}
\frac{d}{du}\int_{{\cal V}_\A} \rho^*d^3w=0\;,
\ea
which is a law of conservation of the number of particles in the interior of body A.

\subsection{Scalar Virial Theorem}
Let us define scalar integrals
\ba\label{ui8hx}
\mathfrak{T}_\A&=&\frac12\int_{{\cal V}_\A} \rho^* \nu^2 d^3w\;,\\\label{kilc6}
\mathfrak{U}_\A&=&-\frac12\int_{{\cal V}_\A}\rho^*U_\A d^3w=-\frac12\int_{{\cal V}_\A}\rho^*\rho'^*|{\bm w}-{\bm w}'|^{-1}d^3wd^3w'\;,\\\label{0ob2}
\mathfrak{P}_\A&=&\int_{{\cal V}_\A} \rho^*\Pi d^3w\;,
\ea
where $\mathfrak{T}_\A$ is the total kinetic energy of the internal motions of matter of body A, $\mathfrak{U}_\A$ is the total energy of gravitational self-interaction of body A, and $\mathfrak{P}_\A$ is the free thermodynamic energy of matter of body A. Making use of microscopic equations of motion \eqref{hn73ax} of matter of body A, it is straightforward to prove a scalar virial theorem relating the above three quantities, by taking time derivatives from each of them and adding the derivatives all together. It yields the law of the time evolution of the general-relativistic mass of body A defined in \eqref{grmass}. After accounting for \eqref{pow37c}, we have
\ba \label{aa55zz77}
\dot{\mathfrak m}_\A=\dot{\mathfrak{T}_\A}+\dot{\mathfrak{U}_\A}+\dot{\mathfrak{P}_\A}&=&\sum_{l=1}^\infty\frac{1}{l!}Q_L\dot{\M}^L_\A\;,
\ea
where the multipole moments $\M^L_\A$ are defined in \eqref{klz2a1}.
This equation is employed in section \ref{iop23az} for computing the time derivative from the inertial mass of body A.

\subsection{Vector Virial Theorem}\label{fcstrevckivt}
We consider time derivatives from two {\it non-canonical} dipoles 
\ba\label{twodip8752}
{\cal N}^i_\A&=&\int_{{\cal V}_\A}\rho^*w^2w^id^3w\;,\\\label{y3v46z31xq}
{\cal R}^i_\A&=&\int_{{\cal V}_\A}\rho^*\nu^jw^jw^id^3w\;.
\ea
Let's introduce the following notations for vector integrals
\ba\label{eeeeeeeee7}
\mathfrak{T}^i_\A&=&\frac12\int_{{\cal V}_\A} \rho^* \nu^2 w^i d^3w\;,\\\label{qqqqqqq9}
\mathfrak{U}^i_\A&=&-\frac12\int_{{\cal V}_\A}\rho^*U_\A w^i d^3w\\\label{hhhhhhhh5}
\mathfrak{S}^i_\A&=&\int_{{\cal V}_\A} \sigma^{kk}w^i d^3w\;.
\ea
Vector virial theorem states \citep[Eq. 169]{k2019PRD} that 
\ba\label{vect46z3a}
\frac65\dot{\cal R}^i_\A-\frac1{10}\ddot{\cal N}^i&=&2\mathfrak{T}^i_\A+\mathfrak{U}^i_\A+\mathfrak{S}^i_\A
+\sum_{l=1}^\infty\frac{1}{(l-1)!}{\cal Q}_L\M^{iL}_\A-\frac12\sum_{l=1}^\infty\frac{1}{(2l+3)l!}{\cal Q}_{iL}{\cal N}^{L}_\A\;.
\ea
This equation is used in computation of an {\it active} dipole moment in section \ref{nhuste432}.

\subsection{Tensor Virial Theorem}
Let us now introduce tensor integrals
\ba\label{ui319}
{\mathfrak I}_\A^{ij}&=&\int_{{\cal V}_\A} \rho^* w^i w^j d^3w\;,\\
\mathfrak{T}_\A^{ij}&=&\frac12\int_{{\cal V}_\A} \rho^* \nu^i \nu^j d^3w\;,\\
\mathfrak{U}_\A^{ij}&=&-\frac12\int_{{\cal V}_\A}\frac{\rho^*\rho'^*(w^i-w'^i)(w^j-w'^j)}{|{\bm w}-{\bm w}'|^3}d^3wd^3w'\;,\\\label{op25xc2}
\mathfrak{S}_\A^{ij}&=&\int_{{\cal V}_\A} \sigma^{ij} d^3w\;.
\ea
Their scalar counterparts obtained by contraction of the two indices are
\ba\label{ui320}
{\cal N}_\A&=&\int_{{\cal V}_\A} \rho^* |{\bm w}|^2d^3w\;,\\\label{ui320bbb}
\mathfrak{S}_\A&=&\int_{{\cal V}_\A} \sigma^{kk} d^3w\;,
\ea
and the remaining two scalars are given above in \eqref{ui8hx}, \eqref{kilc6}.

Now, we take a second time derivative from ${\mathfrak{I}}_\A^{ij}$ and use \eqref{hn73ax}.
We get a tensor virial theorem
\ba\label{ec5}
\frac12\ddot{\mathfrak I}^{ij}_\A&=&\sum_{l=0}^\infty\frac1{l!}Q_L{}^{(i}_{\phantom A}\M^{j)L}_\A+\sum_{l=0}\frac1{l!(2l+3)}Q_{ijL}{\cal N}^{L}_\A+2\mathfrak{T}_\A^{ij}+\mathfrak{U}_\A^{ij}+\mathfrak{S}_\A^{ij}\;,
\ea
Contracting over two indices yields a scalar virial theorem for the central moment of inertia
\ba\label{ec7aa}
\frac12\ddot{\cal N}_\A
&=&\sum_{l=1}^\infty\frac{1}{(l-1)!}Q_L \M^L_\A+2\mathfrak{T}_\A+\mathfrak{U}_\A+\mathfrak{S}_\A\;.
\ea
After substituting the gravitoelectric potential ${\cal Q}_L$ from \eqref{jje8c}, and reshuffling terms in \eqref{ec5} and \eqref{ec7aa} we obtain more explicit tensor and scalar virial theorem
\ba \label{jex61az8}
\frac12\ddot{\mathfrak I}^{ij}_\A-2\mathfrak{T}_\A^{ij}-\mathfrak{U}_\A^{ij}-\mathfrak{S}_\A^{ij}&=&\sum_{\B\neq\A}\sum_{l=0}^\infty\sum_{n=0}^\infty\frac{(-1)^n}{l!n!}\M^N_\B \M^{L(i}_\A\pd^{j)LN}_{\phantom{A}} R^{-1}_{\A\B}+\sum_{\B\neq\A}\sum_{l=0}^\infty\sum_{n=0}^\infty\frac{(-1)^n}{l!n!}\frac{{\cal N}^L_\A \M^N_\B}{2l+3}\pd_{ijLN} R^{-1}_{\A\B}\;,
\\
\label{ec7}
\frac12\ddot{\cal N}_\A-2\mathfrak{T}_\A-\mathfrak{U}_\A-\mathfrak{S}_\A&=&\sum_{\B\neq\A}\sum_{l=1}^\infty\sum_{n=0}^\infty\frac{(-1)^n}{(l-1)!n!} \M^L_\A\M^N_\B\pd_{LN} R^{-1}_{\A\B}\;,
\ea
which connects the internal and external dynamic variables. The scalar and tensor virial theorems are used for calculation of the Lagrangian in section \ref{rsimt129} and in section \ref{mw71cz52} for comparison of the Lagrangians derived by different authors.  
\end{appendix}
\bibliographystyle{unsrtnat}
\bibliography{Lagrangian_references}% Produces the bibliography via BibTeX.
\end{document}